\documentclass[twocolumn,showpacs,prb,aps]{revtex4}
\usepackage{amsmath}
\usepackage{graphicx}
\usepackage{dcolumn}
\newcommand{\be}{\begin{equation}}
\newcommand{\ee}{\end{equation}}
\newcommand{\bea}{\begin{eqnarray}}
\newcommand{\eea}{\end{eqnarray}}

\begin{document}

\title{Crystal Growth and Physical Properties of SrCu$_2$As$_2$, SrCu$_2$Sb$_2$ and BaCu$_2$Sb$_2$}

\author{V. K. Anand}
\email{vanand@ameslab.gov}
\author{P. Kanchana Perera}
\author{Abhishek Pandey}
\author{R. J. Goetsch}
\author{A. Kreyssig}
\author{D. C. Johnston}
\email{johnston@ameslab.gov}
\affiliation {Ames Laboratory and Department of Physics and Astronomy, Iowa State University, Ames, Iowa 50011}

\date{\today}

\begin{abstract}

We report the growth of single crystals of SrCu$_2$As$_2$, SrCu$_2$Sb$_2$, SrCu$_2$(As$_{0.84}$Sb$_{0.16}$)$_2$ and BaCu$_2$Sb$_2$ using the self-flux technique and their structural, magnetic, thermal and transport properties that were investigated by powder x-ray diffraction (XRD), magnetic susceptibility $\chi$, specific heat $C_{\rm p}$ and electrical resistivity $\rho$ measurements versus temperature $T$ from 1.8 to 350~K\@. Rietveld refinements of XRD patterns for crushed crystals confirm that SrCu$_2$As$_2$ crystallizes in the ThCr$_2$Si$_2$-type body-centered tetragonal structure (space group $I4/mmm$) and SrCu$_2$Sb$_2$ crystallizes in the CaBe$_2$Ge$_2$-type primitive tetragonal structure (space group $P4/nmm$). However, as reported previously, BaCu$_2$Sb$_2$ is found to have a large unit cell consisting of three blocks.  Here a ThCr$_2$Si$_2$-type block is sandwiched between two CaBe$_2$Ge$_2$-type blocks along the $c$-axis with an overall symmetry of $I4/mmm$, as reported, but likely with a monoclinic distortion. The $\chi$ data of all these compounds are diamagnetic and reveal nearly $T$-independent anisotropic behavior.  The $\chi$ of SrCu$_2$As$_2$ is found to be larger in the $ab$-plane than along the $c$-axis, as also previously reported for pure and doped BaFe$_2$As$_2$, whereas the $\chi$ values of SrCu$_2$Sb$_2$ and BaCu$_2$Sb$_2$ are larger along the $c$-axis. This difference in anisotropy appears to arise from the differences between the crystal structures.  The finite values of the Sommerfeld linear specific heat coefficients $\gamma$ and the $T$ dependences of $\rho$ reveal metallic character of all four compounds. The electronic and magnetic properties indicate that these compounds are $sp$ metals with Cu in the nonmagnetic $3d^{10}$ electronic configuration corresponding to the oxidation state Cu$^{+1}$, as previously predicted theoretically for SrCu$_2$As$_2$ by D.~J.~Singh [Phys. Rev. B {\bf 79}, 153102 (2009)].  We present a brief review of theoretical and experimental work on the doping character of transition metals for Fe in BaFe$_2$As$_2$.  The \mbox{As--As} covalent interlayer bond distances in the collapsed-tetragonal (Ca,Sr,Ba)Cu$_2$As$_2$ compounds are much shorter than the nonbonding As--As distances in BaFe$_2$As$_2$.  Thus the electronic character of the Cu and the strength of the As--As interlayer bonding are both expected to drastically change between weakly Cu-substituted BaFe$_2$As$_2$ and pure BaCu$_2$As$_2$, perhaps via a first-order lattice instability such as a miscibility gap in the Ba(Fe$_{1-x}$Cu$_x)_2$As$_2$ system.

\end{abstract}


\maketitle

\section{Introduction}

The observation of superconductivity with transition temperatures $T_{\rm c}$ up to 38~K in 122-type iron pnictides (e.g., $A_{1-x}$K$_x$Fe$_2$As$_2$ compounds, $A$ = Ca, Sr, Ba, and Eu)\cite{Rotter, ChenCPL, Jeevan, Sasmal, Wu} having the layered ThCr$_2$Si$_2$-type structure, shortly after the discovery of superconductivity with $T_{\rm c}$ up to to 55~K in 1111-type oxypnictides (e.g., $Ln$FeAsO$_{1-x}$F$_x$ compounds, $Ln$ = La, Ce, Nd and Sm),\cite{Kamihara, RenCPL, ChenXH, ChenPRL, RenEPL} triggered research activities worldwide and it is now believed that the iron pnictide compounds represent a new class of high-$T_c$ superconductors.\cite{Johnston2010, RevCanfield, Mandrus} Superconductivity in these iron arsenides lies in close proximity to antiferromagnetic (AFM) itinerant spin density wave (SDW) transitions that can be easily suppressed by substitutions at the $A$, Fe and/or As sites.\cite{Johnston2010} The generic phase diagram for the emergence of superconductivity
  upon suppression of the SDW transition in iron arsenides is qualitatively similar to that of the high-$T_c$ cuprates, although with the important difference that the cuprate parent compounds are local moment antiferromagnetic insulators whereas the iron arsenide parent compounds are itinerant SDW semimetals. \cite{Johnston2010, RevCanfield, Mandrus, Johnston1997, Orenstein, Damascelli, Lee} Interestingly, as noted above superconductivity in iron arsenides can be induced by substitutions on the Fe site by other transition metals, which is quite different from the high-$T_c$ cuprate superconductors where such substitutions at the Cu site never induce superconductivity and indeed can lead to a rapid suppression of the superconductivity.

In the fiducial 122-type compound BaFe$_2$As$_2$, the formal oxidation states of the atoms are assigned as Ba$^{+2}$, Fe$^{+2}$ and As$^{-3}$ so the Fe atoms are formally in the $3d^6$ electronic configuration as also occurs in the 1111-type parent compounds.\cite{Johnston2010}  It has been observed that partial Co and Ni substitutions at the Fe site in BaFe$_2$As$_2$ induce superconductivity\cite{Sefat2008, WangC, Li}  with $T_{\rm c}$ up to 25~K whereas no superconductivity is induced by Mn (Refs.~\onlinecite{Kasinathan2009, Liu, Kim2010}) or Cr (Refs.~\onlinecite{Sefat2009a}, \onlinecite{Marty}) substitutions. These findings suggest that the superconductivity might have some relation with the average number of 3$d$ conduction electrons of the transition metal atom.  That is, hole-doping with a smaller number of 3$d$ electrons (Mn$^{+2}$, Cr$^{+2}$) than that of Fe$^{+2}$ does not induce superconductivity, whereas electron-doping with more 3$d$ electrons than Fe$^{+2}$ (Co$^{+2}$ and Ni$^{+2}$) does induce superconductivity.  From this point of view divalent copper Cu$^{+2}$ with the 3$d^9$ electronic configuration and three more $d$-electrons than Fe should be a strong electron dopant for iron arsenide superconductors.  However, Cu-doping for Fe in BaFe$_2$As$_2$ has been found to yield strongly suppressed $T_{\rm c} \lesssim 2$~K in Ba(Fe$_{1-x}$Cu$_x$)$_2$As$_2$, and then only in a very limited concentration range near $x=0.044$, even though the Cu doping suppresses the structural/SDW transition of the parent compound BaFe$_2$As$_2$.\cite{Ni2010}  Simultaneous Co- and Cu-doping for Fe in BaFe$_2$As$_2$ leads to superconductivity with higher values of $T_{\rm c}$ for $y > 0$ in Ba(Fe$_{1-x-y}$Co$_y$Cu$_x$)$_2$As$_2$ compounds than for $y=0$ (Ref.~\onlinecite{Ni2010}).  On the other hand, the application of pressure to $A$Fe$_2$As$_2$ ($A$ = Sr or Ba) compounds or isoelectronic substitutions of Ru for Fe or P for As in BaFe$_2$As$_2$, which nominally do not result in charge doping, can also induce high-$T_{\rm c}$ superconductivity.\cite{Johnston2010}

A +2 oxidation state of copper also plays a key role in high-$T_c$ cuprates such as in the antiferromagnetic insulator parent compound ${\rm La_2CuO_4}$ with a N\'eel temperature of 325~K where the Cu$^{+2}$ $3d^9$ ion carries a local magnetic moment with spin $S = 1/2$ due to the single hole in the Cu $3d$-shell.\cite{Johnston1997, Orenstein, Damascelli, Lee}  Then hole doping such as in La$_{2-x}$Sr$_x$CuO$_4$ leads to high-$T_{\rm c}$ superconductivity where the localized Cu magnetic moment is retained and the antiferromagnetic correlations between them are likely involved with the superconducting mechanism.\cite{Johnston1997}  Considering this feature, we speculated that a 122-type compound containing Cu$^{+2}$ local moments with $S$ = 1/2 completely replacing the Fe in the Fe-sites of the 122-type compounds might bridge the gap between the two families of high-$T_c$ superconductors and enrich our knowledge of the physics of high-$T_c$ superconductivity in general. This would complement our recent discovery that Ba$_{1-x}$K$_x$Mn$_2$As$_2$ is such an antiferromagnetic local-moment metal where the Mn$^{+2}$ ions have a large spin $S = 5/2$.\cite{Johnston_PRB, Pandey_arxiv}

Electronic structure calculations for SrCu$_2$As$_2$ and BaCu$_2$As$_2$ by D.~J.~Singh predicted that the Cu 3$d$ bands are narrow and are located about 3~eV below the Fermi energy $E_{\rm F}$.\cite{Singh}  These calculations also found that there is little contribution to the density of states at $E_{\rm F}$ from the Cu~3$d$ orbitals.  Therefore these compounds are predicted to be $sp$-band metals with the Cu atoms having a formal oxidation state of Cu$^{+1}$ and a nonmagnetic and chemically inert $3d^{10}$ electronic configuration.\cite{Singh}

To investigate these issues, we synthesized single crystals of the previously known\cite{Pfisterer, Dunner, Cordier, Pfisterer1983} Cu-based compounds SrCu$_2$As$_2$, SrCu$_2$Sb$_2$ and BaCu$_2$Sb$_2$ and measured their structural, magnetic, thermal and electronic transport properties, and report here our results.  We also synthesized and studied crystals of SrCu$_2$(As$_{0.84}$Sb$_{0.16}$)$_2$.  To our knowledge, detailed investigations of the physical properties of these compounds have not been previously carried out.  Pfisterer and Nagorsen reported in 1983 that the $\chi$ of a polycrystalline sample of SrCu$_2$As$_2$ was diamagnetic and nearly independent of $T$ from 80 to 400~K.\cite{Pfisterer1983}

Our measurements confirm the above theoretical prediction\cite{Singh} that SrCu$_2$As$_2$ is an $sp$-band metal and we find that it also applies to the other three compounds studied here as well.  Thus the Cu ions have a formal oxidation state of Cu$^{+1}$ with a nonmagnetic $3d^{10}$ electronic configuration.  Therefore these compounds cannot be considered to be a bridge to the high-$T_{\rm c}$ cuprates.

A matter of great current interest is the nature, extent and even existence of current carrier doping on substituting various transition metals for Fe in the high-$T_{\rm c}$ parent compounds ${\rm (Ca,Sr,Ba)Fe_2As_2}$ because of their relevance to understanding the changes in the magnetic and superconducting transition temperatures with concentration of the substituting elements.  We will give a brief  overview of this topic and then discuss how our results on SrCu$_2$As$_2$ relate to the previous results.  We suggest that the electronic character of the Cu and the strength of the As--As interlayer bonds both strongly change between weakly Cu-substituted BaFe$_2$As$_2$ and pure BaCu$_2$As$_2$, perhaps via a first-order  lattice instability such as a miscibility gap in the Ba(Fe$_{2-x}$Cu$_x)_2$As$_2$ system.

The remainder of this paper is organized as follows. The experimental details are described in Sec.~\ref{ExpDetails}. Section~\ref{Crystallography} describes the crystallographic studies. The physical properties of SrCu$_2$As$_2$, SrCu$_2$Sb$_2$, SrCu$_2$(As$_{0.84}$Sb$_{0.16}$)$_2$ and BaCu$_2$Sb$_2$ are described in Secs.~\ref{SrCu2As2}, \ref{SrCu2Sb2}, \ref{SrCu2AsSb} and \ref{BaCu2Sb2}, respectively.  A discussion of our results in the context of previous work is given in Sec.~\ref{Discussion}.  In Sec.~\ref{Sec:PnicBonding} we discuss the occurrence or not of $X$--$X$ interlayer covalent bonding in (Ca,Sr,Ba)$M_2X_2$ compounds with the ThCr$_2$Si$_2$ structure, where $M$ = Cr, Mn, Fe, Co, Ni and Cu and $X$ = As or P, and conclude that SrCu$_2$As$_2$ is in the collapsed tetragonal phase with short As--As interlayer bonds.  In Sec.~\ref{Sec:MagPropCorr} we discuss the relationships between the $X$--$X$ bonding and the magnetic properties.  A brief review of the doping properties that occur upon substituting other transition metals for Fe in BaFe$_2$As$_2$ is given in Sec.~\ref{Sec:CuDoping}, together with a discussion of how our results on SrCu$_2$As$_2$ impact this discussion.  A summary and our conclusions are given in Sec.~\ref{Conclusion}.

\section{\label{ExpDetails} EXPERIMENTAL DETAILS}

\begin{figure}
\includegraphics[width=3in]{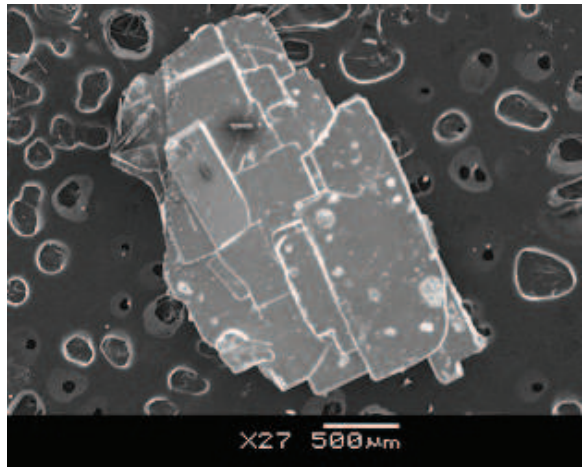}
\caption{\label{fig:SrCuAsSb_SEM} (Color online) Scanning electron micrograph of a cluster of intergrown SrCu$_2$(As$_{0.84}$Sb$_{0.16}$)$_2$ crystals.}
\end{figure}

\begin{table}
\caption{\label{tab:table1} The average chemical compositions (atomic~\%) obtained from EDX analyses of SrCu$_2$As$_2$, SrCu$_2$Sb$_2$, SrCu$_2$(As$_{0.84}$Sb$_{0.16}$)$_2$ and BaCu$_2$Sb$_2$ crystals. The systematic error in the EDX measurements is estimated to be $\sim 2$--3 atomic \%.}
\begin{ruledtabular}
\begin{tabular}{lcccc}

Compound &  Sr or Ba  &	 Cu 	&	As	&	Sb	 \\	
\hline
SrCu$_2$As$_2$ & 18.3 & 41.5 & 40.2\\		
SrCu$_2$Sb$_2$ & 19.4 & 38.9 & & 41.8 \\
SrCu$_2$(As$_{0.84}$Sb$_{0.16}$)$_2$ & 18.8 & 40.9 & 34.0 & 6.3\\
BaCu$_2$Sb$_2$  & 19.5 & 39.4 & & 41.1\\	
\end{tabular}
\end{ruledtabular}
\end{table}

Single crystals of SrCu$_2$As$_2$, SrCu$_2$Sb$_2$, SrCu$_2$(As$_{0.84}$Sb$_{0.16}$)$_2$ and BaCu$_2$Sb$_2$ were grown by the self-flux growth technique using the high purity elements Sr (99.95\%) and Ba (99.99\%) from Sigma Aldrich, and Cu (99.999\%),  As (99.99999\%) and Sb (99.999\%) from Alfa Aesar.  Prereacted CuAs or CuSb were used as flux. For the growth of SrCu$_2$As$_2$ and SrCu$_2$Sb$_2$ crystals, the Sr and flux (CuAs or CuSb) were taken in a 1:5 molar ratio, placed in alumina crucibles and sealed inside evacuated quartz tubes. The crystal growth was carried out by heating the samples to 1100~$^\circ$C at a rate of 60~$^\circ$C/h, holding there for 12~h and then cooling to 800~$^\circ$C at a rate of 2.5~$^\circ$C/h. The crystals were separated from the flux by decanting the flux with a centrifuge at 800~$^\circ$C, yielding shiny plate-like crystals. While the sizes of the SrCu$_2$Sb$_2$ crystals were typically $3 \times 2.5 \times 0.3$~mm$^3$, one of the SrCu$_2$As$_2$ crystals was larger ($6 \times 3 \times 1$~mm$^3$)\@. In our attempt to grow crystals of the mixed compound SrCu$_2$AsSb starting with Sr:CuAs:CuSb in a 1:2.5:2.5 molar ratio with the same heating profile as above, we instead obtained SrCu$_2$(As$_{0.84}$Sb$_{0.16}$)$_2$ crystals as determined from energy-dispersive x-ray (EDX) analysis. In this case scanning electron microscope (SEM) images revealed intergrowths of shiny plate-like crystals stacked on top of each other (Fig.~\ref{fig:SrCuAsSb_SEM}). The physical properties of this composition were measured on such an assembly of stacked crystals.

For the growth of BaCu$_2$Sb$_2$ crystals, Ba and CuSb were taken in a 1:5 molar ratio and heated to 1000~$^\circ$C at a rate of 80~$^\circ$C/h, held there for 15 h, cooled to 700~$^\circ$C at a rate of 2.5~$^\circ$C/h, and then the flux was decanted at 700~$^\circ$C to obtain crystals of typical size $2.5 \times 2 \times 0.5$~mm$^3$\@. We did not succeed in synthesizing single crystals of BaCu$_2$As$_2$ using the self-flux growth or with Sn as a flux with similar heating profiles.

The chemical compositions of the crystals were determined using a JEOL SEM equipped with an EDX analyzer, which showed that these are close to the respective stoichiometric compositions SrCu$_2$As$_2$, SrCu$_2$Sb$_2$, SrCu$_2$(As$_{0.84}$Sb$_{0.16}$)$_2$ and BaCu$_2$Sb$_2$. The average chemical compositions obtained from the EDX analyses of the single crystals are presented in Table~\ref{tab:table1}.  The crystal structures of the samples were determined by powder x-ray diffraction (XRD) measurements that were carried out on powdered crystals using a Rigaku Geigerflex x-ray diffractometer and Cu~K$_\alpha$ radiation.

The specific heat measurements were performed by the relaxation method in zero applied magnetic field in a Quantum Design physical properties measurement system (PPMS). The electrical resistivity was measured in the $ab$-plane of each crystal in zero applied magnetic field by the standard four-probe ac technique using the ac transport option of the PPMS\@. The 50~$\mu$m diameter platinum wire electrical leads were attached to the crystals using silver epoxy.

Magnetization and magnetic susceptibility measurements were performed using a Quantum Design superconducting quantum interference device magnetic properties measurement system (MPMS).  The single crystals were mounted either on a small piece of plastic or on a thin quartz rod using a small amount of GE 7031 varnish. We separately measured the magnetic moment of the sample holder (quartz rod/plastic and GE varnish) and subtracted it from the measured magnetic moment to obtain the sample contribution.

\begin{figure}
\includegraphics[width=3in]{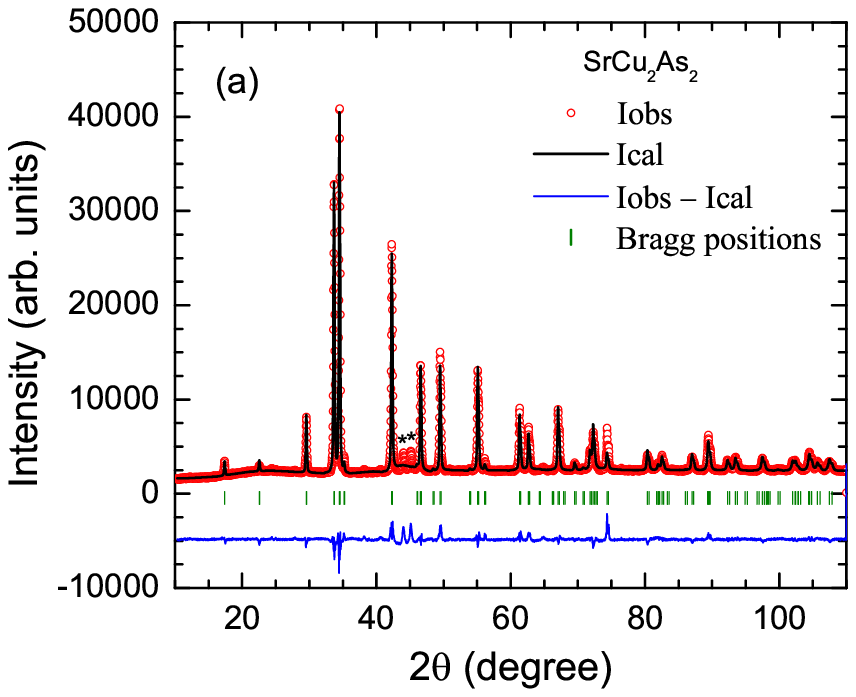}\vspace{0.05in}
\includegraphics[width=3in]{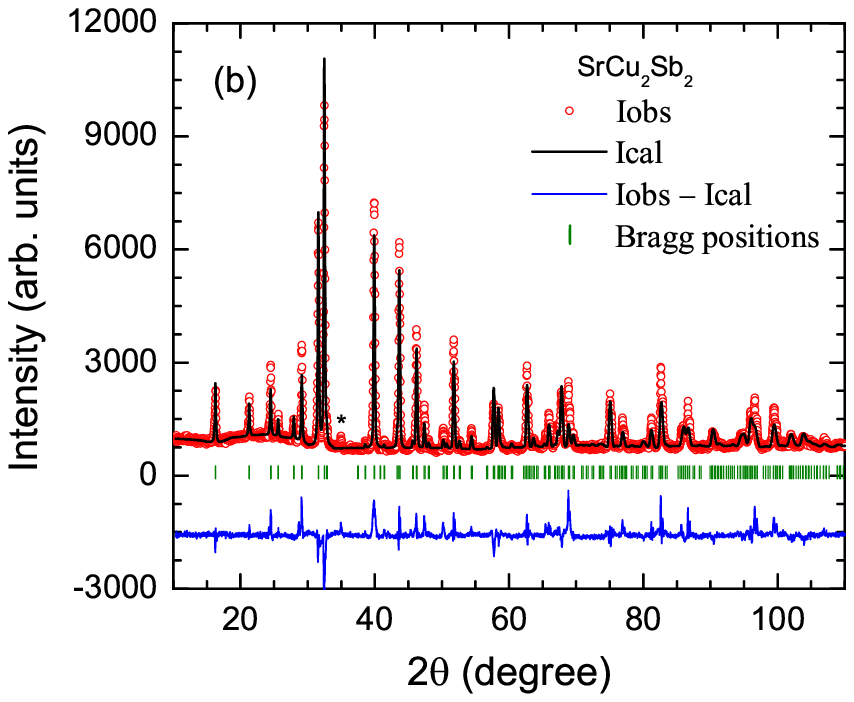}\vspace{0.05in}
\includegraphics[width=3in]{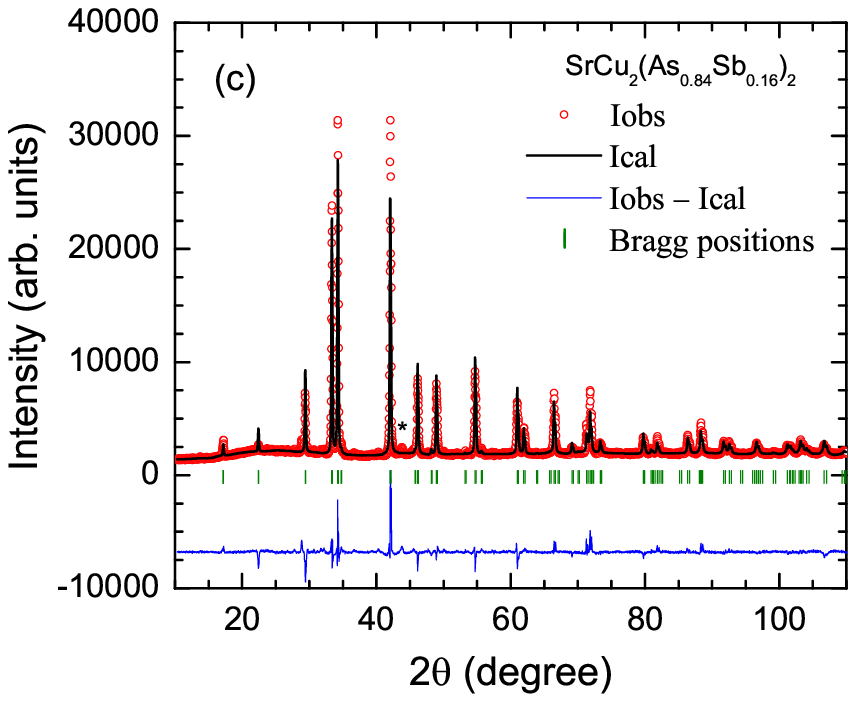}
\caption{\label{fig:SrCu2AsSb_XRD} (Color online) Powder x-ray diffraction patterns of (a) SrCu$_2$As$_2$, (b) SrCu$_2$Sb$_2$ and (c) SrCu$_2$(As$_{0.84}$Sb$_{0.16}$)$_2$ recorded at room temperature. In (a) and (c), the solid lines through the experimental points are the Rietveld refinement profiles calculated for the ThCr$_2$Si$_2$-type body-centered tetragonal structure (space group $I4/mmm$), and in (b) the solid line through the data points is the Rietveld refinement profile for the CaBe$_2$Ge$_2$-type primitive tetragonal structure (space group $P4/nmm$).  In (a), (b) and (c), the short vertical bars mark the fitted Bragg peak positions. The lowermost curves represent the differences between the experimental and calculated intensities. The unindexed peaks marked with stars correspond to peaks from the flux.}
\end{figure}

\begin{figure}
\includegraphics[width=2.9in]{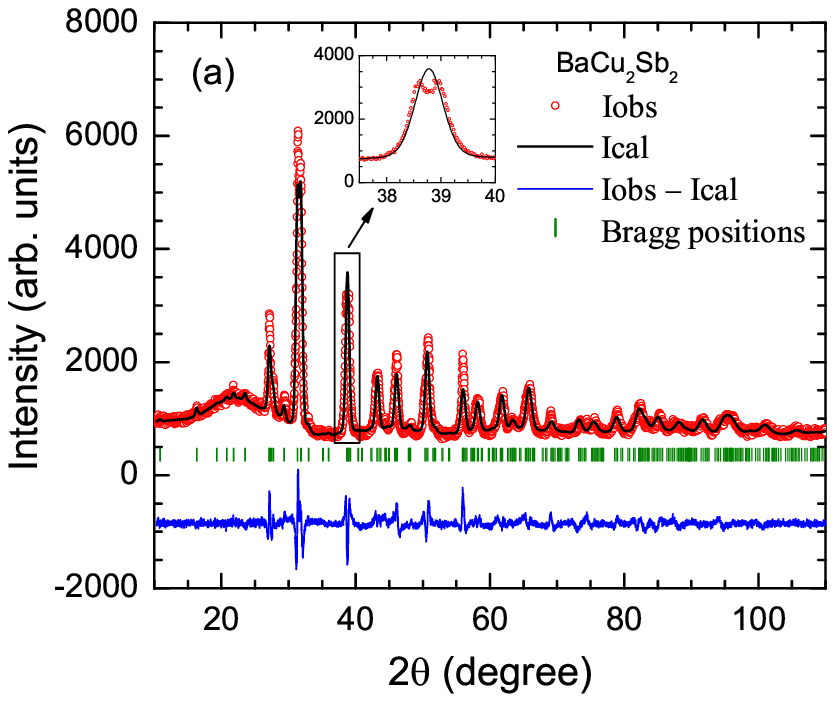}\vspace{0.05in}
\includegraphics[width=2.9in]{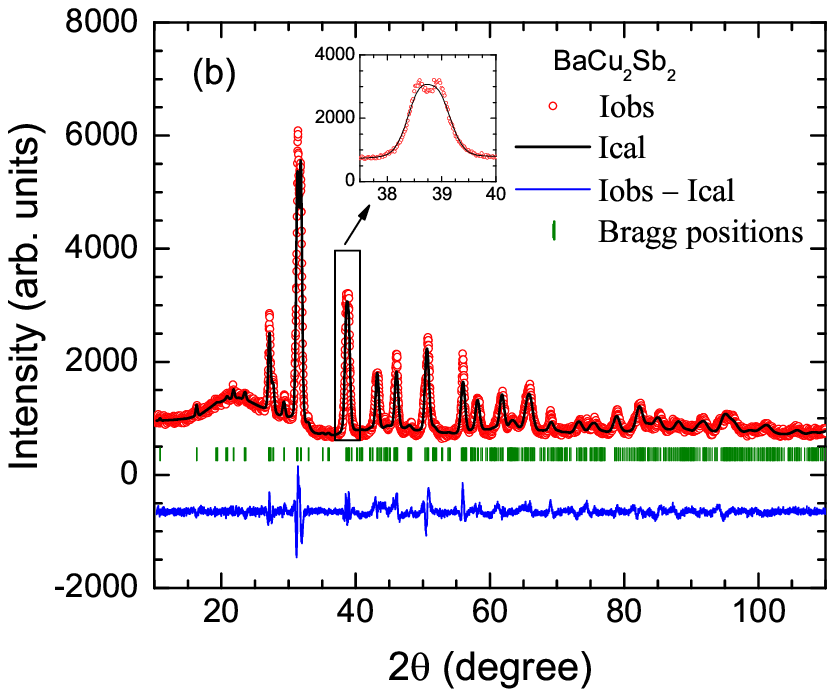}\vspace{0.05in}
\includegraphics[width=2.9in]{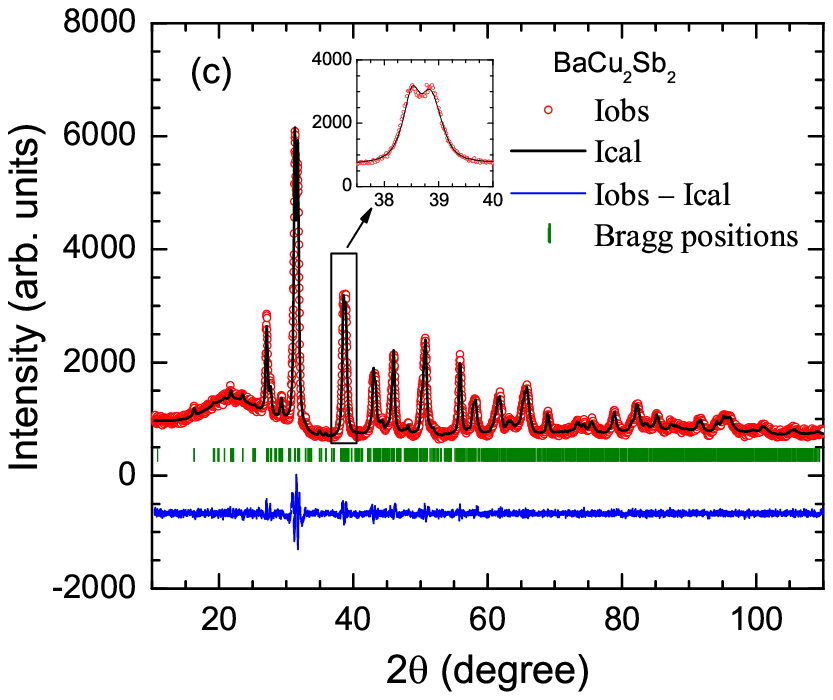}
\caption{(Color online) Powder x-ray diffraction pattern of BaCu$_2$Sb$_2$ recorded at room temperature. (a) The solid line through the experimental points is the Rietveld refinement profile calculated for the tetragonal structure (space group $I4/mmm$).  (b) Rietveld refinement profile calculated for an orthorhombic  structure (space group \emph{Immm}). (c) Le Bail profile fit for a monoclinic structure (space group $P2_1/c$). In (a), (b) and (c), the short vertical bars mark the fitted Bragg peak positions, the lowermost curves represent the differences between the experimental and calculated intensities, and the insets show a small section of the XRD pattern highlighting an exemplary splitting and fitting of one of the peaks.}
\label{fig:BaCu2Sb2_XRD}
\end{figure}

\begin{figure}
\includegraphics[width=3.2in]{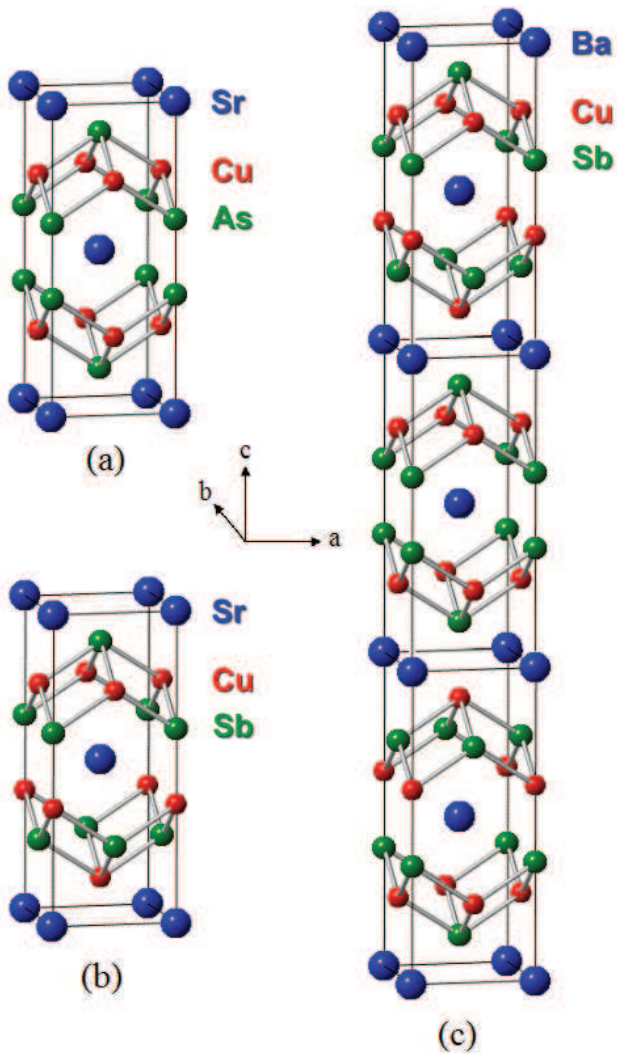}
\caption{(Color online) (a)  ${\rm ThCr_2Si_2}$-type body-centered tetragonal  crystal structure ($I4/mmm$) of SrCu$_2$As$_2$. (b) CaBe$_2$Ge$_2$-type primitive tetragonal crystal structure ($P4/nmm$) of SrCu$_2$Sb$_2$. The order of the Cu and Sb layers in the lower half of the unit cell is reversed in (b) compared with (a). In order to compare the structures, the origin of the SrCu$_2$Sb$_2$ unit cell is shifted by (1/4, 1/4, 1/4) from that for the $P4/nmm$ space group. (c) Body-centered tetragonal representation of BaCu$_2$Sb$_2$ (space group $I4/mmm$), consisting of a central ${\rm ThCr_2Si_2}$-type block sandwiched between two CaBe$_2$Ge$_2$-type blocks, resulting in a large $c$-axis parameter $c = 32.586$~\AA\ (see Table~\ref{tab:table2}).}
\label{fig:structure_fig} 
\end{figure}

\begin{table*}
\caption{\label{tab:table2} Crystallographic and Rietveld refinement parameters obtained from powder XRD data for crushed crystals of SrCu$_2$As$_2$, SrCu$_2$Sb$_2$, SrCu$_2$(As$_{0.84}$Sb$_{0.16}$)$_2$ and BaCu$_2$Sb$_2$.  If no value for the lattice parameter $b$ is listed, then $b=a$.  Error bars for the last digit of a quantity are given in parentheses and literature references are given in square brackets.}
\begin{ruledtabular}
\begin{tabular}{llllll}
 & SrCu$_2$As$_2$ & SrCu$_2$Sb$_2$ & SrCu$_2$(As$_{0.84}$Sb$_{0.16}$)$_2$ & BaCu$_2$Sb$_2$ & BaCu$_2$Sb$_2$ \\
\hline
Structure & ThCr$_2$Si$_2$-type & CaBe$_2$Ge$_2$-type & ThCr$_2$Si$_2$-type  \\
 & tetragonal & tetragonal & tetragonal & tetragonal & orthorhombic\\
Space group & $I4/mmm$ & $P4/nmm$ & $I4/mmm$ & $I4/mmm$ & $Immm$\\

Lattice parameters\\

  \hspace{0.8 cm}  $a$ (\AA)            			&  4.2725(1) 	&  4.5162(2)   & 4.2940(1)  & 4.6438(5)  & 4.6224(5)   \\	
  \hspace{0.8 cm}  			          			&  4.279(1) [\onlinecite{Dunner}] 	& 	4.51(1) [\onlinecite{Cordier}]	     & 		  & 4.655(1) [\onlinecite{Dunner}] & 		  \\	
  \hspace{0.8 cm}  			          			&  4.271(1) [\onlinecite{Pfisterer}]	& 		     & 		  &   		& 		  \\	
  \hspace{0.8 cm}  $b$ (\AA)            			&   	&     &   &   & 4.6679(5)  \\	
  \hspace{0.8 cm}  $c$ (\AA)          				&  10.2000(3)	&  10.9008(5)  & 10.3190(4) & 32.600(4) & 32.606(4)  \\
  \hspace{0.8 cm}  			          			&  10.215(1) [\onlinecite{Dunner}]	&  10.91(2) [\onlinecite{Cordier}]		  & 			 & 32.709(6) [\onlinecite{Dunner}]		 & 		  \\
  \hspace{0.8 cm}  			            			&  10.18(2) [\onlinecite{Pfisterer}]	&  		   & 		  & 		  & 		  \\	
  \hspace{0.8 cm}  $V_{\rm cell}$  ({\AA}$^{3}$) 	&  186.19(1) 	&  222.33(2)   & 190.26(1)  & 703.0(1) & 703.5(1) \\

Refinement quality \\

  \hspace{0.8 cm}    $\chi^2$	   & 9.72 & 9.13 & 12.3 & 4.98 & 4.00 \\	
  \hspace{0.8 cm}    $R_{\rm p}$ (\%)  & 3.70 & 6.81 & 4.59 & 5.13 & 4.77  \\
  \hspace{0.8 cm}    $R_{\rm wp}$ (\%) & 5.71 & 9.52 & 7.32 & 6.98 & 6.26 \\

\end{tabular}
\end{ruledtabular}
\end{table*}

\begin{table}
\caption{\label{tab:table3} Atomic coordinates obtained from the Rietveld refinements of powder XRD data for powdered crystals of SrCu$_2$As$_2$, SrCu$_2$Sb$_2$, SrCu$_2$(As$_{0.84}$Sb$_{0.16}$)$_2$ and BaCu$_2$Sb$_2$.}
\begin{ruledtabular}
\begin{tabular}{lclll}
   \hspace{0.6cm}  Atom & Wyckoff   &	 $x$ 	&	$y$	&	$z$	  \\	
   & symbol & \\
\hline
SrCu$_2$As$_2$ ($I4/mmm$) \\				
   \hspace{0.8cm}     Sr & 2a  	&	 0 	&	0	&   0		  \\
   \hspace{0.8cm}     Cu & 4d	&	 0 	&	1/2	&	1/4 	  \\
   \hspace{0.8cm}     As & 4e 	&	 0 	&   0 	&	0.3789(1) \\

SrCu$_2$Sb$_2$ ($P4/nmm$) \\		
   \hspace{0.8cm}     Sr  & 2c &	 1/4 	&	1/4		&   0.2353(5)	 \\
   \hspace{0.8cm}     Cu1 & 2a &	 3/4 	&	1/4		&	0			 \\
   \hspace{0.8cm}     Cu2 & 2c &	 1/4 	&   1/4 	&	0.6330(8) 	 \\
   \hspace{0.8cm}     Sb1 & 2b &	 3/4 	&	1/4		&	1/2 		 \\
   \hspace{0.8cm}     Sb2 & 2c &	 1/4 	&   1/4 	&	0.8707(4)	 \\

SrCu$_2$(As$_{0.84}$Sb$_{0.16}$)$_2$ \\				
   \hspace{0.8cm}     Sr & 2a  	&	 0 	&	0	&   0		  \\
   \hspace{0.8cm}     Cu & 4d	&	 0 	&	1/2	&	1/4 	  \\
   \hspace{0.8cm}     As/Sb & 4e 	&	 0 	&   0 	&	0.3781(2) \\

BaCu$_2$Sb$_2$ ($I4/mmm$)\\		
   \hspace{0.8cm}     Ba1 & 2a &	 0  	&	0	    &   0		     \\
   \hspace{0.8cm}     Ba2 & 4e &	 0  	&	0   	&   0.6654(4)	 \\
   \hspace{0.8cm}     Cu1 & 4e &	 0  	&	0   	&	0.7947(7)	 \\
   \hspace{0.8cm}     Cu2 & 8g &	 0  	&   1/2 	&	0.0830(6) 	 \\
   \hspace{0.8cm}     Sb1 & 4d &	 0  	&	1/2 	&	1/4 		 \\
   \hspace{0.8cm}     Sb2 & 4e &	 0  	&   0   	&	0.1234(3)	 \\
   \hspace{0.8cm}     Sb3 & 4e &	 0  	&   0   	&	0.5428(3)	 \\

BaCu$_2$Sb$_2$ ($Immm$)\\		
   \hspace{0.8cm}     Ba1 & 2a &	 0  	&	0	    &   0		     \\
   \hspace{0.8cm}     Ba2 & 4i &	 0  	&	0   	&   0.6650(3)	 \\
   \hspace{0.8cm}     Cu1 & 4i &	 0  	&	0   	&	0.7933(6)	 \\
   \hspace{0.8cm}     Cu2 & 4j &	 1/2  	&   0 	    &	0.0808(9) 	 \\
   \hspace{0.8cm}     Cu3 & 4j &	 1/2  	&   0 	    &	0.5880(8) 	 \\
   \hspace{0.8cm}     Sb1 & 4j &	 1/2  	&	0   	&	0.7484(8)	 \\
   \hspace{0.8cm}     Sb2 & 4i &	 0  	&   0   	&	0.1230(3)	 \\
   \hspace{0.8cm}     Sb3 & 4i &	 0  	&   0   	&	0.5426(3)	 \\
\end{tabular}
\end{ruledtabular}
\end{table}

\section{\label{Crystallography} Crystallography}

The XRD data on crushed single crystals of SrCu$_2$As$_2$, SrCu$_2$Sb$_2$, SrCu$_2$(As$_{0.84}$Sb$_{0.16}$)$_2$ and BaCu$_2$Sb$_2$ were analyzed by Rietveld refinement using the {\tt FullProf} \cite{Rodriguez} software. The room-temperature XRD patterns along with the Rietveld fit profiles are presented in Fig.~\ref{fig:SrCu2AsSb_XRD} for SrCu$_2$As$_2$, SrCu$_2$Sb$_2$ and SrCu$_2$(As$_{0.84}$Sb$_{0.16}$)$_2$ and in Fig.~\ref{fig:BaCu2Sb2_XRD} for BaCu$_2$Sb$_2$. The XRD data indicate that the crushed crystals are single-phase. The unindexed peaks marked with stars arise from small amounts of flux attached to the surfaces of the crystals. The Rietveld refinements confirmed that SrCu$_2$As$_2$ crystallizes in the ThCr$_2$Si$_2$-type body-centered tetragonal (bct) structure (space group $I4/mmm$) and SrCu$_2$Sb$_2$ crystallizes in the CaBe$_2$Ge$_2$-type primitive tetragonal structure (space group $P4/nmm$).  As seen in Fig.~\ref{fig:structure_fig}(a), the structure of SrCu$_2$As$_2$ contains stacked square lattices of Cu atoms, as in the high $T_{\rm c}$ cuprates, but the coordination of copper by the anions is quite different in the two types of material.  On the other hand, the structure of SrCu$_2$Sb$_2$ contains two distinct types of Cu square lattice with different lattice parameters that are rotated by 45$^\circ$ from each other and stacked in a specific sequence, as seen in Fig.~\ref{fig:structure_fig}(b).

Both the ThCr$_2$Si$_2$-type and CaBe$_2$Ge$_2$-type structures are ternary derivatives of the BaAl$_4$ structure \cite{Parthe} and consist of layers of $A$ (Th, Ca), $T$ (Cr, Be) and $X$ (Si, Ge) atoms stacked along the crystallographic $c$-axis. While the $A$ atoms form a bct sublattice in both cases, the two structures differ in the arrangement of layers of $T$ and $X$ atoms as can be seen from Figs.~\ref{fig:structure_fig}(a) and~\ref{fig:structure_fig}(b).  In the ThCr$_2$Si$_2$-type structure the layers of atoms are in the order Th-Si-Cr-Si-Th-Si-Cr-Si-Th along the $c$-axis, whereas in the CaBe$_2$Ge$_2$-type structure the order of atomic layers is  Ca-Ge-Be-Ge-Ca-Be-Ge-Be-Ca. Thus in between the consecutive Ca-planes the Ge-Be-Ge slab changes into a Be-Ge-Be slab which leads to a loss of mirror symmetry about the $A$-plane in the CaBe$_2$Ge$_2$-type structure that is present in the ThCr$_2$Si$_2$-type structure. SrCu$_2$(As$_{0.84}$Sb$_{0.16}$)$_2$ is found to crystallize in the bct ThCr$_2$Si$_2$-type structure.

On the other hand the XRD data for BaCu$_2$Sb$_2$ could not be described by either a ThCr$_2$Si$_2$-type or a CaBe$_2$Ge$_2$-type structure. Instead, as reported by D\"{u}nner et al., \cite{Dunner} the structure of BaCu$_2$Sb$_2$ is an ordered intergrowth of ThCr$_2$Si$_2$- and CaBe$_2$Ge$_2$-types of unit cells. The unit cell of BaCu$_2$Sb$_2$ consists of three of these unit cell blocks stacked along the $c$-axis, as shown in Fig.~\ref{fig:structure_fig}(c), leading to a large $c$ = 32.6 {\AA}, with a ThCr$_2$Si$_2$-type block sandwiched between two CaBe$_2$Ge$_2$-type blocks resulting in an overall symmetry of $I4/mmm$.  A Rietveld refinement of the XRD data of BaCu$_2$Sb$_2$ with this structural model is shown in Fig.~\ref{fig:BaCu2Sb2_XRD}(a). The overall agreement between the experimental XRD data and the calculated pattern is reasonable. However, we observe splittings of some of the XRD peaks, as illustrated in the inset of Fig.~\ref{fig:BaCu2Sb2_XRD}(a), which suggest that the symmetry is lower than the reported one.

We attempted to refine the XRD data for BaCu$_2$Sb$_2$ with a lower orthorhombic ($Immm$) symmetry, which is a subgroup of the $I4/mmm$ space group. The Rietveld refinement profile with this structural model is shown in Fig.~\ref{fig:BaCu2Sb2_XRD}(b).  In terms of the goodness of fit, this structural model proved better as deduced from the respective $\chi^2$ values in Table~\ref{tab:table2}, but it could not account for the magnitudes of the splittings of the peaks, as illustrated in the inset of Fig.~\ref{fig:BaCu2Sb2_XRD}(b).  Further lowering of the symmetry to monoclinic space group $C2/m$, which is a subgroup of $Immm$, gave a better refinement quality but could not account for all of the peaks present.  Further lowering of the symmetry to space group $P2_1/c$, which is a subgroup of $C2/m$, allowed us to fit all the peak positions as illustrated in the inset of Fig.~\ref{fig:BaCu2Sb2_XRD}(c).  A Le Bail profile fit for this $P2_1/c$ monoclinic structure with $a = 4.6414(3)$, $b = 4.6418(3)$, $c = 32.680(2)$~{\AA}\ and $\beta  = 90.616^\circ$ is shown in Fig.~\ref{fig:BaCu2Sb2_XRD}(c). A similar lowering of symmetry to $P2_1/c$ was necessary for the 122-type compound ${\rm SrRh_2As_2}$ which exhibits polymorphism from the high-$T$ tetragonal ${\rm ThCr_2Si_2}$-type structure to a low-$T$ $P2_1/c$ monoclinic structure below 350~$^\circ$C\@.\cite{Zinth2011}  We conclude that the most likely space group is the monoclinic $P2_1/c$ that we have identified.  The detailed crystal structure (determination of the precise atomic positions) remains to be determined. Our attempt to solve the structure using single crystal x-ray diffraction failed because the crystal quality was insufficient to obtain reliable data.

The crystallographic and refinement parameters of all four compounds are listed in Tables~\ref{tab:table2} and \ref{tab:table3}.  The lattice parameters obtained from our refinements are in reasonable agreement with available literature values,\cite{Pfisterer, Dunner, Cordier} which are also listed in Table~\ref{tab:table2}.  During the refinements the thermal parameters $B$ were kept fixed to $B \equiv 0$ since the lattice parameters and $z$ parameters were insensitive to change in $B$ within the error bars.  Furthermore, the occupancies of atomic positions were kept fixed at the value of unity as there were no improvements in the qualities of fit on varying the occupancies by small amounts from unity. The $z$ coordinate of the As atoms in SrCu$_2$As$_2$, $z_{\rm As}$ = 0.3789(1), is very close to the reported values $z_{\rm As} = 0.3785(1)$ (Ref.~\onlinecite{Dunner}) and $z_{\rm As} = 0.377$.\cite{Pfisterer1983}  The $z$-coordinates $z_{\rm Sr} = 0.2353(5)$, $z_{\rm Cu2} = 0.6330(8)$ and $z_{\rm Sb2} = 0.8708(4)$ of SrCu$_2$Sb$_2$ are also very close to the reported values $z_{\rm Sr} = 0.2369(6)$, $z_{\rm Cu2} = 0.6380(6)$ and $z_{\rm Sb2} = 0.8695(6)$.\cite{Cordier}

\section{\label{SrCu2As2} Physical Properties of S\lowercase{r}C\lowercase{u}$_2$A\lowercase{s}$_2$ Crystals}

\subsection{Heat Capacity}
\begin{figure}
\includegraphics[width=3in]{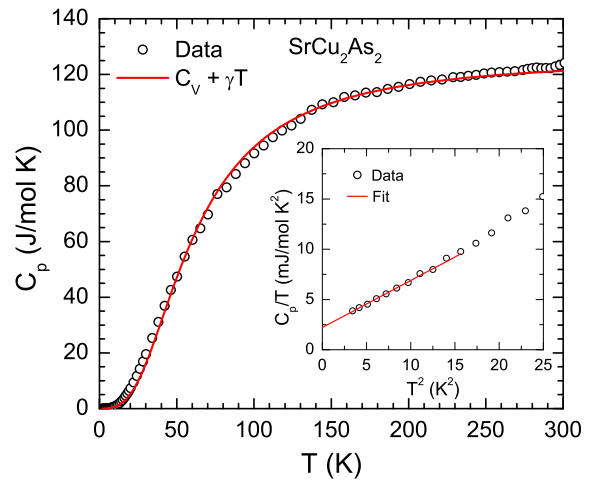}
\caption{\label{fig:fig_HC_SrCu2As2}(Color online) Heat capacity $C_{\rm p}$ of a SrCu$_2$As$_2$ single crystal versus temperature $T$. The solid red curve is a fit of the data by the sum of the contributions from the Debye lattice heat capacity $C_{\rm V\,Debye}(T)$ and predetermined electronic heat capacity $\gamma T$ according to Eq.~(\ref{eq:Debye_HC-fit}). Inset: $C_{\rm p}/T$ vs.\ $T^2$ below 5~K. The straight line is a fit of the data for 1.85~K~$\leq T \leq$~3.5~K by Eq.~(\ref{Eq:CTT2}).}
\end{figure}

The heat capacity $C_{\rm p}$ of SrCu$_2$As$_2$ as a function of $T$ measured at constant pressure~p is presented in Fig.~\ref{fig:fig_HC_SrCu2As2}. No evidence for any phase transitions was observed down to 1.85~K\@. At 300~K the heat capacity attains a value of 123.5~J/mol\,K which is  close to the expected classical high-$T$ Dulong-Petit lattice heat capacity value $C_{\rm V} = 3nR = 15R$ = 124.7~J/mol\,K at constant volume V, \cite{Kittel,Gopal} where $R$ is the molar gas constant and $n = 5$ is the number of atoms per formula unit (f.u.)\ of SrCu$_2$As$_2$. The inset in Fig.~\ref{fig:fig_HC_SrCu2As2} shows the low-$T$ data plotted as $C_{\rm p}/T$ versus $T^2$ allowing a conventional fit by
\be
\frac{C_{\rm p}(T)}{T}=\gamma + \beta T^2,
\label{Eq:CTT2}
\ee
where $\gamma$ is the Sommerfeld electronic linear specific heat coefficient and $\beta$ is the coefficient of the Debye $T^3$ lattice heat capacity at low $T$. A linear fit of the data below 3.5~K according to Eq.~(\ref{Eq:CTT2}), shown as the straight line in the inset of Fig.~\ref{fig:fig_HC_SrCu2As2}, gives the parameters
\begin{equation}
\begin{split}
\gamma & = 2.2(2)~{\rm mJ/mol\,K^2}  \\
\beta  & = 0.47(3)~{\rm mJ/mol\,K^4}.
\label{eq:gamma_beta}
\end{split}
\end{equation}
We obtained the value of the Debye temperature $\Theta_{\rm D}$ from $\beta$ using the relation \cite{Kittel}
\begin{equation}
\Theta_{\rm D} = \left( \frac{12 \pi^{4} N_{\rm A} k_{\rm B} n}{5 \beta} \right)^{1/3},
 \label{eq:Debye-Temp}
\end{equation}
\noindent where $N_{\rm A}$ is Avogadro's number and $k_{\rm B}$ is Boltzmann's constant, yielding
\begin{equation}
\Theta_{\rm D} = 274(6)~{\rm K}.
\label{eq:thetaD}
\end{equation}

The Sommerfeld coefficient $\gamma$ can be used to estimate the density of states at the Fermi level ${\cal D}(E_{\rm F})$ for both spin directions according to \cite{Kittel}
\begin{equation}
\gamma = \frac{\pi^2 k_{\rm B}^2}{3} {\cal D}(E_{\rm F}) (1 + \lambda_{\rm {e-ph}}),
\label{eq:DOS}
\end{equation}
\noindent where $\lambda_{\rm e-ph}$ is the electron-phonon coupling constant which we set to zero and obtain a value
\begin{equation}
{\cal D}(E_{\rm F}) = 0.93~{\rm states/eV\,f.u.}
\label{eq:DOS-value}
\end{equation}
for both spin directions. This value of ${\cal D}(E_{\rm F})$ is less than the value 1.53~states/eV\,f.u.\ for both spin directions predicted from the band structure calculations for SrCu$_2$As$_2$.\cite{Singh}

The ${\cal D}(E_{\rm F})$ can be used to estimate the band effective mass $m^*$ using the relation
\begin{equation}
{\cal D}(E_{\rm F}) = N_{\rm band} \left( \frac {m^* V_{\rm f.u.} k_{\rm F}}{\hbar^2 \pi^2} \right),
\label{eq:DOS_m}
\end{equation}
where ${\cal D}(E_{\rm F})$ is in units of states/erg\,f.u.\ with the quantities on the right-hand side in cgs units, $k_{\rm F}$ is the Fermi wave vector, $V_{\rm f.u.}$ is the volume per formula unit and $N_{\rm band}$ is the number of degenerate bands at $E_{\rm F}$ with the same $m^*$ and Fermi velocity $v_{\rm F}$. The quantity in parentheses on the right-hand side corresponds to the single-band relation. \cite{Kittel} Further, using $k_{\rm F} = m^*v_{\rm F}/\hbar$, Eq.~(\ref{eq:DOS_m}) can be rearranged to yield
\begin{equation}
\left(\frac {m^*}{m_{\rm e}} \right)^2 =  \frac {\hbar^3 \pi^2}{V_{\rm f.u.} m_{\rm e}^2 v_{\rm F}} \frac {{\cal D}(E_{\rm F})}{N_{\rm band}},
\label{eq:eff_mass}
\end{equation}
\noindent where $m_{\rm e}$ is the free electron mass. Using the average value
\be
\langle v_{\rm F} \rangle = 2.7 \times 10^7~{\rm cm/s}
\label{Eq:vFave}
\ee
from the band structure calculation \cite{Singh} and $V_{\rm f.u.} = V_{\rm cell}/Z$ where $Z$ = 2 is the number of formula units per unit cell, we obtain
\begin{equation}
\frac {m^*}{m_{\rm e}} =  1.86 \left[ \frac {{\cal D}(E_{\rm F})}{N_{\rm band}} \right]^{1/2},
\label{eq:eff-mass}
\end{equation}
\noindent where ${\cal D}(E_{\rm F})$ is now in units of states/eV\,f.u.\ for both spin directions. The band structure calculations show that $N_{\rm band} = 2$ for SrCu$_2$As$_2$, but there are three distinct Fermi surfaces arising from the two bands.\cite{Singh} Hence taking $N_{\rm band} = 2$ or~3 and using ${\cal D}(E_{\rm F})$ = 0.93 states/eV\,f.u.\ for both spin directions from Eq.~(\ref{eq:DOS-value}), the effective mass estimated from Eq.~(\ref{eq:eff-mass}) is
\be
\begin{split}
m^* &= 1.27\, m_{\rm e}\hspace{0.3in}(N_{\rm band} = 2)\\
m^* &= 1.04\, m_{\rm e}\hspace{0.3in}(N_{\rm band} = 3).
\label{eq:m_e}
\end{split}
\ee
The ratio $m^* / m_{\rm e} \sim 1$ is consistent with the prediction \cite{Singh} that SrCu$_2$As$_2$ is an $sp$-band metal.

\begin{table*}
\caption{\label{tab:table4} The linear specific heat coefficients $\gamma$ and the coefficient $\beta $ of the $T^3$ term in the low-$T$ heat capacity, and the density of states at the Fermi energy ${\cal D}(E_{\rm F})$ for both spin directions for SrCu$_2$As$_2$, SrCu$_2$Sb$_2$, SrCu$_2$(As$_{0.84}$Sb$_{0.16}$)$_2$ and BaCu$_2$Sb$_2$ single crystals. The Debye temperatures $\Theta_{\rm D}$ obtained at low~$T$ and for all~$T$ from heat capacity measurements and the Debye temperature $\Theta_{\rm R}$ obtained from fitting electrical resistivity data, respectively, are also listed.}
\begin{ruledtabular}
\begin{tabular}{lcccccc}

Compound 							& $\gamma $  		& $\beta $  		&	${\cal D}(E_{\rm F})$ 	&	$\Theta_{\rm D}$ (K)	&  $\Theta_{\rm D}$ (K)		& $\Theta_{\rm R}$ (K) \\	
								& (mJ/mol\,K$^2$)	& (mJ/mol\,K$^4$)	& (states/eV\,f.u.) 		& from low-$T$  			&  from all~$T$			& from $\rho(T)$ \\
\hline
SrCu$_2$As$_2$   					& 2.2(2) 			&  0.47(3) 		& 0.93 					&  	274(6)				& 246(1)  				&  245(4) \\		

SrCu$_2$Sb$_2$   					& 3.9(2) 			&  0.76(3) 		& 1.65 					& 	234(3)				& 214(1)  				&  153(1) \\

SrCu$_2$(As$_{0.84}$Sb$_{0.16}$)$_2$ 	& 2.3(2) 			& 0.83(2) 		& 0.97 					&  	227(2)				& 237(1)  				&  225(6) \\

BaCu$_2$Sb$_2$   					& 3.5(2) 			&  0.65(2) 		& 1.49  					&	246(3)				& 204(1)  				&  160(2)  \\	

\end{tabular}
\end{ruledtabular}
\end{table*}

The heat capacity data of a nonmagnetic metal over an extended $T$ range can be approximated by
\begin{equation}
C_{\rm p}(T) = \gamma T + n C_{\rm{V\,Debye}}(T),
\label{eq:Debye_HC-fit}
\end{equation}
where $\gamma$ is fixed to be the Sommerfeld electronic specific heat coefficient found above, $C_{\rm{V\,Debye}}(T)$ is the Debye lattice heat capacity due to acoustic phonons at constant volume V  per mole of atoms, and $n$ is the number of atoms per f.u.\ ($n = 5$ here).  The Debye function $C_{\rm{V\,Debye}}(T)$ is  given by \cite{Gopal}
\begin{equation}
C_{\rm{V\,Debye}}(T) = 9 R \left( \frac{T}{\Theta_{\rm{D}}} \right)^3 {\int_0^{\Theta_{\rm{D}}/T} \frac{x^4 e^x}{(e^x-1)^2}\,dx}.
\label{eq:Debye_HC}
\end{equation}

The heat capacity data in Fig.~\ref{fig:fig_HC_SrCu2As2} were least-squares fitted by Eqs.~(\ref{eq:Debye_HC-fit}) and (\ref{eq:Debye_HC}) over the $T$ range 2--300~K using the Pad\'{e} approximant function for $C_{\rm{V\,Debye}}(T)$ developed by us in Ref.~\onlinecite{Ryan}, with $\Theta_{\rm D}$ being the only adjustable parameter. The solid curve in Fig.~\ref{fig:fig_HC_SrCu2As2} represents the best fit of the experimental data in Fig.~\ref{fig:fig_HC_SrCu2As2} by Eqs.~(\ref{eq:Debye_HC-fit}) and (\ref{eq:Debye_HC}), yielding $\Theta_{\rm D}$ = 246(1)~K\@. This value of $\Theta_{\rm D}$ is different from the value of $\Theta_{\rm D}$ = 274(6)~K obtained in Eq.~(\ref{eq:thetaD}) from the coefficient $\beta$ deduced from the low-$T$ heat capacity data. The difference between the two values may arise due to the approximations made in Debye theory that result in a $T$-dependent $\Theta_{\rm D}$, so the fit by Eq.~(\ref{eq:Debye_HC-fit}) gives a value of $\Theta_{\rm D}$ that is averaged over $T$. A contribution to the difference also comes from the fact that the heat capacity was measured at constant pressure whereas the Debye theory describes the lattice heat capacity at constant volume.  The parameters obtained from our heat capacity fits are summarized in Table~\ref{tab:table4}.  Also listed is the Debye temperature $\Theta_{\rm R}$ obtained from fitting our experimental $\rho(T)$ data for SrCu$_2$As$_2$ by the Bloch-Gr\"uneisen model in Sec.~\ref{SecSrCu2As2Rho} below.

\subsection{\label{Sec:M(H,T)SrCu2As2} Magnetization and Magnetic Susceptibility}

\begin{figure}
\includegraphics[width=3in]{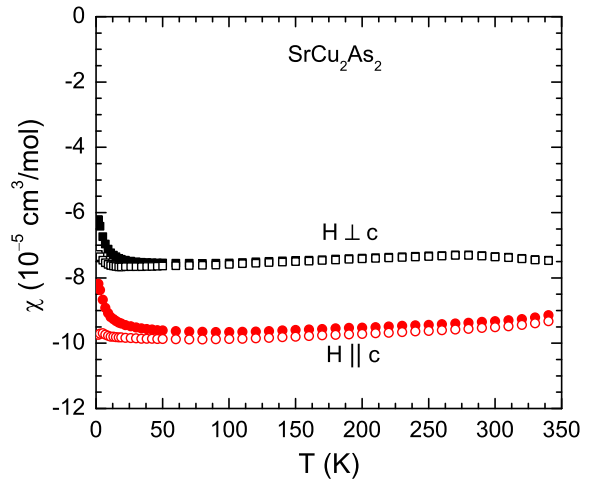}
\caption{\label{fig:fig_MT_SrCu2As2} (Color online) Zero-field-cooled magnetic susceptibility $\chi$ of a SrCu$_2$As$_2$ single crystal versus temperature $T$ in the temperature range 1.8 -- 350~K measured in a magnetic field of 3.0~T applied along the $c$-axis ($\chi_c,\ {\bf H} \parallel {\bf c}$) and in the $ab$-plane ($\chi_{ab},\ {\bf H} \perp  {\bf c}$) (solid symbols). The open symbols represent the intrinsic susceptibility of SrCu$_2$As$_2$ after correcting for the ferromagnetic and paramagnetic impurity contributions as described in the text.}
\end{figure}

The zero-field-cooled (ZFC) magnetic susceptibilities $\chi  \equiv M/H$ of a SrCu$_2$As$_2$ single crystal as a function of $T$ from 1.8 to 340~K in an applied magnetic field  $H = 3.0$~T aligned along the $c$-axis ($\chi_c,\ {\bf H} \parallel  {\bf c}$) and in the $ab$-plane ($\chi_{ab},\ {\bf H} \perp {\bf c}$) are presented in Fig.~\ref{fig:fig_MT_SrCu2As2}, where $M$ is the measured magnetization of the crystal. The $\chi$ has a negative sign and is anisotropic with $\chi_{ab} > \chi_{c}$ over the whole $T$ range.  This kind of anisotropy is common in the iron arsenide family; e.g., both BaFe$_2$As$_2$ (Ref.~\onlinecite{Wang}) and SrFe$_2$As$_2$ (Ref.~\onlinecite{Yan}) exhibit $\chi_{ab} > \chi_{c}$.  The powder and temperature average of our anisotropic $\chi$ values in Fig.~\ref{fig:fig_MT_SrCu2As2} is significantly more diamagnetic (negative) than the $T$-independent value between 80 and 400~K of $\chi = -5.4 \times 10^{-5}$~cm$^3$/mol reported previously for a polycrystalline sample.\cite{Pfisterer1983}  This difference may be due to a saturation magnetization of ferromagnetic (FM) impurities in the latter sample that was incompletely corrected for.

Upturns are observed in the $\chi(T)$ data in Fig.~\ref{fig:fig_MT_SrCu2As2} at low $T$ for both $\chi_{ab}$ and $\chi_{c}$ that are likely due to the presence of a small amount of saturable paramagnetic (PM) impurities as deduced from the $M(H)$ isotherm data discussed below.  Since SrCu$_2$As$_2$ is weakly diamagnetic, even trace amounts of magnetic impurities can make significant contributions to the $\chi$ and $M$ data. Therefore, a correction for the magnetic impurity contribution is necessary to determine the intrinsic magnetic behavior of the compound.

\begin{figure}
\includegraphics[width=3in]{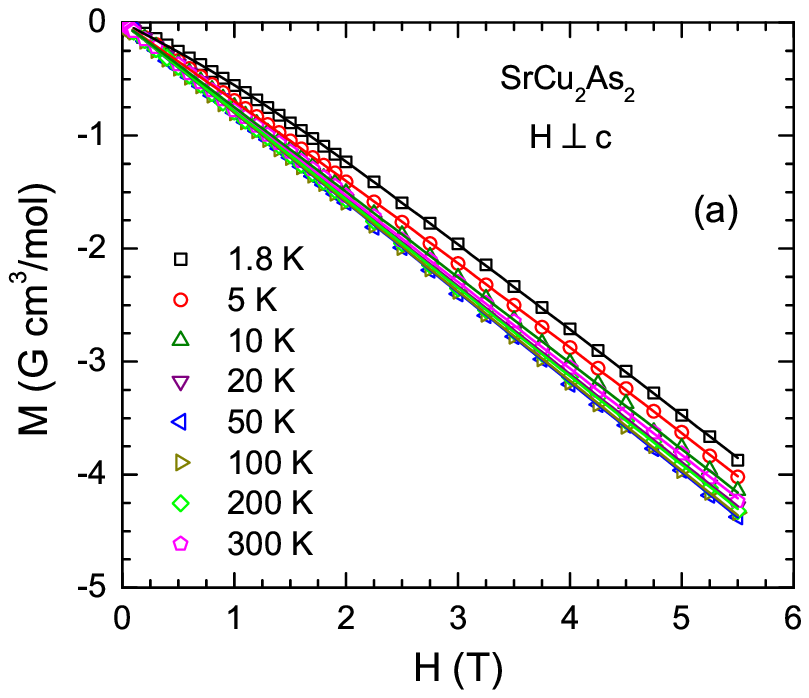}\vspace{0.1in}
\includegraphics[width=3in]{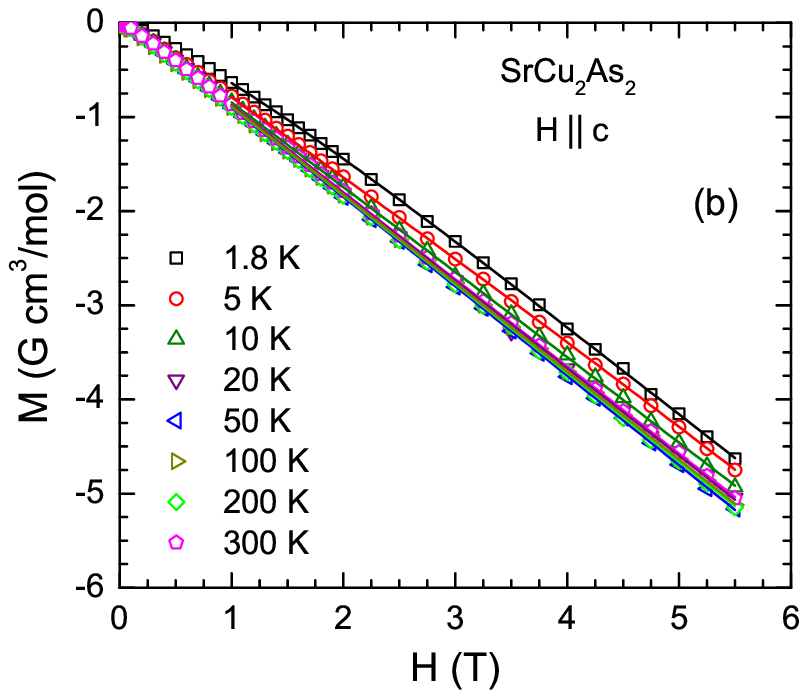}
\caption{\label{fig:fig_MH_SrCu2As2} (Color online) Isothermal magnetization $M$ of a SrCu$_2$As$_2$ single crystal versus magnetic field $H$ measured at the indicated temperatures for {\bf H} applied (a) perpendicular to the $c$-axis  ($M_{ab},\ {\bf H} \perp  {\bf c}$) and (b) along the $c$-axis ($M_c,\ {\bf H} \parallel {\bf c}$). The solid curves are fits by Eq.~(\ref{eq:MH_fit}).}
\end{figure}

The isothermal $M$ data for SrCu$_2$As$_2$ as a function of $H$ measured at eight temperatures between 1.8 and 300~K with {\bf H} applied along the $c$-axis ($M_c,\ {\bf H} \parallel {\bf c}$) and in the $ab$-plane ($M_{ab},\ {\bf H} \perp  {\bf c}$) are presented in Fig.~\ref{fig:fig_MH_SrCu2As2}. The $M(H)$ data are diamagnetic and reveal anisotropic behavior with $M_{ab}(H) > M_c(H)$, which is consistent with the above observation $\chi_{ab} > \chi_{c}$.  The presence of slight nonlinearities in the $M(H)$ isotherms at low fields, particularly at low $T$, reveals the presence of saturable ferromagnetic FM and/or paramagnetic PM impurities in the sample.

Since the $M(H)$ contribution of a FM impurity is typically nonlinear only at low fields $H\lesssim 1$--2~T one can estimate the FM impurity saturation magnetization $M_{\rm s}$ from the high-field $M(H)$ data. Furthermore, typical FM impurities have Curie temperatures significantly above room temperature, and saturable PM impurities do not show nonlinear $M(H)$ behaviors above $\sim 25$~K\@.  We therefore determined the $M_{\rm s}$ of the FM impurities by fitting the $M(H)$ data at $H \geq 2.0$~T for $T \geq 25$~K by
\begin{equation}
M(H) = M_{\rm s} + \chi H,
\label{eq:MH_linear-fit}
\end{equation}
where $\chi$ is the susceptibility of the sample for a particular field orientation that can still contain the contribution from paramagnetic PM impurities as discussed next. The $M_{\rm s}$ values obtained from the fits were found to be independent of $T$ for $T\geq 25$~K to within the error bars.  The values are listed in Table~\ref{tab:tableMH}. The $M_{\rm s}^{c}$ value corresponds to the saturation magnetization of only 4 molar ppm of Fe metal impurities.

The PM impurity contribution to $M$ can be estimated by fitting the $M(H)$ data for each field direction by
\begin{equation}
M(T,H)=M_{\rm s}+\chi_0 H + f_{\rm imp} M_{\rm{{s_{imp}}}} B_{S_{\rm imp}}(x),
\label{eq:MH_fit}
\end{equation}
where $M_{\rm s}$ is the FM impurity saturation magnetization value found above, $\chi_0$ is the intrinsic susceptibility of the compound, $f_{\rm imp}$ is the molar fraction of PM impurities, $M_{\rm {s_{imp}}} = N_{\rm A} g_{\rm imp} \mu_{\rm B} S_{\rm imp}$ is the PM impurity saturation magnetization, $N_{\rm A}$ is Avogadro's number, $\mu_{\rm B}$ is the Bohr magneton, and $g_{\rm imp}$ and $S_{\rm imp}$ are the spectroscopic splitting factor ($g$-factor) and the spin of the impurities, respectively.  The Brillouin function $B_{S_{\rm imp}}$ for the PM impurities is given by
\bea
B_{S_{\rm imp}}(x) & =& \frac{1}{2S_{\rm imp}}\Bigg\{\left( 2S_{\rm imp}+1 \right) \coth \left[(2S_{\rm imp}+1)\frac{x}{2} \right]  \nonumber\\
				   && \hspace{2cm} -\ \coth \left( \frac{x}{2} \right)\Bigg\},\label{eq:Brillouin}
\eea
where
\be
x \equiv \frac{g_{\rm imp} \mu_{\rm{B}} H}{k_{\rm{B}} (T-\theta_{\rm imp})}.
\ee
A Weiss temperature $\theta_{\rm imp}$ is included in the Brillouin function in order to take weak interactions between PM impurity spins into account in a mean-field way.  In particular, when $B_{S_{\rm imp}}(x)$ is expanded to first order in $x$ for $x\ll 1$, one obtains a Curie-Weiss law for the low-field susceptibility of the impurity spins with Weiss temperature $\theta_{\rm imp}$.  In order to reduce the number of fitting parameters, the impurity $g$-factor was set to $g_{\rm imp}$ = 2 for this compound as well as for ${\rm SrCu_2Sb_2}$ in Sec.~\ref{Sec:SrCu2Sb2MH} below.

The solid curves in Fig.~\ref{fig:fig_MH_SrCu2As2} are the fits of the $M(H)$ data by Eq.~(\ref{eq:MH_fit}) that were obtained using the above $M_{\rm s}^{ab}$ and $M_{\rm s}^{c}$ values for ${\bf H} \perp  {\bf c}$ and ${\bf H} \parallel {\bf c}$, respectively. Since $M_{\rm s}^{ab}$ is zero we fitted the $M(H)$ data for ${\bf H} \perp  {\bf c}$ by Eq.~(\ref{eq:MH_fit}) in the whole range $0 \leq H \leq 5.5$~T\@.  However, since $M_{\rm s}^{c}$ is nonzero we fitted the $M(H)$ data for ${\bf H} \parallel {\bf c}$ only in the range $1.0 \leq H \leq 5.5$~T\@. We found that setting $\theta_{\rm imp}$ = 0 gives a good fit for both field directions, so in the final fits we set $\theta_{\rm imp} \equiv  0$.

\begin{table}
\caption{\label{tab:tableMH} Parameters obtained for SrCu$_2$As$_2$ and SrCu$_2$Sb$_2$ from fitting $M(H)$ isotherms at 1.8~K by Eqs.~(\ref{eq:MH_linear-fit}) and (\ref{eq:MH_fit}), where $\theta_{\rm imp} \equiv 0$.  Here $M_s$ is the saturation magnetization of ferromagnetic impurities, $\chi_0$ is the intrinsic susceptibility, and $f_{\rm imp}$ and $S_{\rm imp}$ are the molar fraction and spin of the paramagnetic impurities, respectively.}
\begin{ruledtabular}
\begin{tabular}{lcccc}
Compound & field  & $M_{\rm s}$  &  $\chi_0$   &	$f_{\rm imp} S_{\rm imp}$ \\
		& direction & (${\rm \frac{G\,cm^3}{mol}}$) & (${\rm 10^{-5}~\frac{cm^3}{mol}}$) & ($10^{-5}$)\\
\hline
SrCu$_2$As$_2$ & $H \perp c$      &  0.00(2)  & $-7.62(2)$ & 3.05(8)  \\				
			   & $H \parallel c$  &  0.05(3)  & $-9.32(4)$ & 4.1(2)   \\				
SrCu$_2$Sb$_2$ & $H \perp c$      &  0.00(5)  & $-17.1(1)$ & 10.9(3)  \\	
	           & $H \parallel c$  &  0.02(3)  & $-5.56(3)$ & 2.2(1)  \\				
\end{tabular}
\end{ruledtabular}
\end{table}

The parameter $\chi_0$ and the product of the parameters $f_{\rm imp}$ and $S_{\rm imp}$ obtained from the fits of the $M(H)$ isotherms at 1.8~K for ${\bf H} \parallel {\bf c}$ and ${\bf H} \perp {\bf c}$ in Fig.~\ref{fig:fig_MH_SrCu2As2} by Eq.~(\ref{eq:MH_fit}) are listed in Table~\ref{tab:tableMH}.  From Table~\ref{tab:tableMH}, both $M_{\rm s}$ and the product $f_{\rm imp} S_{\rm imp}$ are different for ${\bf H} \parallel  {\bf c}$ compared with ${\bf H} \perp {\bf c}$, which indicates that the magnetizations of the FM and PM impurities are both anisotropic. The origins of these anisotropies are not clear.

From Eq.~(\ref{eq:MH_fit}), the magnetic impurity contribution to the magnetization, for fields above the saturation field of about 1~T for the FM impurities, is
\[
M_{\rm imp}(H,T) = M_{\rm s} + f_{\rm imp} M_{\rm{{s_{imp}}}} B_{S_{\rm imp}}(x).
\]
The $\chi(T)\equiv M(T)/H$ data in Fig.~\ref{fig:fig_MT_SrCu2As2} were taken with $H = 3$~T.  Therefore, this field was above the saturation field of the FM impurities and one can correct for the contributions of both the FM and PM impurities to obtain the intrinsic susceptibility versus $T$ according to
\[
\chi_0(T) = \frac{M(T)-M_{\rm imp}(H,T)}{H}.
\]
In this way we obtained the intrinsic $\chi_0(T)$ data shown as the open symbols in Fig.~\ref{fig:fig_MT_SrCu2As2}.  The magnetic impurities are seen to have little influence on the measured susceptibility except below about 50~K where the PM impurity contribution to $M(T)$ becomes significant.

The largest value of $f_{\rm imp}S_{\rm imp} = 1.09 \times 10^{-4}$ in Table~\ref{tab:tableMH}, which is for $\chi_{ab}$ of ${\rm SrCu_2Sb_2}$ (see below), corresponds to a very small concentration of paramagnetic impurities.  For the worst case assumption $S_{\rm imp} = 1/2$, the impurity concentration is $f_{\rm imp} = 2.2 \times 10^{-4} = 0.022$~mol\% = 220~molar ppm.  If $S_{\rm imp} = 5/2$, one obtains the five-times smaller value $f_{\rm imp} = 4.4 \times 10^{-5} = 0.0044$~mol\% = 44~molar ppm.  The values of $M_{\rm s} \leq 0.05$~G\,cm$^3$/mol in Table~\ref{tab:tableMH} correspond to $M_{\rm s} \leq 9.0 \times  10^{-6}~\mu_{\rm B}$/f.u\@.  Thus if $S = 1/2$ and $g = 2$, this corresponds to a ferromagnetic molar impurity concentration of $\leq 9.0$~molar ppm.  On the other hand, if $S = 5/2$ and $g = 2$, this corresponds to a ferromagnetic molar impurity concentration of $\leq 1.8$~molar ppm.  As stated in Sec. II, the purity of the Sr used in the synthesis of the ${\rm SrCu_2As_2}$ and ${\rm SrCu_2Sb_2}$ crystals was 99.95~mol\% (metals basis), i.e., containing 0.05~mol\% of elemental metal impurities.  Thus our values of $f_{\rm imp}S_{\rm imp}$ and $M_{\rm s}$ in Table~\ref{tab:tableMH} are consistent with (less than) the elemental metal impurity level supplied by the source of our Sr starting material.  The nature of these magnetic impurities is unknown, and this information was not provided by the supplier of the Sr starting material.

The residual low-$T$ upturns in $\chi_{ab}$ and $\chi_c$ in Fig.~\ref{fig:fig_MT_SrCu2As2} (and in Fig.~\ref{fig:fig_MT_SrCu2Sb2} below) after correction for the contribution of the paramagnetic impurities are very small and are most likely due to imperfect subtractions of the paramagnetic impurity contributions.  The small changes in the magnitudes of $\chi$  between 300 and 350~K in Fig.~\ref{fig:fig_MT_SrCu2As2} (and in Figs.~\ref{fig:fig_MT_SrCu2Sb2} and \ref{fig:fig_MT_SrCu2AsSb} below) are believed to be experimental artifacts of our SQUID magnetometer.  Such small changes are seen in many samples of different types that we have measured on this instrument in this $T$ range.  The origin of this systematic error is unknown.

The contributions to $\chi_0$ of a nonmagnetic metal  are
\begin{equation}
\chi_0 = \chi_{\rm {core}}+\chi_{\rm {VV}}+\chi_{\rm {L}} + \chi_{\rm {P}},
\label{eq:chi}
\end{equation}
where the first three terms comprise the orbital susceptibility contributions.  Here $\chi_{\rm {core}}$ is the diamagnetic susceptibility due to the localized core electrons, $\chi_{\rm {VV}}$ is the paramagnetic Van Vleck susceptibility, and $\chi_{\rm {L}}$ is the diamagnetic Landau susceptibility of the conduction carriers.  The last term $\chi_{\rm {P}}$ is the paramagnetic Pauli spin susceptibility of the conduction carriers.

The powder- and temperature-average (from 1.8 to 350~K) of $\chi_0$ is $\langle \chi_0 \rangle = (2 \chi_{ab} + \chi_{c})/3 = -8.25 \times 10^{-5}$~cm$^3$/mol. Since our compounds are closer to being covalent metals than ionic metals, we used  atomic diamagnetic values\cite{Mendelsohn} for $\chi_{\rm core}$ to obtain $\chi_{\rm core} = -1.75 \times 10^{-4}$~cm$^3$/mol.  The $\chi_{\rm {P}}$ can be estimated from ${\cal D}(E_{\rm F})$ using the relation \cite{Ashcroft}
\begin{equation}
\chi_{\rm {P}} =  \mu_{\rm B}^2 {\cal D}(E_{\rm F}).
\label{eq:Chi-Pauli}
\end{equation}
We then obtain $\chi_{\rm {P}} = 3.00 \times 10^{-5}$~cm$^3$/mol using ${\cal D}(E_{\rm F})$ = 0.98~states/eV\,f.u.\ for both spin directions from Eq.~(\ref{eq:DOS-value}). The $\chi_{\rm {L}}$ can be estimated from the relation \cite{Ashcroft, Elliott}
\begin{equation}
 \chi_{\rm {L}} = - \frac{1}{3} \left( \frac {m_{\rm e}}{m^*} \right)^2 \chi_{\rm {P}}.
 \label{eq:Chi-Landau}
\end{equation}
Using $m^* = 1.27\,m_{\rm e}$ from Eq.~(\ref{eq:m_e}), one obtains $\chi_{\rm {L}} = -0.62 \times 10^{-5}$~cm$^3$/mol. One can then estimate the angle- and temperature-averaged $\chi_{\rm {VV}}$ using Eq.~(\ref{eq:chi}) and the above values of $\chi_{\rm {core}}$, $\chi_{\rm {P}}$ and $\chi_{\rm {L}}$, which give $\chi_{\rm {VV}} = 6.87 \times 10^{-5}$~cm$^3$/mol, a physically reasonable value.  The above contributions to the average intrinsic magnetic susceptibility of SrCu$_2$As$_2$ are summarized in Table~\ref{tab:table5}.

\begin{table}
\caption{\label{tab:table5} Estimated contributions to the intrinsic angle- and temperature-averaged magnetic susceptibilities $\langle \chi_0 \rangle$ of SrCu$_2$As$_2$, SrCu$_2$Sb$_2$, SrCu$_2$(As$_{0.84}$Sb$_{0.16}$)$_2$ and BaCu$_2$Sb$_2$ crystals in units of 10$^{-5}$\,cm$^3$/mol.  Here $\chi_{\rm P}$ is the Pauli spin susceptibility of the conduction carriers, and the orbital susceptibility contributions are the  diamagnetism $\chi_{\rm core}$ of the atomic electron cores, the Landau diamagnetism $\chi_{\rm L}$ of the conduction carriers and the Van Vleck paramagnetism $\chi_{\rm VV}$.}
\begin{ruledtabular}
\begin{tabular}{lccccc}

Compound & $\langle \chi_0 \rangle$ & $\chi_{\rm core}$  &	 $\chi_{\rm {P}}$ 	&	$\chi_{\rm L}$	&	$\chi_{\rm VV}$ \\	
\hline
SrCu$_2$As$_2$   & $-$8.25  &  $-$17.50 & 3.00 & $-$0.62 & 6.87 \\		
SrCu$_2$Sb$_2$   & $-$15.58 &  $-$20.79 & 5.33 & $-$1.78 & 1.66 \\
SrCu$_2$(As$_{0.84}$Sb$_{0.16}$)$_2$ & $-$6.95 & $-$18.03 & 3.14 & $-$1.05 & 8.99 \\
BaCu$_2$Sb$_2$   &$-$8.02 & $-$23.10 & 4.81 & $-$1.60 & 11.9 \\	

\end{tabular}
\end{ruledtabular}
\end{table}

\subsection{\label{SecSrCu2As2Rho} Electrical Resistivity Measurements and the Bloch-Gr\"uneisen Model}

\begin{figure}
\includegraphics[width=3in]{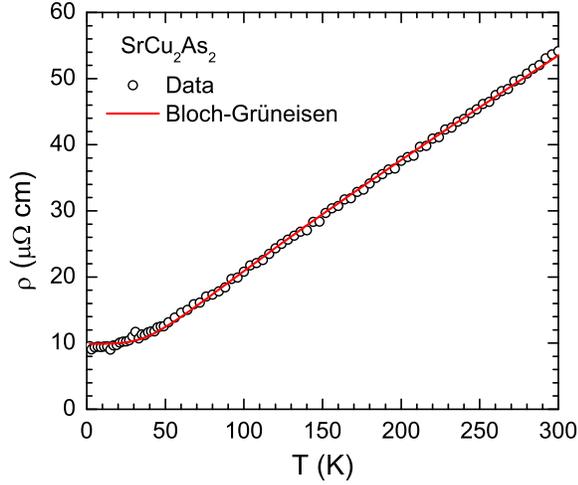}
\caption{\label{fig:fig_rho_SrCu2As2} (Color online) In-plane electrical resistivity $\rho$ of a SrCu$_2$As$_2$ crystal versus temperature $T$. The red solid curve is the fit by the Bloch-Gr\"{u}neisen model.}
\end{figure}

The in-plane electrical resistivity $\rho$ data for SrCu$_2$As$_2$ as a function of $T$ are presented in Fig.~\ref{fig:fig_rho_SrCu2As2}. Metallic character is indicated from the positive temperature coefficient of $\rho(T)$. The relatively low value (for 122-type compounds) of the residual resistivity
\[
\rho_0\equiv \rho(T\to0) = 9.3~\mu \Omega\,{\rm cm}
\]
and a reasonably large residual resistivity ratio
\[
{\rm RRR} \equiv \frac{\rho(300\,{\rm K})}{\rho(1.8\,{\rm K})}\approx 6
\]
indicate that our crystal is of good quality.

We fitted the $\rho(T)$ data of SrCu$_2$As$_2$ by the Bloch-Gr\"{u}neisen (BG) model that describes the electrical resistivity $\rho_{\rm {BG}}(T)$ due to  scattering of the conduction electrons by acoustic lattice vibrations in a monatomic metal according to\cite{Blatt}
\begin{equation}
\rho_{\rm {BG}}(T)= 4 \mathcal{R}(\Theta _{\rm R}) \left( \frac{T}{\Theta _{\rm R}}\right)^5 \int_0^{\Theta_{\rm{R}}/T}{\frac{x^5}{(e^x-1)(1-e^{-x})}dx},
 \label{eq:Bloch-Gruneisen}
\end{equation}
where $\Theta_{\rm R}$ is the Debye temperature obtained from a fit of experimental resistivity data by the BG theory and the prefactor function $\mathcal{R}(\Theta _{\rm R})$ is given by
\begin{equation}
\mathcal{R}(\Theta _{\rm R}) = \frac {\hbar}{e^2} \left [ \frac {\pi^3(3\pi^2)^{1/3} \hbar^2}{4 n_{\rm cell}^{2/3} a M k_{\rm B} \Theta_{\rm{R}}} \right ].
 \label{eq:BG_coupling}
\end{equation}
Here $\hbar$ is Planck's constant divided by $2\pi$, $e$ is the fundamental electric charge, $n_{\rm cell}$ is the number of conduction (valence) electrons per unit cell of volume $a^3$ containing one atom, $M$ is the mass of the one atom per unit cell and $k_{\rm B}$ is Boltzmann's constant. The prefactor in Eq.~(\ref{eq:BG_coupling}) is $\hbar/e^2$ = 4108.24 $\Omega$ in SI units.  Thus $\mathcal{R}(\Theta _{\rm R})$ can be found in the favored units of $\Omega$\,cm by expressing the quantities in square brackets in cgs units.  The expression for $\mathcal{R}(\Theta _{\rm R})$ in Eq.~(\ref{eq:BG_coupling}) needs to be modified slightly according to Eq.~(\ref{eq:R_mod}) below in order to apply the BG theory to calculating the magnitude of the  resistivity of polyatomic metals.

Even for simple $s$- or $sp$-metals it is usually not possible to accurately fit both the magnitude and $T$ dependence of $\rho$ by the BG model with $\Theta_{\rm R}$ as the only adjustable
parameter.\cite{Blatt}  To fit the $T$ dependence of $\rho$ using the BG model we first calculate the BG prediction for $\rho(T=\Theta_{\rm R})$ from Eq.~(\ref{eq:Bloch-Gruneisen}) as\cite{Ryan}
\begin{equation}
\rho_{\rm BG}(T=\Theta_{\rm R})=0.9\,464\,635\,{\cal R}(\Theta _{\rm R}).
\label{eq:theta_R}
\end{equation}
Then the $T$ dependence $\rho_{\rm BG}(T)/\rho_{\rm BG}(\Theta_{\rm R})$ is a function $f$ of the ratio $y=T/\Theta_{\rm R}$ and is given by Eqs.~(\ref{eq:Bloch-Gruneisen})--(\ref{eq:theta_R}) as
\begin{equation}
f(y) = 4.226\,259 \,y^5 \int_0^{1/y}\frac{x^5}{(e^x - 1)(1-e^{-x})}\,dx.
\label{eq:Bloch-Gruneisen_theta_R}
\end{equation}
Now we can fit the experimental $\rho(T)$ data by the $T$ dependence of the BG prediction according to\cite{Ryan}
\begin{equation}
\rho(T) = \rho_0 + \rho(\Theta_{\rm R}) f(T/\Theta_{\rm R}),
\label{eq:BG_fit}
\end{equation}
where $\rho_0$, $\rho(\Theta_{\rm R})$ and $\Theta_{\rm R}$ are all independently adjustable fitting parameters. To carry out the fit of the experimental data by Eq.~(\ref{eq:BG_fit}), we use an analytic Pad\'e approximant function (a ratio of two polynomials in $y^{-1}$) that accurately fits $f(y)$ in Eq.~(\ref{eq:Bloch-Gruneisen_theta_R}).  The numerical coefficients in the approximant are obtained by a high-accuracy nonlinear-least-squares fit of numerical values of $f(y)$ obtained from Eq.~(\ref{eq:Bloch-Gruneisen_theta_R}) by the Pad\'e approximant as described in Ref.~\onlinecite{Ryan}.

\begin{table}
\caption{\label{Tab:RhoFitParams} Parameters associated with Bloch-Gr\"uneisen fits to the resistivities $\rho$ within the $ab$-plane of the listed single crystals.  Here $\rho_0$ is the residual resistivity for $T\to0$, $\Theta_{\rm R}$ is the Debye temperature determined from resistivity measurements, $\rho(\Theta_{\rm R})$ is the fitted value of $\rho$ at $T = \Theta_{\rm R}$, and ${\cal R}(\Theta_{\rm R})$ is the quantitiy defined in Eq.~(\ref{eq:BG_coupling}) and is obtained from the fitted value of $\rho(\Theta_{\rm R})$ using Eq.~(\ref{eq:theta_R}).}
\begin{ruledtabular}
\begin{tabular}{lcccc}
Compound 		& $\rho_0$ 		&  $\Theta_{\rm R}$ 	&	$\rho(\Theta_{\rm R})$  	& ${\cal R}(\Theta_{\rm R})$\\
			& ($\mu\Omega$\,cm) & (K)				& ($\mu\Omega$\,cm)			& ($\mu\Omega$\,cm)\\
\hline
SrCu$_2$As$_2$ &  9.90(7)  		&  245(4)  			& 34.9(6) 				&  36.9 \\				
SrCu$_2$Sb$_2$ &  13.64(1)  		&  153(1)  			& 13.8(1) 				&  14.6 \\
SrCu$_2$(As$_{0.84}$Sb$_{0.16}$)$_2$		&  17.65(5)		&  225(6)				& 16.5(4)					&  17.4 \\
BaCu$_2$Sb$_2$	&  2.59(3)		&  160(2)				& 17.9(2)					&  18.9  \\
\end{tabular}
\end{ruledtabular}
\end{table}

The solid red curve in Fig.~\ref{fig:fig_rho_SrCu2As2} is the least-squares fit of the $\rho(T)$ data in Fig.~\ref{fig:fig_rho_SrCu2As2} by Eqs.~(\ref{eq:Bloch-Gruneisen_theta_R}) and (\ref{eq:BG_fit}) using the Pad\'{e} approximant function\cite{Ryan} in place of Eq.~(\ref{eq:Bloch-Gruneisen_theta_R}).  The fit parameters obtained are $\rho_0$ = 9.90(7)~$\mu \Omega$\,cm, $\rho(\Theta_{\rm R}) = 34.9(6)\,\mu \Omega$\,cm, and $\Theta_{\rm{R}}$ = 245(4)~K\@.  The value of $\Theta_{\rm R}$ is identical within the errors to the value $\Theta_{\rm{D}}$ = 246(1)~K obtained from the fit of heat capacity data over the whole measured $T$ range  by the Pad\'e approximant for the Debye function (Table~\ref{tab:table4}). The value of ${\cal R}(\Theta_{\rm R})$ can be calculated from the value of $\rho(\Theta_{\rm R})$ using Eq.~(\ref{eq:theta_R}), which gives $\mathcal{R}(\Theta_{\rm R})$ = 36.9~$\mu \Omega$\,cm.  The parameters obtained from the fit of $\rho(T)$ for SrCu$_2$As$_2$ by the BG model are summarized in Table~\ref{Tab:RhoFitParams}.

To compare our experimental value of $\rho$(300~K) for SrCu$_2$As$_2$ with the value predicted by the Bloch-Gr\"uneisen theory, it is necessary to slightly modify the expression for $\mathcal{R}(\Theta _{\rm R})$ in Eq.~(\ref{eq:BG_coupling}), which applies to monatomic metals, to allow application to polyatomic metals.  Thus one obtains\cite{Ryan}
\begin{equation}
 \mathcal{R}=\frac{\hbar}{e^2} \left[ \frac{\pi^3 (3 \pi^2)^{1/3} \hbar^2}{4 n_{\rm{cell}}^{2/3} a k_{\rm{B}} \Theta_{\rm{R}}} \left(\frac{1}{M}\right)_{\rm ave} \right],
 \label{eq:R_mod}
\end{equation}
where the variables have the same meaning as before, except that
\begin{equation}
\left( \frac{1}{M} \right)_{\rm ave} = \frac{N_{\rm A}}{n}\sum\limits_{i=1}^n \frac{1}{M_i}
\end{equation}
is the average inverse mass of the atoms in the compound, $n$ is the number of atoms per f.u.\ and $M_i$ is the atomic weight of element~$i$.  In this case, the equivalent lattice parameter of a primitive cubic unit cell containing one atom is $a=[V_{\rm cell}/(nZ)]^{1/3}$ ($nZ=10$ here) because there are $Z=2$ formula units of $A{\rm Cu_2}X_2$, with $n=5$ atoms per formula unit, per body-centered tetragonal unit cell with volume $V_{\rm cell}$ given in Table~\ref{tab:table2}.

The number of current carriers per atom $n_{\rm cell}$ in Eq.~(\ref{eq:R_mod}) can be estimated from the band theory value\cite{Singh} of $v_{\rm F}$. Since $m^*v_{\rm F} = \hbar k_{\rm F}$, and using the one-band relation \cite{Kittel} $k_{\rm F} = (3\pi^2n^*)^{1/3}$ where $n^*$ is the carrier density per band, one obtains
\begin{equation}
n^* = \frac {1}{3\pi^2} \left( \frac {m^*v_{\rm F}}{\hbar} \right)^3 = 8.78 \times 10^{20}~{\rm cm}^{-3},
\label{eq:n*}
\end{equation}
where to obtain the second equality we used the value of $\langle v_{\rm F}\rangle$ given in Eq.~(\ref{Eq:vFave}) and the value of $m^*$ given in Eq.~(\ref{eq:m_e}).
This gives
\begin{equation}
n_{\rm cell} = N_{\rm band} n^* \frac {V_{\rm cell}}{nZ} = 0.0327.
\label{eq:ncell}
\end{equation}
Inserting this value of $n_{\rm cell}$ and the value $\Theta_{\rm{R}}$ = 245~K from Table~\ref{tab:table4} into Eq.~(\ref{eq:R_mod}) we obtain $\mathcal{R}(\Theta_{\rm R})$ = 10.0~$\mu \Omega$\,cm.  The value of $\rho(300$~K) predicted by the BG model is then numerically calculated from Eq.~(\ref{eq:Bloch-Gruneisen}) to be $\rho(300~{\rm K}) = 21.7~\mu\Omega$\,cm.   The measured value $\rho(300~{\rm K}) = 54~\mu\Omega$\,cm from Fig.~\ref{fig:fig_rho_SrCu2As2} is a factor of 2.5 larger than this theoretical value.  However, this should be considered to be satisfactory agreement in view of the even larger factors of 3 to 8 discrepancy between experiment and theory for other $sp$-metals such as Au, Be, Cd, Zn, Mg, Al and In.\cite{Blatt}  Such discrepancies can be attributed at least in part to deviations of the shapes of the Fermi surfaces from the spherical shape assumed in the Bloch-Gr\"uneisen theory.

We conclude that both the magnitude and $T$ dependence of the in-plane electrical resistivity of SrCu$_2$As$_2$ are adequately fitted over the entire $T$ range 1.8--300~K by the Bloch-Gr\"uneisen model in which the $T$-dependence of the resistivity arises solely from electron-phonon scattering.

\section{\label{SrCu2Sb2} Physical Properties of S\lowercase{r}C\lowercase{u}$_2$S\lowercase{b}$_2$ Crystals}

\subsection{Heat Capacity}

\begin{figure}
\includegraphics[width=3in]{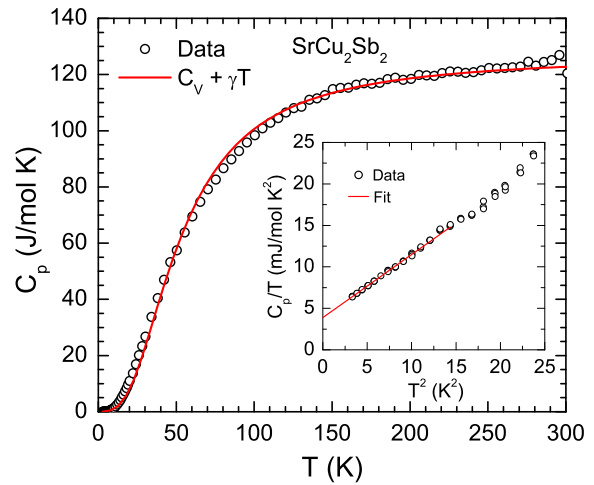}
\caption{\label{fig:fig_HC_SrCu2Sb2}(Color online)  Heat capacity $C_{\rm p}$ of a SrCu$_2$Sb$_2$ single crystal versus temperature $T$. The solid red curve is a fit of the data by the sum of the contributions from the Debye lattice heat capacity $C_{\rm V\,Debye}(T)$ and predetermined electronic heat capacity $\gamma T$ according to Eq.~(\ref{eq:Debye_HC-fit}). Inset: $C_{\rm p}/T$ vs.\ $T^2$ below 5~K. The straight line is the fit by $C_{\rm p}/T = \gamma + \beta T^2$ [Eq.~(\ref{Eq:CTT2})] for 1.8~K~$\leq T \leq$~3.5~K.}
\end{figure}

The $C_{\rm p}(T)$ data for SrCu$_2$Sb$_2$ are presented in Fig.~\ref{fig:fig_HC_SrCu2Sb2} for $1.8 \leq T \leq 300$~K. The $C_{\rm p}(T)$ data do not show any evidence for any phase transitions over this $T$ range. Like SrCu$_2$As$_2$, the value of $C_{\rm p}(T = 300~{\rm K})\approx 124$~J/mol\,K for SrCu$_2$Sb$_2$ is very close to the expected classical Dulong-Petit value $C_{\rm V}$ = 124.7~J/mol\,K\@. A fit of the low-$T$ $C_{\rm p}(T)$ data below 3.5~K by Eq.~(\ref{Eq:CTT2}) (solid line in the inset of Fig.~\ref{fig:fig_HC_SrCu2Sb2}) yields $\gamma$ = 3.9(2)~mJ/mol\,K$^2$ and $\beta$ = 0.76(3)~mJ/mol\,K$^4$. Using Eq.~(\ref{eq:DOS}), the $\gamma$ value gives ${\cal D}(E_{\rm F})$ = 1.65~states/eV\,f.u.\ for both spin directions which is larger than that of SrCu$_2$As$_2$. The Debye temperature determined using Eq.~(\ref{eq:Debye-Temp}) is $\Theta_{\rm D}$ = 234(3)~K\@.

A fit of the $C_{\rm p}(T)$ data over the whole $T$ range 1.8--300 K by the sum of an electronic contribution $\gamma T$, with $\gamma$ fixed at the above value, and the Debye lattice heat capacity contribution via Eqs.~(\ref{eq:Debye_HC-fit}) and (\ref{eq:Debye_HC}) gives $\Theta_{\rm D}$ = 214(1)~K which is somewhat smaller than the above value  $\Theta_{\rm D} = 234(3)$~K obtained from a fit to the low-$T$ $C_{\rm p}(T)$ data only.  The solid curve in Fig.~\ref{fig:fig_HC_SrCu2Sb2} is the fit of the data by the sum of electronic $\gamma T$ and Debye lattice heat capacity terms where for the latter we used our Pad\'{e} approximant function.\cite{Ryan}  The parameters obtained from our fits to $C_{\rm p}(T)$ for SrCu$_2$Sb$_2$ are summarized in Table~\ref{tab:table4}.

\subsection{\label{Sec:SrCu2Sb2MH} Magnetization and Magnetic Susceptibility}

\begin{figure}
\includegraphics[width=3in]{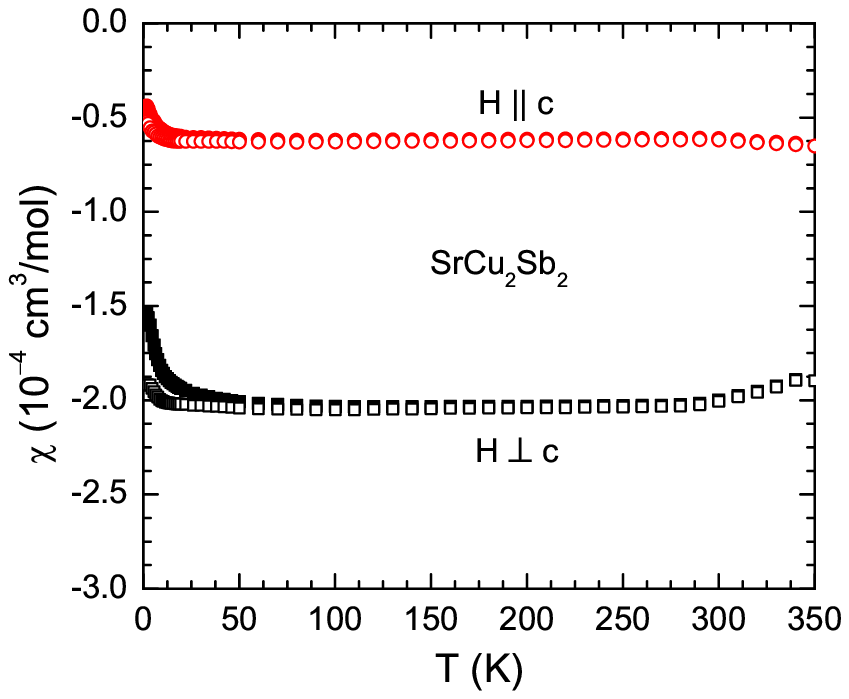}
\caption{\label{fig:fig_MT_SrCu2Sb2} (Color online) Zero-field-cooled magnetic susceptibility $\chi\equiv M/H$ of a SrCu$_2$Sb$_2$ single crystal versus temperature $T$ in the temperature range 1.8--350~K measured in a magnetic field of 3.0~T applied along the $c$-axis ($\chi_c,\ {\bf H} \parallel {\bf c}$) and in the $ab$-plane ($\chi_{ab},\ {\bf H} \perp  {\bf c}$) (solid symbols). The open symbols are the intrinsic susceptibility of SrCu$_2$Sb$_2$ after correcting for ferromagnetic and paramagnetic impurity contributions.}
\end{figure}

The ZFC anisotropic $\chi(T)$ data of a SrCu$_2$Sb$_2$ single crystal in $H$ = 3.0~T are presented in Fig.~\ref{fig:fig_MT_SrCu2Sb2} together with the intrinsic $\chi$ values after correction for PM and FM impurity contributions as described below.  Similar to the case of SrCu$_2$As$_2$, the $\chi$ has negative values. However, the $\chi$ anisotropy in SrCu$_2$Sb$_2$ is different from that of SrCu$_2$As$_2$. Here it is clear that $\chi_{c} > \chi_{ab}$ instead of $\chi_{c} < \chi_{ab}$.  The reason for this large qualitative difference in the magnetic anisotropy between the two compounds is not known in detail but is probably associated with the differences between the crystal structures of the two compounds.

\begin{figure}
\includegraphics[width=3in]{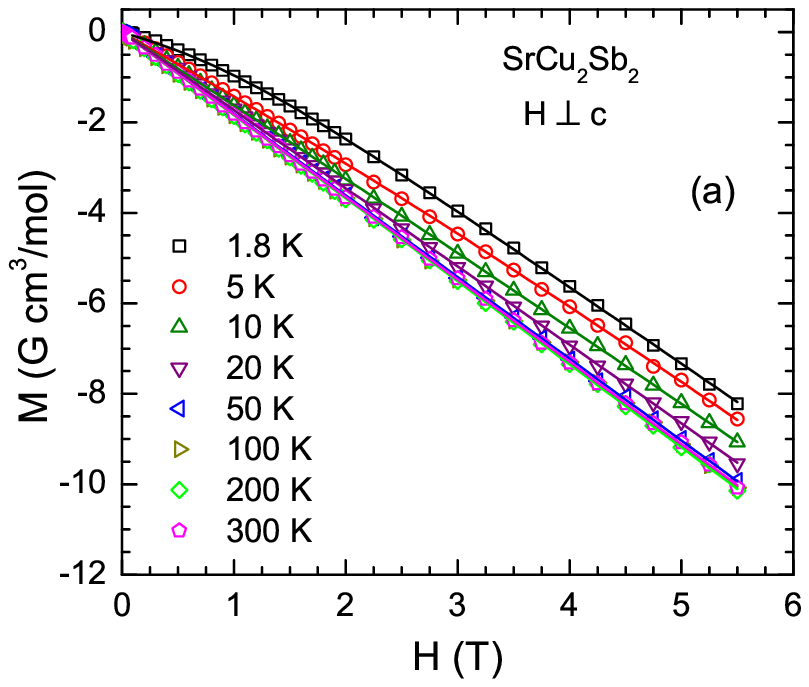}\vspace{0.1in}
\includegraphics[width=3in]{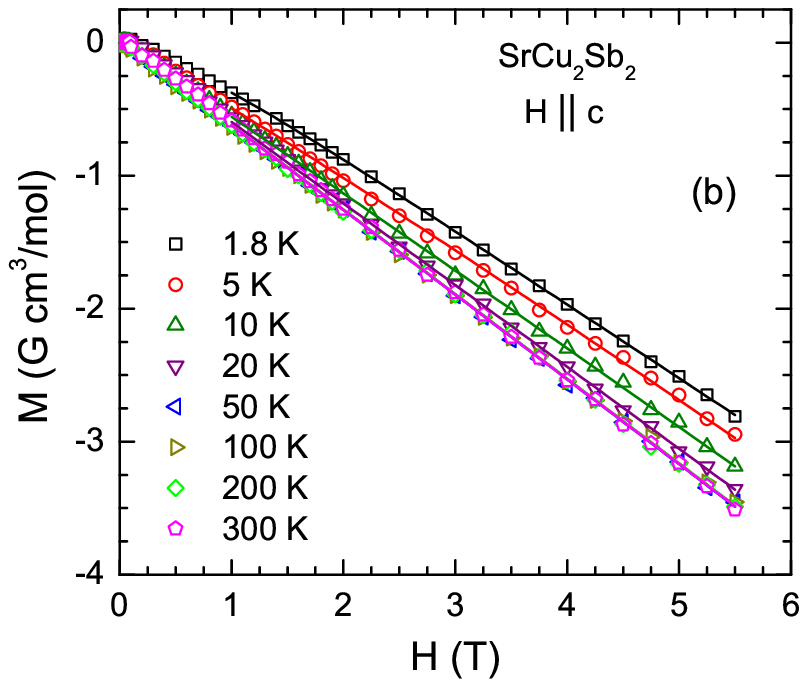}
\caption{\label{fig:fig_MH_SrCu2Sb2} (Color online) Isothermal magnetization $M$ of a SrCu$_2$Sb$_2$ single crystal versus magnetic field $H$ measured at the indicated temperatures for {\bf H} applied (a) in the $ab$-plane ($M_{ab},\ {\bf H} \perp  {\bf c}$) and (b) along the $c$-axis ($M_c,\ {\bf H} \parallel {\bf c}$). The solid curves represent fits by Eq.~(\ref{eq:MH_fit}) with the parameters in Table~\ref{tab:tableMH}, where $\theta_{\rm imp}=0$ was used for the fits.}
\end{figure}

The anisotropic isothermal $M(H)$ data for SrCu$_2$Sb$_2$ at different $T$ are presented in Fig.~\ref{fig:fig_MH_SrCu2Sb2}. The $M(H)$ curves exhibit anisotropic diamagnetic behaviors with $M_{c}(H) > M_{ab}(H)$, as expected from the $\chi(T)\equiv M(T)/H$ data in Fig.~\ref{fig:fig_MT_SrCu2Sb2}. The contributions from FM and PM impurities that cause the slight nonlinearities in the $M(H)$ isotherms at low fields were estimated by fitting the $M(H)$ data by Eqs.~(\ref{eq:MH_linear-fit}) and~(\ref{eq:MH_fit}). The $M_{\rm s}$ values were determined by fitting the $M(H)$ data above 2.0~T and at $T \geq 25$~K by Eq.~(\ref{eq:MH_linear-fit}). These values were found to be independent of $T$\@.   The fitting parameters for both $M_{ab}(H)$ and $M_{c}(H)$ isotherms at 1.8~K are listed in Table~\ref{tab:tableMH} where we found that good fits are obtained using $\theta_{\rm imp} = 0$ for both data sets.  The fits by Eq.~(\ref{eq:MH_fit}) are shown as the solid curves in Fig.~\ref{fig:fig_MH_SrCu2Sb2}.  Similar to the case of SrCu$_2$As$_2$, both $M_{\rm s}$ and the product $f_{\rm imp} S_{\rm imp}$ have different values for ${\bf H} \perp  {\bf c}$ and ${\bf H} \parallel {\bf c}$, again indicating that the magnetizations of the FM and PM impurities in SrCu$_2$Sb$_2$ are anisotropic. The $M(H)$ isotherms at the other (higher) $T$ were fitted using the values of $f_{\rm imp}$ and $S_{\rm imp}$ obtained from the 1.8~K $M(H)$ fits in order to extract the intrinsic anisotropic susceptibilities. We have corrected the measured susceptibility data for these FM and PM impurity contributions, and the resulting intrinsic susceptibilities are shown by open symbols in Fig.~\ref{fig:fig_MT_SrCu2Sb2}.  There are no significant differences between the uncorrected and corrected data above 50~K\@.  For $T \lesssim 50$~K, the PM impurity Curie-Weiss-like contribution is nearly eliminated in the corrected data, yielding an intrinsic susceptibility that is nearly independent of $T$, as expected. There is a 10\% discrepancy between the measured value of $\chi_{ab}$ (Fig.~\ref{fig:fig_MT_SrCu2Sb2}) and that obtained from the high-field ($H \geq 2$~T) slopes of the $M(H)$ isotherms for ${\bf H} \perp  {\bf c}$ [Fig.~\ref{fig:fig_MH_SrCu2Sb2}(b)]. The reason for this discrepancy is not known, but it has no significant influence on our discussions or conclusions.

The powder- and temperature-average of the intrinsic magnetic susceptibility is $\langle \chi_0\rangle = -1.56 \times 10^{-4}$~cm$^3$/mol. The different contributions to this averaged intrinsic susceptibility were estimated in the same way as for SrCu$_2$As$_2$ in Sec.~\ref{Sec:M(H,T)SrCu2As2} above. The diamagnetic core susceptibility is estimated as $\chi_{\rm {core}} = -2.08 \times 10^{-4}$~cm$^3$/mol using the atomic diamagnetic susceptibilities.\cite{Mendelsohn}  From Eq.~(\ref{eq:Chi-Pauli}), the Pauli susceptibility is $\chi_{\rm {P}} = 5.33 \times 10^{-5}$~cm$^3$/mol using ${\cal D}(E_{\rm F})$ = 1.65~states/eV\,f.u.\ for both spin directions determined above from the $\gamma$ value.  The $\chi_{\rm {L}} = -1.78 \times 10^{-5}$~cm$^3$/mol was obtained by taking $m^* =  m_{\rm e}$ in Eq.~(\ref{eq:Chi-Landau}). The Van Vleck susceptibility is then calculated from Eq.~(\ref{eq:chi}) to be $\chi_{\rm {VV}} = 1.66 \times 10^{-5}$~cm$^3$/mol.  These magnetic susceptibility contributions are summarized in Table~\ref{tab:table5}.

The powder-averaged $\chi$ of SrCu$_2$Sb$_2$ is significantly more negative than that of SrCu$_2$As$_2$, probably due at least in part to the larger $\chi_{\rm {core}}$ of Sb compared to that of As.  On the other hand, the magnitude of the anisotropy of $\chi$ is much larger in SrCu$_2$Sb$_2$ than in SrCu$_2$As$_2$, for reasons that are not clear to us.

\subsection{Electrical Resistivity}

\begin{figure}
\includegraphics[width=3in]{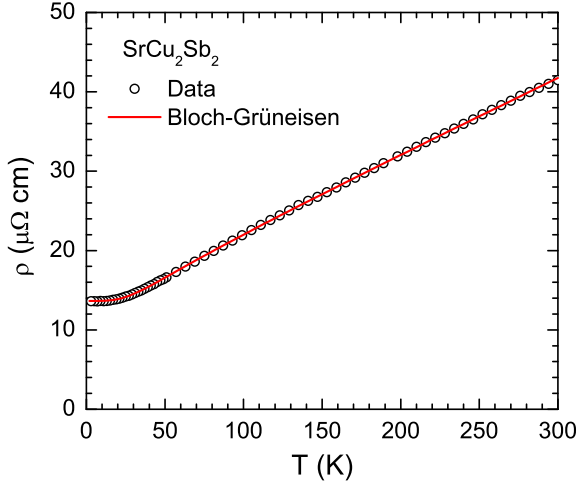}
\caption{\label{fig:fig_rho_SrCu2Sb2} (Color online) In-plane electrical resistivity $\rho$ of a SrCu$_2$Sb$_2$ single crystal versus temperature $T$. The solid curve is a fit by the Bloch-Gr\"{u}neisen model.}
\end{figure}

The in-plane $\rho(T)$ data for SrCu$_2$Sb$_2$ are presented in Fig.~\ref{fig:fig_rho_SrCu2Sb2}. The $T$ dependence of $\rho$ indicates metallic behavior with a residual resistivity $\rho_0$ = 13.6~$\mu \Omega$\,cm at 1.8~K and RRR~$\approx 3$.  The $\rho(T)$ of SrCu$_2$Sb$_2$ was analyzed using the Bloch-Gr\"{u}neisen model. A least-squares fit by Eqs.~(\ref{eq:Bloch-Gruneisen_theta_R}) and (\ref{eq:BG_fit})  using our Pad\'{e} approximant function\cite{Ryan} is shown as the red solid curve in Fig.~\ref{fig:fig_rho_SrCu2Sb2}. The fit parameters are $\rho_0$ = 13.64(1)~$\mu \Omega$\,cm, $\rho(\Theta_{\rm R})$ = 13.8(1)~$\mu \Omega$\,cm and $\Theta _{\rm{R}}$ = 153(1)~K\@.  Using Eq.~(\ref{eq:theta_R}), one obtains $\mathcal{R}(\Theta_{\rm R})$ = 14.6~$\mu \Omega$\,cm.  These parameters are summarized in Table~\ref{Tab:RhoFitParams}.

\section{\label{SrCu2AsSb} Physical Properties of Oriented S\lowercase{r}C\lowercase{u}$_2$(A\lowercase{s}$_{0.84}$S\lowercase{b}$_{0.16}$)$_2$ Crystals}

\subsection{Heat Capacity}

\begin{figure}
\includegraphics[width=3in]{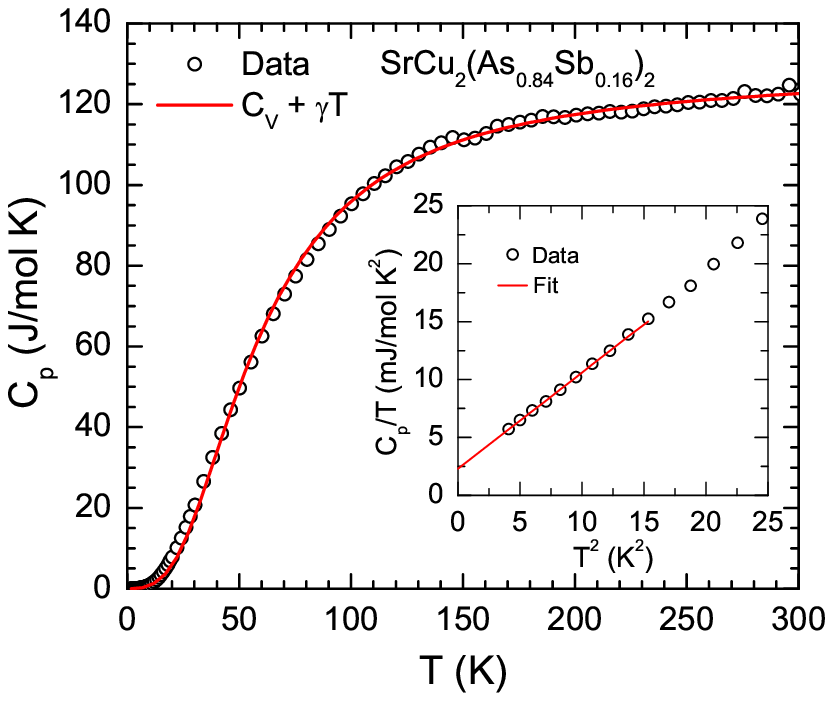}
\caption{\label{fig:fig_HC_SrCu2AsSb}(Color online) Heat capacity $C_{\rm p}$ of SrCu$_2$(As$_{0.84}$Sb$_{0.16}$)$_2$ versus temperature $T$. The solid red curve is a fit of the data by the sum of the contributions from the Debye lattice heat capacity $C_{\rm V\,Debye}(T)$ and predetermined electronic heat capacity $\gamma T$ according to Eq.~(\ref{eq:Debye_HC-fit}). Inset: $C_{\rm p}/T$ vs.\ $T^2$ below 5~K. The straight line is a fit by $C_{\rm p}/T = \gamma + \beta T^2$ [Eq.~(\ref{Eq:CTT2})] for 2.0~K~$\leq T \leq$~3.5~K.}
\end{figure}

The $C_{\rm p}(T)$ data for SrCu$_2$(As$_{0.84}$Sb$_{0.16}$)$_2$ are presented in Fig.~\ref{fig:fig_HC_SrCu2AsSb}. Like SrCu$_2$As$_2$ and SrCu$_2$Sb$_2$, the $C_{\rm p}(T)$ data of SrCu$_2$(As$_{0.84}$Sb$_{0.16}$)$_2$ reveal no evidence of any phase transitions. The $C_{\rm p}(T = 300~{\rm K}) = 122.8$~J/mol\,K is again close to the Dulong-Petit value $C_{\rm V}$ = 124.7~J/mol\,K\@. From a linear fit of $C_{\rm p}/T$ vs. $T^2$ data below 3.5~K by Eq.~(\ref{Eq:CTT2}) (solid red line in the inset of Fig.~\ref{fig:fig_HC_SrCu2AsSb}), we obtained $\gamma$ = 2.3(2)~mJ/mol\,K$^2$ and $\beta$ = 0.83(2)~mJ/mol\,K$^4$. The value of $\gamma$ yields ${\cal D}(E_{\rm F})$ = 0.97~states/eV\,f.u.\ for both spin directions from Eq.~(\ref{eq:DOS}) and the value of $\beta$ gives $\Theta_{\rm D}$ = 227(2)~K from Eq.~(\ref{eq:Debye-Temp}).  A value $\Theta_{\rm D}$ = 237(1)~K is obtained from a fit of the $C_{\rm p}(T)$ data by the Debye lattice heat capacity model [Eqs.~(\ref{eq:Debye_HC-fit}) and (\ref{eq:Debye_HC})] over the $T$ range $1.8 \leq T \leq 300$~K as shown by the solid red curve in Fig.~\ref{fig:fig_HC_SrCu2AsSb}. This value is similar to the value obtained from the low-$T$ fit.  The above specific heat parameters are compared in Table~\ref{tab:table4} with those of the other three compounds studied here.

\subsection{Magnetization and Magnetic Susceptibility}

\begin{figure}
\includegraphics[width=3in]{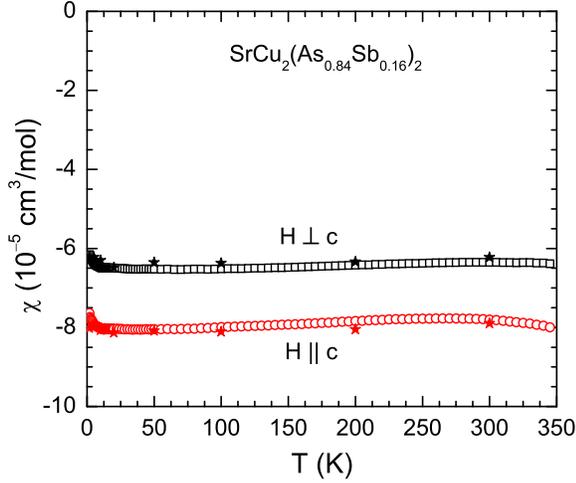}
\caption{\label{fig:fig_MT_SrCu2AsSb} (Color online) Zero-field-cooled magnetic susceptibility $\chi\equiv M/H$ of  SrCu$_2$(As$_{0.84}$Sb$_{0.16}$)$_2$ versus temperature $T$ in a magnetic field $H = 3.0$~T applied along the $c$-axis ($\chi_c,\ {\bf H} \parallel {\bf c}$) and in the $ab$-plane ($\chi_{ab},\ {\bf H} \perp  {\bf c}$). The solid stars represent the intrinsic susceptibility of SrCu$_2$(As$_{0.84}$Sb$_{0.16}$)$_2$ obtained from the high-field slopes of $M(H)$ isotherms.}
\end{figure}

The anisotropic ZFC $\chi(T)$ data of ${\rm SrCu_2(As_{0.84}Sb_{0.16})_2}$ in $H$ = 3.0~T are presented in Fig.~\ref{fig:fig_MT_SrCu2AsSb}.  Like SrCu$_2$As$_2$ and SrCu$_2$Sb$_2$, the $\chi $ of SrCu$_2$(As$_{0.84}$Sb$_{0.16}$)$_2$ is diamagnetic.  As can be seen from Fig.~\ref{fig:fig_MT_SrCu2AsSb}, $\chi_{ab} > \chi_{c}$, similar to the case of SrCu$_2$As$_2$ but in contrast to that of SrCu$_2$Sb$_2$. Thus the anisotropy seems to be a characteristic of the crystal structure, since the structure of ${\rm SrCu_2(As_{0.84}Sb_{0.16})_2}$ is the same as that of SrCu$_2$As$_2$ but different from that of SrCu$_2$Sb$_2$.

\begin{figure}
\includegraphics[width=3in]{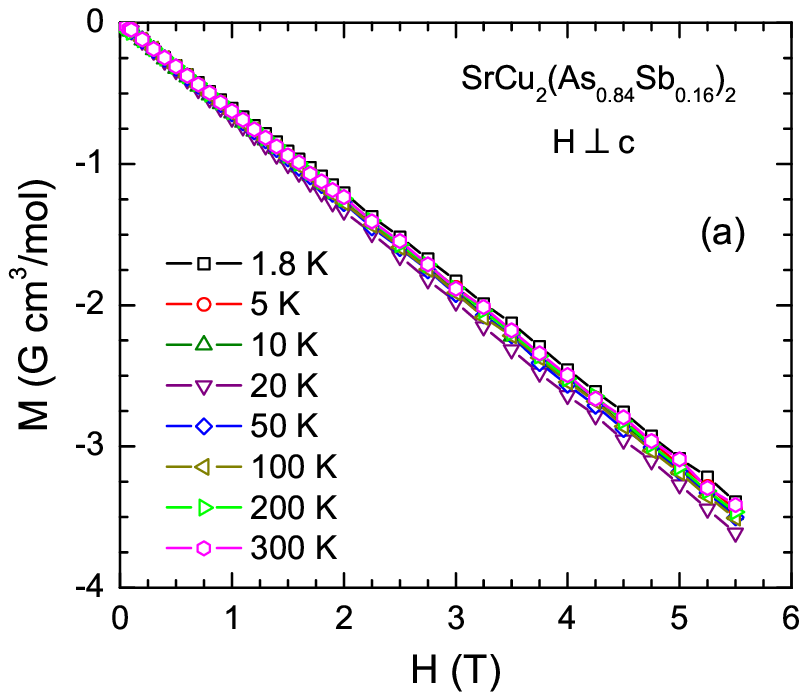}\vspace{0.1in}
\includegraphics[width=3in]{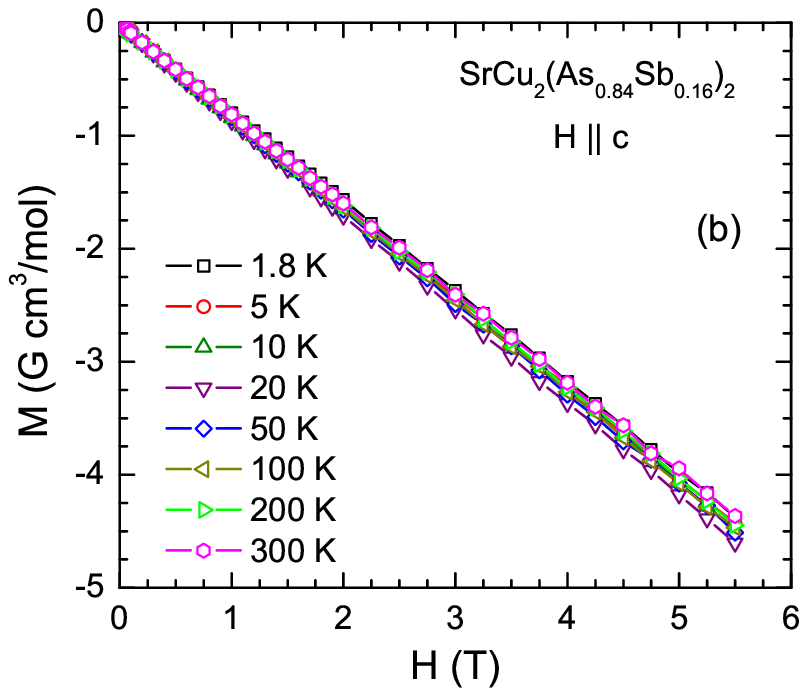}
\caption{\label{fig:fig_MH_SrCu2AsSb} (Color online) Isothermal magnetization $M$ of SrCu$_2$(As$_{0.84}$Sb$_{0.16}$)$_2$ versus magnetic field $H$ measured at the indicated temperatures for ${\bf H}$ applied (a) in the $ab$-plane ($M_{ab}, {\bf H} \perp  {\bf c}$) and (b) along the $c$-axis ($M_c, {\bf H} \parallel {\bf c}$).}
\end{figure}

The anisotropic isothermal $M(H)$ data for SrCu$_2$(As$_{0.84}$Sb$_{0.16}$)$_2$ at different $T$ are presented in Fig.~\ref{fig:fig_MH_SrCu2AsSb}. The $M(H)$ curves exhibit anisotropic behavior with $M_{ab} > M_{c}$, consistent with the $\chi(T) \equiv M(T)/H$ data discussed above. Furthermore, we observe that the $M$ is almost proportional to $H$ at each $T$, indicating only very small amounts of FM and PM impurities in the sample. The intrinsic $\chi_0$ values obtained from the high-field ($H \geq 2$~T) slopes of the $M(H)$ isotherms are shown as solid stars in Fig.~\ref{fig:fig_MT_SrCu2AsSb}.  The powder- and temperature-average $\langle \chi_0\rangle$ of $\chi_0$  and the different contributions $\chi_{\rm core}$, $\chi_{\rm P}$, $\chi_{\rm L}$ and $\chi_{\rm VV}$ to $\langle \chi_0\rangle$ estimated in the same ways as in section~\ref{SrCu2As2} are presented in Table~\ref{tab:table5}.  Interestingly, the $\langle \chi_0\rangle$ of SrCu$_2$(As$_{0.84}$Sb$_{0.16}$)$_2$ is not intermediate between those of SrCu$_2$As$_2$ and SrCu$_2$Sb$_2$.

\subsection{Electrical Resistivity}

\begin{figure}
\includegraphics[width=3in]{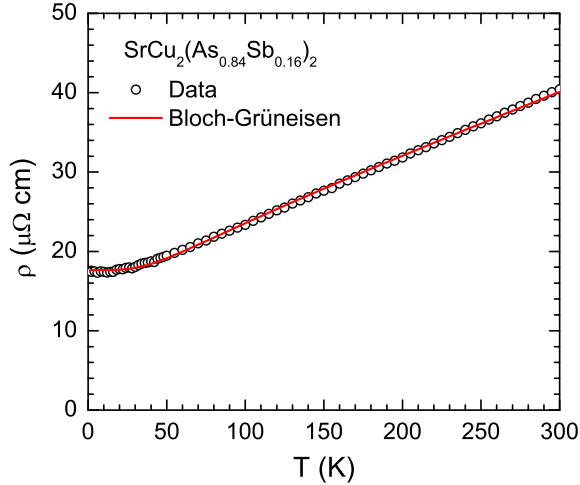}
\caption{\label{fig:fig_rho_SrCu2AsSb} (Color online) In-plane electrical resistivity $\rho$ of SrCu$_2$(As$_{0.84}$Sb$_{0.16}$)$_2$ versus temperature $T$. The solid curve is a fit by the Bloch-Gr\"uneisen model.}
\end{figure}

The in-plane $\rho(T)$ data for SrCu$_2$(As$_{0.84}$Sb$_{0.16}$)$_2$ are presented in Fig.~\ref{fig:fig_rho_SrCu2AsSb}. Like SrCu$_2$As$_2$ and SrCu$_2$Sb$_2$, the $\rho(T)$ of SrCu$_2$(As$_{0.84}$Sb$_{0.16}$)$_2$ exhibits metallic behavior with $\rho_0$ = 17.6~$\mu \Omega$\,cm, which is the largest value among the four compounds studied here (see Table~\ref{Tab:RhoFitParams}).  The ${\rm RRR}\approx 2$ is smaller than those of the end-point compounds SrCu$_2$As$_2$ (${\rm RRR} \approx 6$) and SrCu$_2$Sb$_2$ (${\rm RRR} \approx 3$) as expected from the atomic disorder on the As/Sb sublattices, and also of BaCu$_2$Sb$_2$ (${\rm RRR} \approx 15$, see below) with yet another structure.  The large value of $\rho_0$ and the low RRR value for SrCu$_2$(As$_{0.84}$Sb$_{0.16}$)$_2$ are consistent with this compound being an $sp$-metal because the conduction carriers evidently strongly sense the atomic disorder on the As/Sb sublattices as they hop along these sublattices.  An analysis of the $\rho(T)$ data of SrCu$_2$(As$_{0.84}$Sb$_{0.16}$)$_2$ using the Bloch-Gr\"{u}neisen model in Sec.~\ref{SecSrCu2As2Rho} (solid curve in Fig.~\ref{fig:fig_rho_SrCu2AsSb}) gives $\rho_0$ = 17.65(5)~$\mu \Omega$\,cm, $\rho(\Theta_{\rm R})$ = 16.5(4)~$\mu \Omega$\,cm, $\Theta _{\rm{R}}$ = 225(6)~K, and using Eq.~(\ref{eq:theta_R}), $\mathcal{R}(\Theta_{\rm R})$ = 17.4~$\mu\Omega$\,cm.  These parameters are compared  in Table~\ref{Tab:RhoFitParams} with those obtained for the other three compounds studied here.

\section{\label{BaCu2Sb2} Physical Properties of B\lowercase{a}C\lowercase{u}$_2$S\lowercase{b}$_2$ Crystals}

\subsection{Heat Capacity}

\begin{figure}
\includegraphics[width=3in]{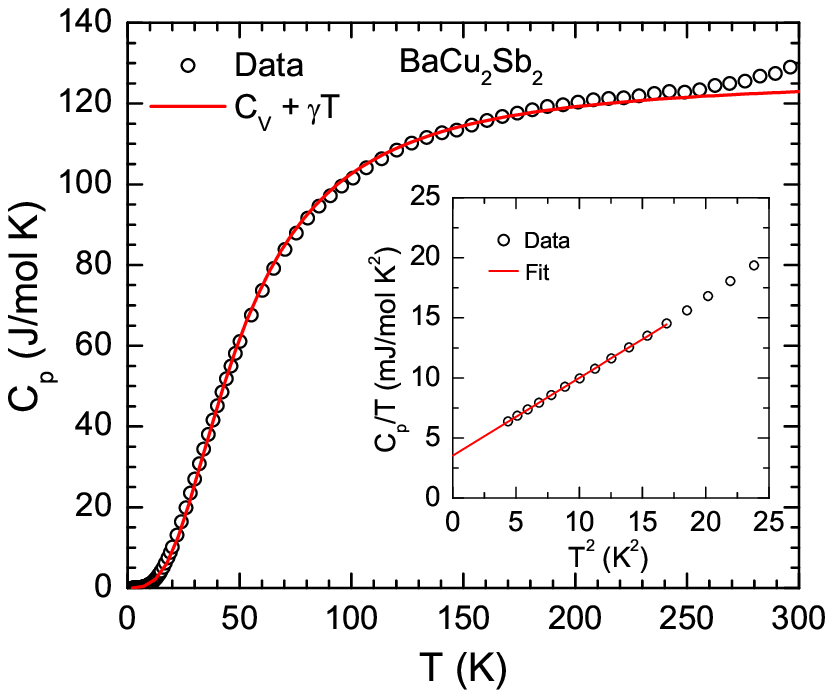}
\caption{\label{fig:fig_HC_BaCu2Sb2}(Color online) Heat capacity $C_{\rm p}$ of a BaCu$_2$Sb$_2$ single crystal versus temperature $T$. The solid red curve is a fit of the data by the sum of the contributions from the Debye lattice heat capacity $C_{\rm V\,Debye}(T)$ and predetermined electronic heat capacity $\gamma T$ according to Eq.~(\ref{eq:Debye_HC-fit}). Inset: $C_{\rm p}/T$ vs.\ $T^2$ below 5~K\@. The straight line is a fit by Eq.~(\ref{Eq:CTT2}) for 2.0~K~$\leq T \leq$~3.5~K.}
\end{figure}

The $C_{\rm p}(T)$ data for a BaCu$_2$Sb$_2$ crystal are presented in Fig.~\ref{fig:fig_HC_BaCu2Sb2}. The data reveal no evidence for any phase transitions.  The data exhibit a saturation value of $\sim$ 125~J/mol\,K near room temperature which is close to the expected classical Dulong-Petit value of $C_{\rm V}$ = 124.7~J/mol\,K\@. The origin of the weak upturn in $C_{\rm p}(T)$ above 250~K is unknown and may be spurious, perhaps arising from a small inaccuracy in the high-$T$ addenda calibration.  We obtained the coefficients $\gamma$ = 3.5(2)~mJ/mol\,K$^2$ and $\beta$ = 0.65(2)~mJ/mol\,K$^4$ from a linear fit of the low-$T$ $C_{\rm p}/T $ vs. $T^2$ data below 3.5~K by Eq.~(\ref{Eq:CTT2}) which is shown as the solid straight line in the inset of Fig.~\ref{fig:fig_HC_BaCu2Sb2}. The $\gamma$ value and Eq.~(\ref{eq:DOS}) yield ${\cal D}(E_{\rm F})$ = 1.49~states/eV\,f.u.\ for both spin directions.  The $\beta$ and Eq.~(\ref{eq:Debye-Temp}) yield $\Theta_{\rm D}$ = 246(3)~K\@. From a fit of the $C_{\rm p}(T)$ data over the full $T$ range by a sum of electronic and lattice contributions via Eqs.~(\ref{eq:Debye_HC-fit}) and (\ref{eq:Debye_HC}), we obtained a good fit using $\Theta_{\rm D}$ = 204(1)~K as shown by the solid red curve in Fig.~\ref{fig:fig_HC_BaCu2Sb2}.  These specific heat parameters are summarized in Table~\ref{tab:table4}.

\subsection{Magnetization and Magnetic Susceptibility}

\begin{figure}
\includegraphics[width=3in]{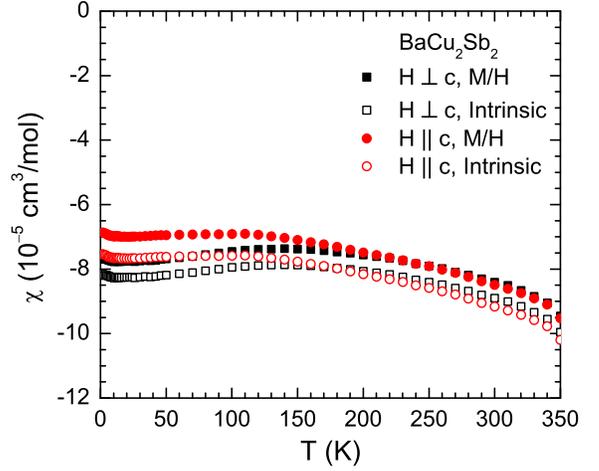}
\caption{\label{fig:fig_MT_BaCu2Sb2} (Color online) Zero-field-cooled magnetic susceptibility $\chi\equiv M/H$ of a BaCu$_2$Sb$_2$ single crystal versus temperature $T$ measured in $H= 3.0$~T applied along the $c$-axis ($\chi_c,\ {\bf H} \parallel {\bf c}$) and in the $ab$-plane ($\chi_{ab},\ {\bf H} \perp  {\bf c}$) (solid symbols). The intrinsic susceptibilities after correcting for the small ferromagnetic impurity contributions are shown as open symbols.}
\end{figure}

The anisotropic $\chi(T)$ data of a BaCu$_2$Sb$_2$ single crystal in $H$ = 3.0~T are plotted in Fig.~\ref{fig:fig_MT_BaCu2Sb2}. Like the other three compounds, the susceptibility is diamagnetic over the whole $T$ range.  The data are weakly $T$-dependent.  A weak anisotropy is seen below $\sim 170$~K with $\chi_{c} > \chi_{ab}$ as in SrCu$_2$Sb$_2$.

\begin{figure}
\includegraphics[width=3in]{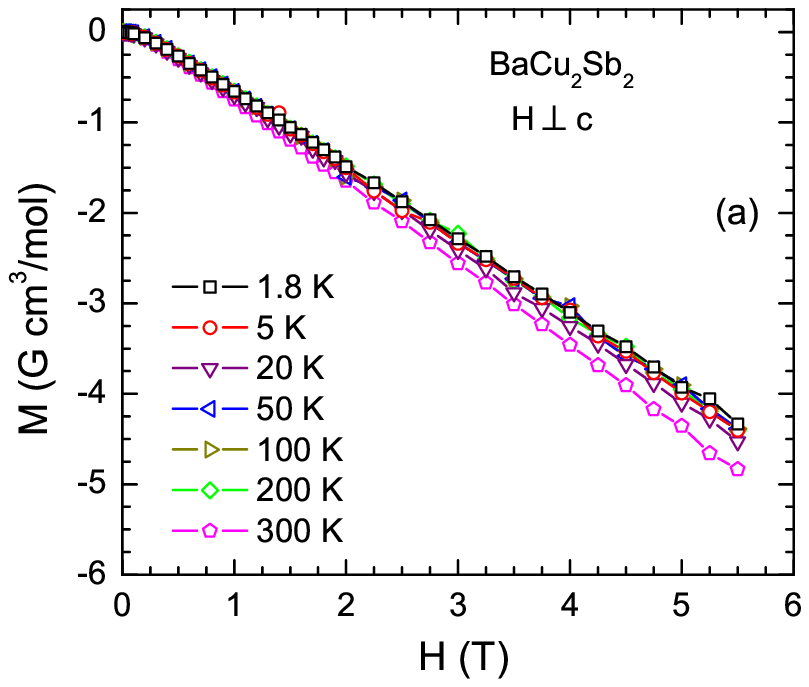}\vspace{0.1in}
\includegraphics[width=3in]{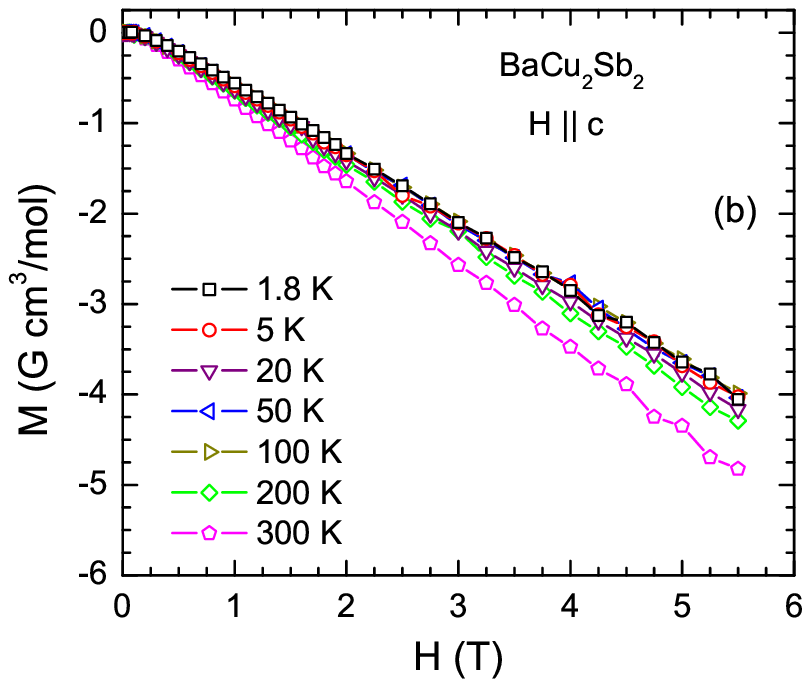}
\caption{\label{fig:fig_MH_BaCu2Sb2} (Color online) Isothermal magnetization $M$ of a BaCu$_2$Sb$_2$ single crystal versus magnetic field $H$ measured at the indicated temperatures for {\bf H} applied (a) in the $ab$-plane ($M_{ab},\ {\bf H} \perp  {\bf c}$) and (b) along the $c$-axis ($M_c,\ {\bf H} \parallel {\bf c}$).}
\end{figure}

The $M(H)$ isotherms at different $T$ are shown for BaCu$_2$Sb$_2$ in Fig.~\ref{fig:fig_MH_BaCu2Sb2}. Consistent with the $\chi(T)$ results in Fig.~\ref{fig:fig_MT_BaCu2Sb2}, the $M(H)$ curves of BaCu$_2$Sb$_2$ exhibit weakly anisotropic diamagnetic behavior with $M_{c}(H) > M_{ab}(H)$).  The isotherms provide evidence for the presence of FM impurities but no evidence for saturable PM impurities in the sample.  In order to obtain the intrinsic $\chi$ the $M(H)$ isotherms at each $T$ were fitted by Eq.~(\ref{eq:MH_linear-fit}) at high fields $H \geq 2$~T  and the resulting $T$-independent FM impurity contributions ($M_{\rm s}^{ab}$ = 0.15 G\,cm$^3$/mol and $M_{\rm s}^{c}$ = 0.20 G\,cm$^3$/mol) were subtracted from the $M(T)$ measured at $H = 3$~T\@. The different values of $M_{\rm s}^{ab}$  and $M_{\rm s}^{c}$ suggest an anisotropic FM impurity contribution. To give the scale of the FM impurity magnetization, the above value $M_{\rm s}^{ab}$ = 0.15 G\,cm$^3$/mol is equivalent to the magnetization contribution from 12 molar ppm of Fe metal impurities.  The intrinsic anisotropic susceptibilities obtained after subtracting the FM impurity contributions are shown by open symbols in Fig.~\ref{fig:fig_MT_BaCu2Sb2}.  The corrections for the FM impurities are seen to be rather small.

We have estimated the different contributions to the intrinsic $\chi$ of BaCu$_2$Sb$_2$ using the same methods as for the previous three compounds. The powder- and temperature-average over the $T$ range 10 to 100~K is $\langle \chi_0\rangle = -8.02 \times 10^{-5}$~cm$^3$/mol, and we infer $\chi_{\rm {core}} = -2.31 \times 10^{-4}$~cm$^3$/mol, $\chi_{\rm {P}} = 4.81 \times 10^{-5}$~cm$^3$/mol, and $\chi_{\rm {L}} = -1.60 \times 10^{-5}$~cm$^3$/mol (using $m^* =  m_{\rm e}$ in Eq.~\ref{eq:Chi-Landau}). We thus obtain $\chi_{\rm {VV}} = 1.19 \times 10^{-4}$~cm$^3$/mol from Eq.~(\ref{eq:chi}). A comparison of $\chi_{\rm {VV}}$ with those of the other compounds (Table~\ref{tab:table5}) suggests that $\chi_{\rm {VV}}$ is largest in BaCu$_2$Sb$_2$.

\subsection{Electrical Resistivity}

\begin{figure}
\includegraphics[width=3in]{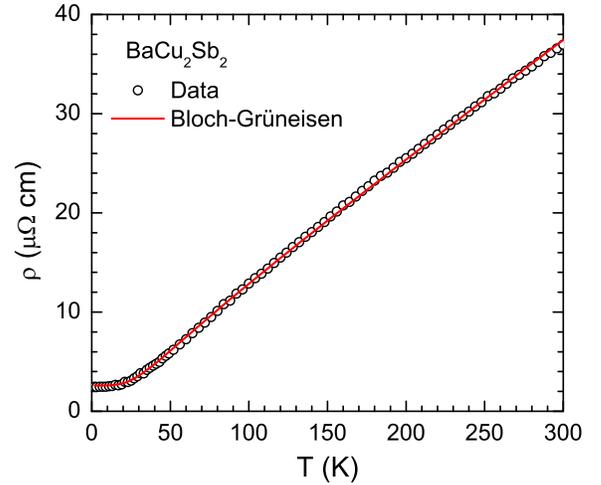}
\caption{\label{fig:fig_rho_BaCu2Sb2} (Color online) In-plane electrical resistivity $\rho$ of a BaCu$_2$Sb$_2$ single crystal versus temperature $T$. The solid red curve is a fit by the Bloch-Gr\"{u}neisen model.}
\end{figure}

The $\rho(T)$ data for BaCu$_2$Sb$_2$ are presented in Fig.~\ref{fig:fig_rho_BaCu2Sb2}. The $\rho(T)$ exhibits metallic behavior with $\rho_0$ = 2.5~$\mu \Omega$\,cm at 1.8~K and ${\rm RRR}\approx 15$, indicating that this crystal has the highest perfection of the four compounds studied here.  An analysis of the $\rho(T)$ data in terms of the Bloch-Gr\"{u}neisen model in Eq.~(\ref{eq:BG_fit}) yielded a good fit with $\rho_0$ = 2.59(3)~$\mu \Omega$\,cm, $\rho(\Theta_{\rm R})$ = 17.9(2)~$\mu \Omega$\,cm, $\Theta _{\rm{R}}$ = 160(2)~K, and using Eq.~(\ref{eq:theta_R}), $\mathcal{R}(\Theta_{\rm R})$ = 18.9~$\mu \Omega$\,cm.  The fit is shown as the solid red curve in Fig.~\ref{fig:fig_rho_BaCu2Sb2}. The fit parameters are summarized in Table~\ref{Tab:RhoFitParams} for comparison with those of the other samples.

\section{\label{Discussion} Discussion}

The electronic and magnetic properties of SrCu$_2$As$_2$, SrCu$_2$Sb$_2$, SrCu$_2$(As$_{0.84}$Sb$_{0.16}$)$_2$ and BaCu$_2$Sb$_2$ indicate that they are all nonmagnetic $sp$-band metals.  The magnetic susceptibilities show temperature-independent diamagnetism and the heat capacities show small Sommerfeld electronic specific heat coefficients, both indicative of small densities of states at the Fermi energy.  The electrical resistivities show no evidence for a $T^2$ term that might indicate significant $d$-electron scattering, and instead are all well-described by the Bloch-Gr\"uneisen theory for scattering of $sp$-band electrons by longitudinal acoustic lattice vibrations.

For ${\rm SrCu_2As_2}$, our results are consistent with the band structure calculations for ${\rm ThCr_2Si_2}$-type ${\rm SrCu_2As_2}$ and ${\rm BaCu_2As_2}$.\cite{Singh}  According to these calculations, the Cu~$3d$ bands are centered about 3~eV below $E_{\rm F}$ and are completely occupied, with little Cu~$3d$ character in the bands at $E_{\rm F}$. The same behavior occurs in Cu metal, where the Cu~$3d$ bands are centered 3.4~eV below $E_{\rm F}$ with a total $3d$-band half-width of 1.4~eV.\cite{Yi2010}  The band calculations thus indicate that ${\rm SrCu_2As_2}$ and ${\rm BaCu_2As_2}$ are $sp$-band metals like Cu metal.\cite{Singh}

In a chemical picture the Cu atoms in these compounds have a $3d^{10}$ electronic configuration with a filled $d$-shell and a formal oxidation state of Cu$^{+1}$, where the Cu~$4s^1$ valence electrons are itinerant.  Assuming Cu$^{+1}$, Sr$^{+2}$ and Ba$^{+2}$ oxidation states in ${\rm (Sr,Ba)Cu_2As_2}$ yields an unusual oxidation state of As$^{-2}$ for the As atoms, in contrast to the oxidation states Fe$^{+2}$ and As$^{-3}$ in (Ca,Sr,Ba)Fe$_2$As$_2$.\cite{Johnston2010}  This in turn suggests the presence of a covalent bond between the As atoms in adjacent layers that are directly above/below each other along the $c$-axis [see Fig.~\ref{fig:structure_fig}(a)], i.e., ${\rm As^{-2}\equiv [As}$--${\rm As]^{-4}/2}$.  Such covalent interlayer pnictogen bonding has been known since the 1980's for transition metal phosphides with the ${\rm ThCr_2Si_2}$ structure.\cite{Hoffman1985, Reehuis1998, Reehuis1990}  The variable degree of P--P covalent bonding in those compounds is reflected in a variable interlayer \mbox{P--P} distance $d_{\rm P-P}$ and a resulting variable $c/a$ ratio.\cite{Jeitschko1985}  Notably, first-order transitions with increasing pressure between non-bonding and partially or fully bonding interplane P--P linkages were found for ${\rm SrRh_2P_2}$ (Refs.~\onlinecite{Wurth1997}, \onlinecite{Huhnt1997}) and ${\rm SrNi_2P_2}$.\cite{Huhnt1997a}

In the recent field of iron arsenide and chalcogenide high-$T_{\rm c}$ superconductivity,\cite{Johnston2010} Kreyssig, Goldman et al.\cite{Kreyssig2008, Goldman2009} discovered a strong decrease in the $c$-axis lattice parameter and unit cell volume of ${\rm ThCr_2Si_2}$-type ${\rm CaFe_2As_2}$ in a first-order transition with increasing pressure $p$ and dubbed this phase a ``collapsed tetragonal'' (cT) phase, as opposed to the ambient $p$ (uncollapsed) tetragonal (T) phase.  The T to cT transition occurs with increasing $p$ at 0.4~GPa at low~$T$, increasing to 1.7~GPa at 300~K.\cite{Goldman2009}  This first-order T to cT phase transition can be viewed as resulting from the discontinuous creation with increasing $p$ of covalent As--As interlayer bonds,\cite{Goldman2009} as occurs at ambient $p$ or with increasing $p$ in some ${\rm ThCr_2Si_2}$-type phosphides as discussed above.  The ``cT'' terminology has been widely adopted by the community working on iron arsenide superconductors.

The work by Kreyssig, Goldman et al.\cite{Kreyssig2008, Goldman2009} motivated many additional experimental and theoretical studies of the occurrence and properties of the cT phase in ${\rm ThCr_2Si_2}$-type iron arsenides.  A synchrotron x-ray diffraction study of ${\rm SrFe_2As_2}$ at room $T$ revealed a \mbox{$p$-induced} second-order transition at 10~GPa from the T to a cT structure for which electronic structure calculations indicated the formation of As--As interlayer bonding as responsible for the lattice collapse.\cite{Kasinathan2011}  Another synchrotron x-ray diffraction study at room $T$ showed that a cT transition occurs in ${\rm BaFe_2As_2}$ at an even higher $p\approx 27$~GPa, much higher than the transition $p$ of 1.7~GPa found for the Ca compound at this $T$.\cite{Mittal2011}   A previous high-$p$ synchrotron x-ray diffraction study of the same compound found that the onset $p$ of the cT phase at room $T$ is 17~GPa under nonhydrostatic conditions and 22~GPa under hydrostatic conditions.\cite{Uhoya2010}  Synchrotron x-ray powder diffraction measurements versus $T$ and $p$ were also carried out for ${\rm Ca_{0.67}Sr_{0.33}Fe_2As_2}$, with results intermediate between those obtained previously for the pure Ca and Sr end-members.\cite{Jeffries2012}  In the Ca(Fe$_{1-x}$Co$_x)_2$As$_2$ system, the onset $p$ for the cT transition at low $T$ for $x = 0.032$, 0.051 and 0.063 decreases with increasing $x$.\cite{Prokes2012}  At ambient $p$, a $T$-induced transition from the high-$T$ T phase to the low-$T$ cT phase occurs below $\sim 90$~K for $x=0.05$--0.09 in the system CaFe$_2$(As$_{1-x}$P$_x)_2$.\cite{Kasahara2011}  The doped compounds Ca(Fe$_{1-x}$Rh$_x)_2$As$_2$ similarly exhibit an ambient pressure cT phase up to 50~K for $x = 0.024$ and up to 300~K for $x = 0.19$.\cite{Danura2011}  The $T$-$p$ phase diagrams of lanthanide-doped Ca$_{1-x}$$R_x$Fe$_2$As$_2$ ($R$ = La, Ce, Pr, Nd) compounds, including T and cT phase regions, were also determined.\cite{Saha2012}

In the following Sec.~\ref{Sec:PnicBonding} we examine more quantitatively the systematics of interlayer P--P and As--As bonding in $3d$~transition metal phospides and arsenides with the ${\rm ThCr_2Si_2}$ structure, and we will find that SrCu$_2$As$_2$ is in  the cT phase.  Then in Sec.~\ref{Sec:MagPropCorr} we discuss the relationship of the magnetic character of the ${\rm ThCr_2Si_2}$-type $3d$~transition metal arsenides and phosphides with the degree of interlayer pnictogen bonding and discuss how the properties of SrCu$_2$As$_2$ fit into this scheme.  In Sec.~\ref{Sec:CuDoping} we discuss doping scenarios when Fe is partially substituted by other transition metals, including Cu, in the ${\rm (Ca,Sr,Ba)Fe_2As_2}$ compounds. 

\subsection{\label{Sec:PnicBonding} Pnictogen Interlayer Bonding}

\begin{table*}
\caption{\label{tab:AT2X2} Magnetic and structural properties of $AM_2X_2$ compounds ($A$ = Ca, Sr, Ba; $M$ = Cr, Mn, Fe, Co, Ni, Cu; $X$ = P, As) with the ThCr$_2$Si$_2$ structure.  $d_{X-X} = (1-2z_X)c$ and $d_{X-X}^{\rm intra} = \left(2z_X-\frac{1}{2}\right)c$ are the interlayer and intralayer distances between the pnictogen atoms, respectively.  Separate references are given for the crystal data and magnetic properties.  For the latter, the abbreviation AFM means antiferromagnetic and SDW means itinerant spin density wave.}
\begin{ruledtabular}
\begin{tabular}{lcccccccl}
 Compound & $a$ (\AA)& $c$ (\AA)& $c/a$ & $z_{\rm X}$ & $d_{X-X}$ (\AA) & $d_{X-X}^{\rm intra}$ (\AA)& Ref. & Magnetic properties\\
\hline
CaFe$_2$As$_2$ (T)  & 3.872(9) & 11.730(2) & 3.03 & 0.3665(9) & 3.13(2) & 2.73(2) & [\onlinecite{Wu}] & AFM SDW [\onlinecite{Wu}] \\
CaFe$_2$As$_2$ (cT)& 3.9792(1) & 10.6379(6) & 2.67 & 0.3687(7) & 2.79(2) & 2.52(2) & [\onlinecite{Kreyssig2008}] & Fe moment quenched [\onlinecite{Kreyssig2008,Goldman2009}] \\
CaCo$_2$As$_2$ & 3.9831(2) & 10.2732(6) & 2.58 & 0.3664(2) & 2.745(4) & 2.392(4) & [\onlinecite{Anand2012}] & AFM [\onlinecite{Anand2012, Cheng2012}] \\
CaNi$_2$As$_2$ & 4.053(6) & 9.90(2) & 2.44 & 0.370 & 2.574 & 2.376 & [\onlinecite{Pfisterer,Pfisterer1983}] & Pauli paramagnetic [\onlinecite{Pfisterer1983}]\\
CaCu$_2$As$_2$ & 4.129(1) & 10.251(1) & 2.48 & 0.3799(2) & 2.462(4) & 2.663(4) & [\onlinecite{Pilchowski1990}] & \\
\hline
SrCr$_2$As$_2$ & 3.918(3) & 13.05(1) & 3.33 & 0.367 & 3.471 & 3.054 & [\onlinecite{Pfisterer,Pfisterer1983}] & AFM [\onlinecite{Pfisterer1983}]\\
SrFe$_2$As$_2$ & 3.9289(3) & 12.3172(12) & 3.13 & 0.36035(5) & 3.440(2) & 2.718(2) & [\onlinecite{Saha2011}] & AFM SDW [\onlinecite{Yan}] \\
SrCo$_2$As$_2$ & 3.935(7) & 11.83(2) & 3.01 & 0.362 & 3.265 & 2.650 & [\onlinecite{Pfisterer,Pfisterer1983}] & Paramagnetic [\onlinecite{Jasper2008}]\\
SrNi$_2$As$_2$ & 4.1374(8) & 10.188(4) & 2.46 & 0.3634(1) & 2.783(2) & 2.311(2) & [\onlinecite{Bauer2008}]& No magnetic order [\onlinecite{Bauer2008}]\\
SrCu$_2$As$_2$ & 4.2725(1) & 10.2000(3) & 2.39 & 0.3789(1) & 2.470(2)& 2.630(2) & & Nonmagnetic [This work]\\
\hline
BaCr$_2$As$_2$ & 3.963(3) & 13.60(1) & 3.43 & 0.361 & 3.781 & 3.019 & [\onlinecite{Pfisterer,Pfisterer1983}] & Itinerant AFM [\onlinecite{Singh2008}] \\
BaMn$_2$As$_2$ & 4.1686(4) & 13.473(3) & 3.23 & 0.3615(3) & 3.732(8) & 3.004(8) & [\onlinecite{Singh2009}] & AFM [\onlinecite{Singh2009}] \\
BaFe$_2$As$_2$ & 3.9633(4) & 13.022(2) & 3.29 & 0.35424(6) & 3.797(2) & 2.715(2) & [\onlinecite{Albenque2010}] & AFM SDW [\onlinecite{Wang}] \\
BaCo$_2$As$_2$ & 3.958(5) & 12.67(2) & 3.20 & 0.361 & 3.522 & 2.813 & [\onlinecite{Pfisterer,Pfisterer1983}] & Renormalized paramagnet [\onlinecite{Sefat2009}] \\
BaNi$_2$As$_2$ & 4.112(4) & 11.54(2) & 2.81 & 0.3476(3) & 3.517(7) & 2.253(7) & [\onlinecite{Ronning2008}] & Pauli paramagnetic [\onlinecite{Pfisterer1983}]\\
BaCu$_2$As$_2$ & 4.446(5)  & 10.07(1) & 2.26 & 0.374 & 2.538 & 2.497 & [\onlinecite{Pfisterer,Pfisterer1983}] & Pauli paramagnetic [\onlinecite{Pfisterer1983}]\\
\hline
CaFe$_2$P$_2$ & 3.855(1) & 9.985(1) & 2.59 & 0.3643(3) & 2.710(6) &  2.283(6) & [\onlinecite{Mewis1980}]  & Pauli paramagnetic? [\onlinecite{Raffius1991}]\\
CaCo$_2$P$_2$ & 3.858(1) & 9.593(1) & 2.49 & 0.3721(4) & 2.454(8) & 2.343(8) & [\onlinecite{Mewis1980}]  & Itinerant AFM [\onlinecite{Reehuis1990,Reehuis1998}]\\
CaNi$_2$P$_2$ & 3.916(1) & 9.363(1) & 2.39 & 0.3774(5) & 2.296(9) & 2.386(9) & [\onlinecite{Mewis1980}]  & Pauli paramagnetic [\onlinecite{Jeitschko1987}]\\
CaCu$_{1.75}$P$_2$ & 4.014(1) & 9.627(1) & 2.40 & 0.3831(4) & 2.251(8) & 2.563(8) & [\onlinecite{Mewis1980}] & \\
\hline
SrFe$_2$P$_2$ & 3.825(1) & 11.612(1) & 3.04 & 0.3521(8)  & 3.43(2) & 2.37(2)&  [\onlinecite{Mewis1980}] & Pauli paramagnetic [\onlinecite{Morsen1988}]\\
SrCo$_2$P$_2$ & 3.794(1) & 11.610(1) & 3.06 & 0.3525(5) & 3.42(2) & 3.38(2) & [\onlinecite{Mewis1980}] & Stoner enhanced paramagnet [\onlinecite{Jia2011}]\\
SrNi$_2$P$_2$ & 3.948(1) & 10.677(3) & 2.70 & 0.3539(3) & 3.120(7) & 2.219(7) & [\onlinecite{Keimes1997}] & Paramagnetic [\onlinecite{Ronning2009}]\\
SrCu$_{1.75}$P$_2$ & 4.166(1) & 9.607(1)  & 2.31 & 0.3805(4) & 2.296(8) & 2.507(8) & [\onlinecite{Mewis1980}]  & \\
\hline
BaMn$_2$P$_2$ & 4.037(1) & 13.061(1) & 3.24 & 0.3570(3) & 3.735(8) & 2.795(8) & [\onlinecite{Mewis1980}] & AFM [\onlinecite{Brock1994}] \\
BaFe$_2$P$_2$ & 3.840(1) & 12.442(1) & 3.24 & 0.3456(4) & 3.84(1) & 2.38(1) & [\onlinecite{Mewis1980}]  & Pauli paramagnetic? [\onlinecite{Raffius1991}] \\
BaCo$_2$P$_2$ & 3.7994(3) & 12.391(1) & 3.26 & 0.3461(3) & 3.814(8) & 2.382(8) & [\onlinecite{Pandey2012}] \\
BaNi$_2$P$_2$ & 3.947(1) & 11.820(1) & 2.99 & 0.3431(3) & 3.709(7) & 2.201(7) & [\onlinecite{Keimes1997}] & Pauli paramagnetic [\onlinecite{Mine2008}] \\

\end{tabular}
\end{ruledtabular}
\end{table*}

\begin{figure}
\includegraphics[width=3in]{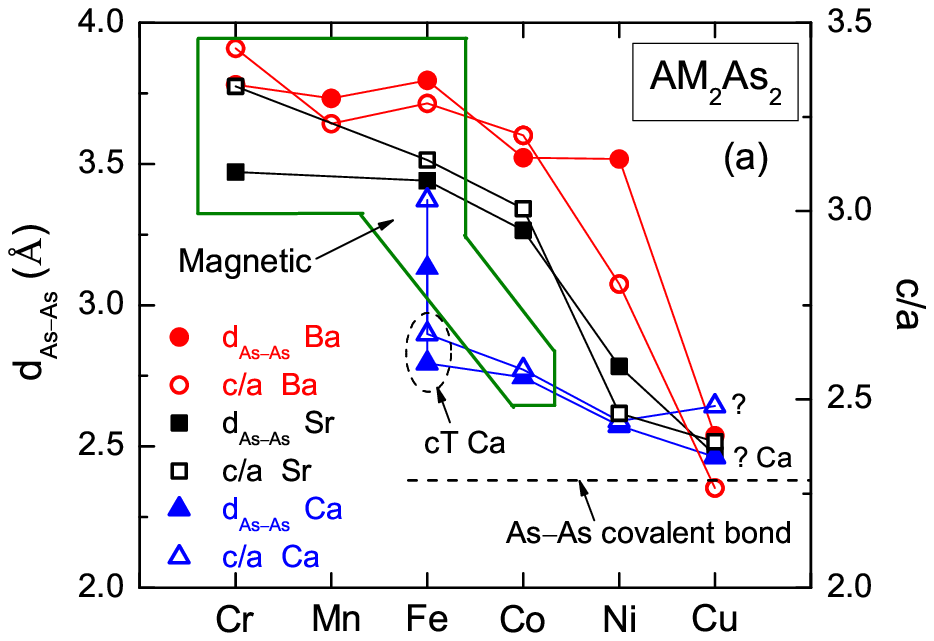}\vspace{0.1in}
\includegraphics[width=3in]{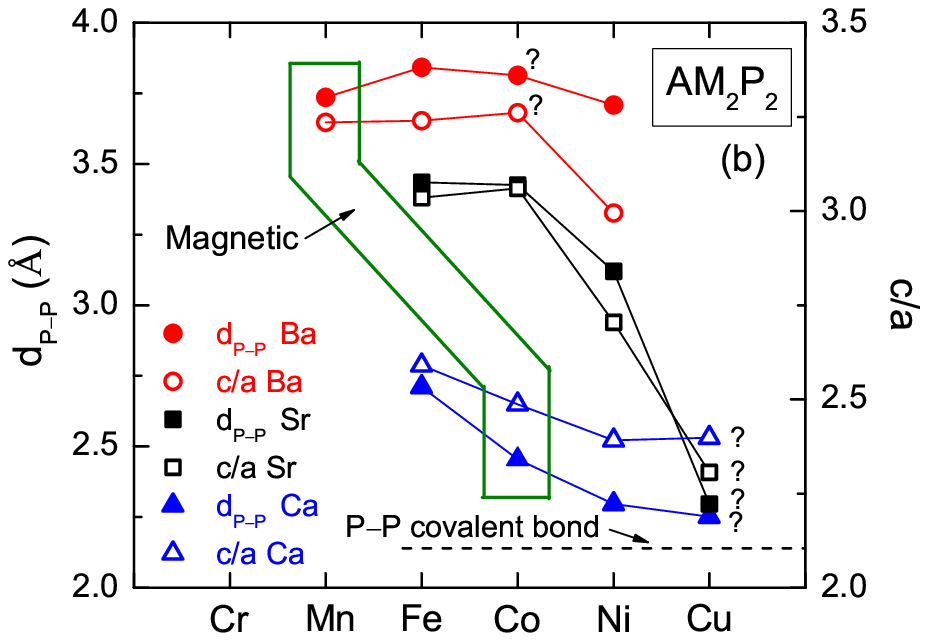}
\caption{(Color online) Interlayer As--As distance $d_{\rm As-As}$ (a) and interlayer P--P distance $d_{\rm P-P}$ (b) (left ordinates) and c/a ratio (right ordinates) versus $3d$~transition metal $M$ for $AM_2{\rm As_2}$ and $AM_2{\rm P_2}$ compounds, respectively.  The compounds with magnetically ordered ground states are enclosed in the green boxes.  The compounds for which the magnetic ground states are not known are marked with ``?''.  The references are given in Table~\ref{tab:AT2X2}.}
\label{fig:fig_AT2X2}
\end{figure}

\begin{figure}
\includegraphics[width=2.75in]{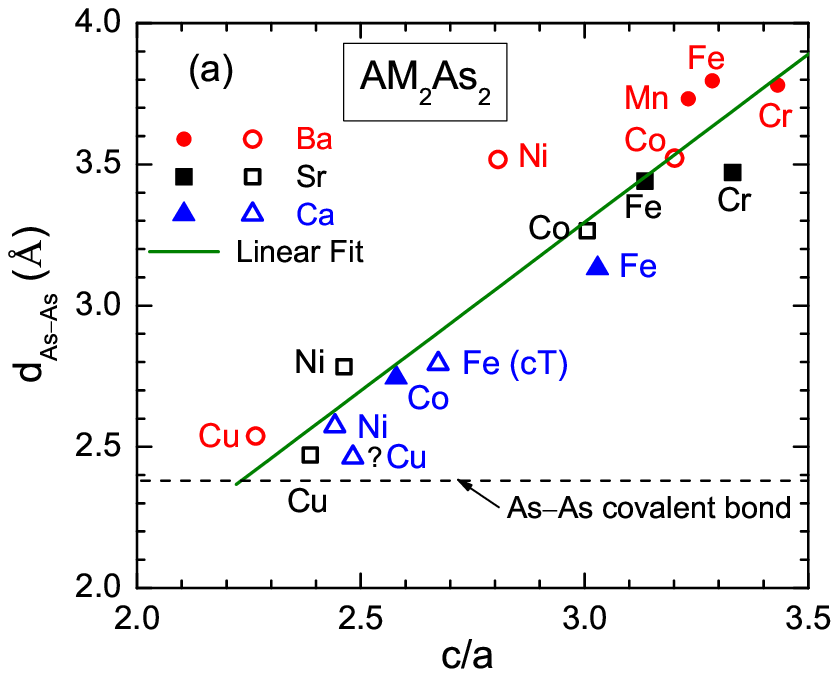}\vspace{0.1in}
\includegraphics[width=2.75in]{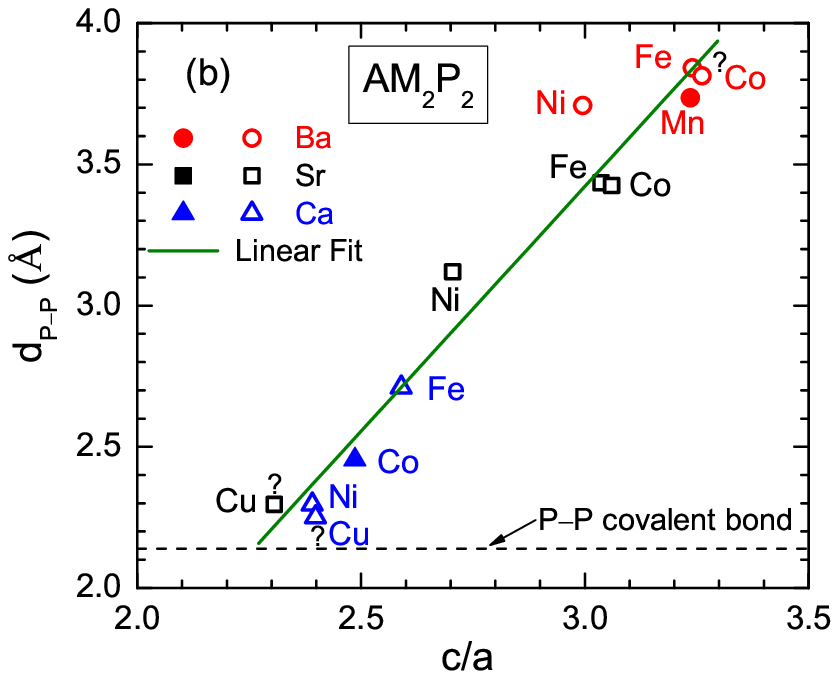}
\caption{(Color online) (a) Interlayer As--As distance $d_{\rm As-As}$ and (b) interlayer P--P distance $d_{\rm P-P}$ versus $c/a$ in $AM_2{\rm As}_2$ and $AM_2{\rm P}_2$ compounds, respectively.  Linear least-squares fits of the data in (a) and (b) by Eq.~(\ref{Eq:dXXvsca}) are plotted as the respective green lines with parameters given in Eqs.~(\ref{FitPars}).  Filled symbols designate  magnetically ordered compounds whereas the open symbols refer to compounds that do not have a magnetically ordered ground state.  The compounds for which the magnetic ground states are not known are marked with ``?''.  The references are given in Table~\ref{tab:AT2X2}.}
\label{fig:fig_AT2X2_2} 
\end{figure}

Several estimates of the (single) covalent bond distances $d_{X-X}$ have been made for $X$ = As and P\@.  In 1988, Von Schnering and H\"onle inferred from a study of polyphosphides that $d_{\rm P-P}=2.23(7)$\,\AA.\cite{VonSchnering1988}  A recent survey in 2008 of the covalent radii of the elements by Cordero et al.\ gave $d_{\rm P-P}=2.14(6)$\,\AA\ and $d_{\rm As-As}=2.38(8)$\,\AA.\cite{Cordero2008}  As will be seen, the difference between the two $d_{\rm P-P}$ values as well as the error bars on any of these three $d_{X-X}$ values are negligible compared with the variations in the interlayer $d_{X-X}$ values across the various $AM_2{\rm (As,P)}_2$ series of compounds.  Here we will use the $d_{\rm P-P}$ and $d_{\rm As-As}$ values of Cordero et al.\ for consistency.

We have compiled in Table~\ref{tab:AT2X2} crystallographic data from our work on ${\rm SrCu_2As_2}$ together with data from the literature on ${\rm ThCr_2Si_2}$-type $AM_2X_2$ compounds ($A$ = Ca, Sr, Ba; $M$ = Cr, Mn, Fe, Co, Ni, Cu) with $X$ = P and As.\cite{Wu, Pfisterer1983, Wang, Yan, Kreyssig2008, Goldman2009, Anand2012, Cheng2012, Pilchowski1990, Saha2011, Jasper2008, Bauer2008, Singh2008, Singh2009, Albenque2010, Sefat2009, Ronning2008, Mewis1980, Raffius1991, Jeitschko1987, Morsen1988, Jia2011, Keimes1997, Ronning2009, Brock1994, Pandey2012, Mine2008}  The $d_{X-X}$ and the $c/a$ ratio for $X = $ As and P are plotted versus $M$ in Figs.~\ref{fig:fig_AT2X2}(a) and~\ref{fig:fig_AT2X2}(b), respectively.  The $d_{\rm P-P}$ and $d_{\rm As-As}$ covalent bond distances of Cordero et al.\cite{Cordero2008} are plotted as the horizontal dashed lines in the respective figures.  The cT phases in each figure are characterized by $d_{\rm As-As}$ or $d_{\rm P-P}$ values that approach the respective interlayer covalent bond distance. 

From Fig.~\ref{fig:fig_AT2X2}, the $d_{X-X}$ and $c/a$ values for the various ${\rm ThCr_2Si_2}$-type compounds are strongly correlated, as previously documented by Jeitschko et al.\ for phosphides.\cite{Jeitschko1985}  The $d_{\rm As-As}$ and $d_{\rm P-P}$ are plotted versus $c/a$ in Figs.~\ref{fig:fig_AT2X2_2}(a) and~\ref{fig:fig_AT2X2_2}(b) for the respective series of $AM_2X_2$ compounds.  An approximate  linear relationship is seen between $d_{X-X}$ and $c/a$ for each class of pnictides, with the phosphides following the linear behavior more precisely.  Linear least squares fits of the data by
\be
d_{X-X} = A + B\,\frac{c}{a}\label{Eq:dXXvsca}
\ee
yield the parameters
\bea
A &=& -0.28\,{\rm \AA},\quad B = 1.19\,{\rm \AA},\quad (X = {\rm As})\label{FitPars}\\*
A &=& -1.78\,{\rm \AA},\quad B = 1.73\,{\rm \AA}.\ \hspace{0.1in}(X = {\rm P})\nonumber
\eea
The fits are shown by the green lines in Figs.~\ref{fig:fig_AT2X2_2}(a) and~\ref{fig:fig_AT2X2_2}(b), respectively.  A plot of $d_{\rm P-P}$ versus $c/a$ showing an approximate linear relationship for a more diverse range of ${\rm ThCr_2Si_2}$-type phosphides is shown in Fig.~1 of Ref.~\onlinecite{Jeitschko1985}.  A plot of $d_{\rm As-As}$ versus $c/a$ for ambient pressure ${\rm (Ca,Sr,Ba)Fe_2As_2}$ and high-pressure cT ${\rm CaFe_2As_2}$ showing a linear relationship was given in Fig.~S1 of  Ref.~\onlinecite{Kimber2009}.

For ${\rm SrCu_2As_2}$, our value $d_{\rm As-As} = 2.470(2)$\,\AA\ is close to the As--As covalent bond distance 2.38(8)~\AA, which confirms that the formal oxidation states of Cu and As atoms in this compound are Cu$^{+1}$ and As$^{-2}$ as discussed above.  The +1 oxidation state of Cu is consistent with our $\rho(T)$, $C_{\rm p}(T)$ and $\chi(T)$ measurements of ${\rm SrCu_2As_2}$ as also previously discussed.  In Sec.~\ref{Conclusion} we suggest that the As--As interlayer covalent bonding in ${\rm SrCu_2As_2}$ may be driven by the high stability of the Cu$^{+1}$ $d^{10}$ electronic configuration rather than the Cu$^{+1}$ oxidation state being a result of the As--As interlayer bonding.

\subsection{\label{Sec:MagPropCorr} Correlations between Pnictogen Interlayer Bonding and Magnetic Properties}

The (Ca,Sr,Ba)$M_2X_2$ compounds that exhibit magnetically ordered ground states are indicated in Fig.~\ref{fig:fig_AT2X2} by a box around them and in Fig.~\ref{fig:fig_AT2X2_2} by filled symbols.  One sees that with the exception of cT phases ${\rm CaCo_2P_2}$ and ${\rm CaCo_2As_2}$ which are  antiferromagnets,\cite{Reehuis1990,Reehuis1998,Anand2012, Cheng2012} magnetic ordering occurs only for the T~phase compounds.  

Reehuis et al.\cite{Reehuis1998} observed for a more extended set of $A{\rm Co_2P_2}$ compounds with the ${\rm ThCr_2Si_2}$ structure, including $A$ = rare earths, that either long-range magnetic ordering of the Co atoms did not occur or FM ordering occurred if there is little or no P--P interlayer covalent bonding, whereas the character of the magnetic ordering changes to AFM if P--P interlayer bonding occurs (these are cT compounds in our terminology).  The latter inference is consistent with our datum for ${\rm CaCo_2P_2}$ in Figs.~\ref{fig:fig_AT2X2} and~\ref{fig:fig_AT2X2_2}.

Kreyssig, Goldman, Pratt et al.\ found that whereas ${\rm CaFe_2As_2}$ exhibits SDW ordering in the T phase as previously reported,\cite{Wu} the pressure-induced transition to the cT phase quenches the magnetic ordering as well as the Fe magnetic moment and associated AFM fluctuations.\cite{Kreyssig2008, Goldman2009, Pratt2009}  Danura et al.\ found in the Ca(Fe$_{1-x}$Rh$_x)_2$As$_2$ system that  after suppressing the SDW with increasing $x$, superconductivity appeared at about 13~K at $x\approx0.02$, but then the superconductivity was quenched at higher $x$ when the compound went into the cT phase.\cite{Danura2011}  Furthermore, they found that converting a superconducting T sample with $x\approx0.02$ to the cT phase under pressure destroyed the superconductivity.  The high-$p$ results of Uhoya et al.\ on ${\rm BaFe_2As_2}$ suggested that the $p$-induced superconductivity with $T_{\rm c} = 34$~K at $p=1$~GPa is suppressed in the higher-$p$ stability range of the cT phase.\cite{Uhoya2010}  These results taken together suggest that the superconductivity in these phases is mediated by spin fluctuations.

The degree of $X-X$ covalent bonding and hence the formal oxidation states of the $M$ atoms in the above $AM_2{\rm As}_2$ compounds are thus directly correlated with their magnetic and superconducting properties.

From first principles calculations, Yildirim  proposed early on that the degree of As--As bonding in ${\rm CaFe_2As_2}$ and the magnitude of the Fe spin are inversely related.\cite{Yildirim2009,Yildirim2009b}  He suggested that the application of pressure on ${\rm CaFe_2As_2}$ causes a reduction in the Fe moment which weakens the Fe-As bonding, and in turn, leads to As--As bonding and hence the cT phase.   This prediction of a $p$-driven transition from a magnetic ground state in the T phase to a nonmagnetic one in the cT phase was confirmed by subsequent first-principles calculations (see, e.g., Ref.~\onlinecite{Tomic2012} in which the effects of nonhydrostatic pressure are also calculated).  Yildirim further suggested that the Fe moment should occur at $p=0$ even in the paramagnetic phase,\cite{Yildirim2009, Yildirim2009b} which was confirmed by the observation of AFM spin fluctuations above the N\'eel temperature $T_{\rm N}$ by inelastic neutron scattering measurements on single crystals of the ${\rm (Ca,Sr,Ba)Fe_2As_2}$ compounds.\cite{Johnston2010, Lumsden2010} 

\subsection{\label{Sec:CuDoping} Transition Metal $M$ Substitutions in Ba(Fe$_{1-x}M_x)_2$As$_2$}

Studies of the effects of partially substituting the Fe atoms in the $A{\rm Fe_2 As_2}$ compounds by other transition metals showed that the long-range SDW ordering and associated lattice distortions in the undoped semimetallic parent compounds had to be largely suppressed before high-$T_{\rm c}$ superconductivity could occur.\cite{Johnston2010}  This was documented, for example, by Canfield and colleagues who determined the $T$-$x$ phase diagrams of Ba(Fe$_{1-x}M_x)_2$As$_2$ ($M$ = Co, Ni, Cu, Pd, Rh) from single crystal studies.\cite{RevCanfield}  They showed that the superconducting transition temperature $T_{\rm c}(x)$, but not the AFM ordering temperature $T_{\rm N}(x)$, approximately overlapped if the $x$-axis is replaced by the average $d$-electron concentration of the transition metal, thus suggesting the importance of this parameter to the superconducting properties.\cite{RevCanfield}  Interestingly, they found that whereas Co and Ni substitutions for Fe induced bulk superconductivity, Cu substitutions did not even though they suppressed the structural and SDW transitions as Co and Ni substitutions do.\cite{RevCanfield}  This has been a conundrum in the FeAs-based superconductivity field.  Angle-resolved photoemission spectroscopy (ARPES) measurements on some of these Ba(Fe$_{1-x}M_x)_2$As$_2$ systems support the contention that such aliovalent $M$-doping increases the electron concentration (see, e.g., the references in Ref.~\onlinecite{Dhaka2011}).

On the other hand, as discussed above, applying pressure to the parent compounds can cause the same changes without doping.  Furthermore, isovalent substitutions of Ru for Fe or P for As in ${\rm BaFe_2 As_2}$ suppress the SDW and induce high-$T_{\rm c}$ superconductivity and thus have $T$-$x$ phase diagrams similar to those of the Ba(Fe$_{1-x}M_x)_2$As$_2$ compounds where $M$ is an aliovalent transition metal.\cite{Johnston2010}  Remarkably, Dhaka et al.\cite{Dhaka2011} found from angle-resolved photoemission spectroscopy (ARPES) measurements on Ba(Fe$_{1-x}$Ru$_x)_2$As$_2$ crystals that the electronic structure near $E_{\rm F}$ does not change with Ru doping.  They speculated that the suppression of the SDW with Ru substitution and concomitant onset of superconductivity arise from a reduction of the Stoner enhancement factor of the static susceptibility at the ordering wave vector.

Various theoretical studies have been carried out to determine whether the carrier concentration changes upon aliovalent transition metal substitutions for Fe in the $A{\rm Fe_2 As_2}$ compounds. Density functional theory (DFT) calculations by Wadati et al.\ of the electronic structures of Ba(Fe$_{1-x}M_x)_2$As$_2$ compounds with $M=$ Co, Ni, Cu, Zn, Ru, Rh and Pd suggested that these dopants do not change the carrier concentration because any additional $d$-electrons are localized on the dopant sites.\cite{Wadati2010}  On the other hand, another DFT study of disordered substitutions of 12.5\% of Co for Fe in BaFe$_2$As$_2$ by Berlijn et al.\cite{Berlijn2012} found that a large chemical potential shift occurs due to the extra $d$-electron of Co as observed in ARPES experiments,\cite{Liu2010,Liu2011,Neupane2011} and that the chemical disorder induces important carrier scattering effects.  They also found that the electron concentration due to the doping tends to pile up around the Co dopant sites as found by Wadati et al.,\cite{Wadati2010} but that these electrons are associated with band states and are not localized.\cite{Berlijn2012}

On the experimental side, local measurements of the electronic environments around the Fe and other atoms in Ba(Fe$_{1-x}M_x)_2$As$_2$  have been carried out.  Fe $K$ near edge x-ray absorption structure (NEXAS) measurements of the electron density around the iron site in Ba(Fe$_{1-x}$Co$_x)_2$As$_2$ indicate that the extra doped \mbox{$d$-electrons} are not concentrated at the Fe sites.\cite{Bittar2011}   Similarly, Merz et al.\ found from $L_{2,3}$ NEXAS measurements on the Fe, Co and As atoms in the Sr(Fe$_{1-x}$Co$_x)_2$As$_2$ system that no observable changes in the Fe, Co and As valences occur on Co-doping.\cite{Merz2012}  Resonant and nonresonant x-ray absorption and emission spectroscopy measurements as well as core and valence level x-ray photoelectron spectroscopy measurements were carried out by McLeod et al.\cite{McLeod2012} on the Fe, Co, Ni and Cu atoms in Ba(Fe$_{1-x}M_x)_2$As$_2$ crystals with $M$ = Co, Ni and Cu.  They found that there is little charge transfer from the dopant atoms to the Fe atoms and that the iron arsenides are not strongly correlated electron systems.  They concluded that the Co and Ni substitutions provide additional conduction electrons, but that the Cu $3d$ electrons are localized in the Cu $3d^{10}$ shell and therefore that Cu-doping reduces the free carrier concentration at the Fermi level.\cite{McLeod2012}  Resonant $L_3$-edge Fe and Co photoelectron spectroscopy measurements of Ca(Fe$_{1-x}{\rm Co}_x)_2$As$_2$ crystals with $x=0.056$ by Levy et al.\ revealed that the Co $3d$ electrons participate in the formation of the Fermi surface, that the Fe and Co atoms have the same +2 oxidation state, and therefore that the extra $d$ electron of Co is donated to the conduction band but whose charge density is associated mainly with the Co dopant atoms.\cite{Levy2012}  The totality of all these local measurements on Co and Fe indicate that the doped electrons of Co are itinerant and participate in the formation of the Fermi surface, but that their spatial densities are located near the Co atoms. 

Ideta et al.\ carried out a comprehensive ARPES study of Ba(Fe$_{1-x}M_x)_2$As$_2$ crystals with $M$ = Ni $(x=0.0375,\ 0.05,\ 0.08)$ and Cu $(x=0.04,\ 0.06,\ 0.08)$ including measurements of both the in-plane $ab$-axis and out-of-plane $c$-axis Fermi surface dispersions and compared the results with previous ARPES measurements on Ba(Fe$_{1-x}$Co$_x)_2$As$_2$.\cite{Ideta2012} They found that volume enclosed by the electron Fermi surface(s) increased and the volume enclosed by the hole Fermi surface(s) decreased with increasing $x$ for Co, Ni and Cu substitutions for Fe, as expected from electron doping in a rigid band picture.  They also found that the $d$-bands associated with the substituted Co, Ni and Cu atoms have increasing binding energy with respect to $E_{\rm F}$, as expected from the increasing impurity potential. However, one also expects in a rigid band approach that the difference $n_{\rm e} - n_{\rm h}$ between the numbers of electron and hole carriers per doped Co, Ni or Cu atom should be equal to the number of extra donated $d$-electrons per atom.  From the difference between the volumes enclosed by the electron and hole Fermi surfaces, they found this prediction to be violated for Ni and Cu dopants, with the itinerant electrons doped per atom by Ni and especially by Cu smaller than the expected values of two and~three, respectively.\cite{Ideta2012}  They suggested that $n_{\rm e} - n_{\rm h}$ is too small for Ni and Cu substitutions either because some of the donated electrons occupy impurity bands that do not cross the Fermi level or they occupy localized states.  When the authors plotted the SDW N\'eel temperature $T_{\rm N}$ and the superconducting $T_{\rm c}$ versus $n_{\rm e} - n_{\rm h}$, they found universal behaviors for both quantities for all three dopants Co, Ni and Cu.  Thus although the rigid band picture is not accurate with respect to electron doping, Ideta et al.\ have identified a quantity that controls both $T_{\rm N}$ and $T_{\rm c}$ in electron-doped ${\rm BaFe_2As_2}$.\cite{Ideta2012}

\subsubsection*{The Divergent Properties of ${\rm SrCu_2As_2}$ and Cu-doped ${\rm BaFe_2As_2}$ and the Role of As--As Interlayer Bonding}

In this paper, we have conclusively demonstrated from physical property measurements that ${\rm SrCu_2As_2}$ is an $sp$-band metal.  We have also demonstrated that As--As interlayer covalent bonding occurs in this compound, which is consistent with the Cu$^{+1}$ oxidation and with the nonmagnetic $d^{10}$ electronic configuration.  In a band picture, the Cu $3d$-bands at $\sim 3$~eV binding energy below $E_{\rm F}$ are fully occupied and the Cu $d$-orbitals contribute little to the bands at $E_{\rm F}$.\cite{Singh}  Therefore, on the basis of our results, the most natural and straightforward expectation for the effect of partially substituting Cu for Fe in the (Ca,Sr,Ba)Fe$_2$As$_2$ compounds is that such Cu substitutions result in strong hole-doping instead of the expected\cite{Ni2010} strong electron doping.  In this case, the hole-doping resulting from replacing Cu for Fe in the $A$Fe$_2$As$_2$ compounds would be qualitatively similar to hole-doping resulting from partial substitutions of Mn or Cr for Fe.  This analogous doping behavior would empirically explain why strongly suppressed and non-bulk superconductivity with $T_{\rm c} \lesssim 2$~K was observed in Cu-substituted BaFe$_2$As$_2$,\cite{Ni2010} since partially substituting Mn or Cr for Fe does not induce superconductivity.\cite{Kasinathan2009, Liu, Kim2010, Sefat2009a, Marty}  The origin of the latter behavior, however, is not understood microscopically at present, especially since hole-doping by partially substituting K for Ba in BaFe$_2$As$_2$ induces superconductivity with $T_{\rm c}$ up to 38~K.\cite{Rotter}  

However, as discussed above, it has been demonstrated both experimentally and theoretically that low-level substitutions of Cu for Fe in ${\rm BaFe_2As_2}$ result in electron doping rather than hole doping.  Furthermore, from Figs.~\ref{fig:fig_AT2X2} and~\ref{fig:fig_AT2X2_2}, interlayer As--As bonding does not occur in BaFe$_2$As$_2$ whereas it does in BaCu$_2$As$_2$.  These differences indicate that the As--As bonding character and associated Cu doping character in Ba(Fe$_{1-x}$Cu$_x)_2$As$_2$ must change with increasing $x$, as possibly suggested from thermoelectric power and Hall effect measurements on single crystals of this system.\cite{Mun2009}  At small $x$, the Cu evidently replaces the $3d^6$ Fe$^{+2}$ atoms as a $3d^9$ Cu$^{+2}$ electron dopant, but at high concentrations would dope as $3d^{10}$ Cu$^{+1}$, which would correspond to a crossover between electron and hole doping with increasing $x$.  However, to accommodate the change in the interlayer As--As bonding from nonbonding to full covalent bonding and considering the very dissimilar chemistry expected for Cu$^{+1}$ and Fe$^{+2}$ with increasing $x$, phase-separation (a miscibility gap) might occur in the Ba(Fe$_{1-x}$Cu$_x)_2$As$_2$ system, as we have demonstrated to occur in the Ba(Fe$_{1-x}$Mn$_x)_2$As$_2$ system.\cite{Pandey2011}  Our preliminary studies of powder samples of Ba(Fe$_{1-x}$Cu$_x)_2$As$_2$ indeed suggest the presence of a miscibility gap in this system for $0.7\lesssim x \lesssim 0.9$.\cite{Anand2012}

\begin{figure}
\includegraphics[width=3.in]{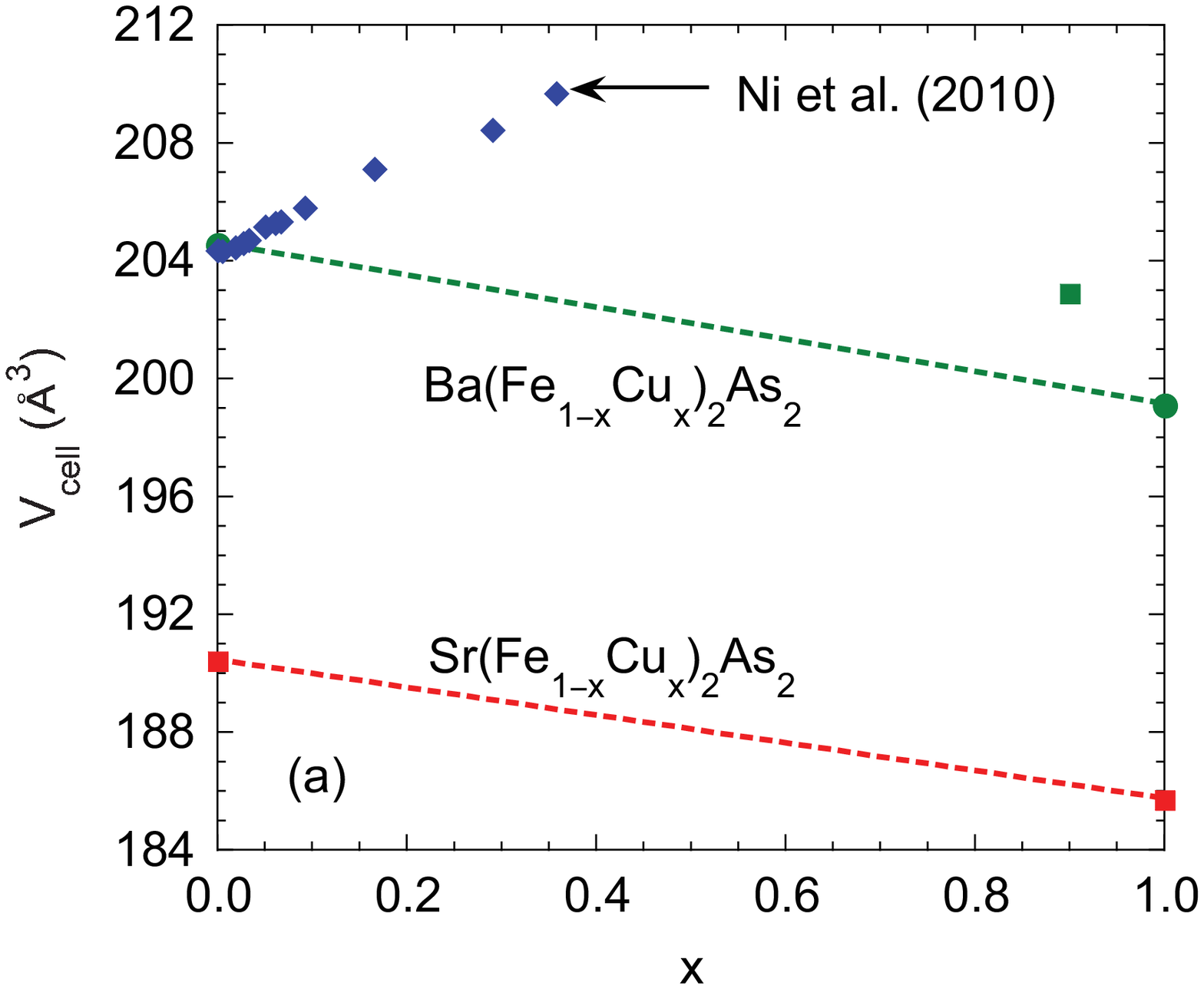}
\includegraphics[width=2.9in]{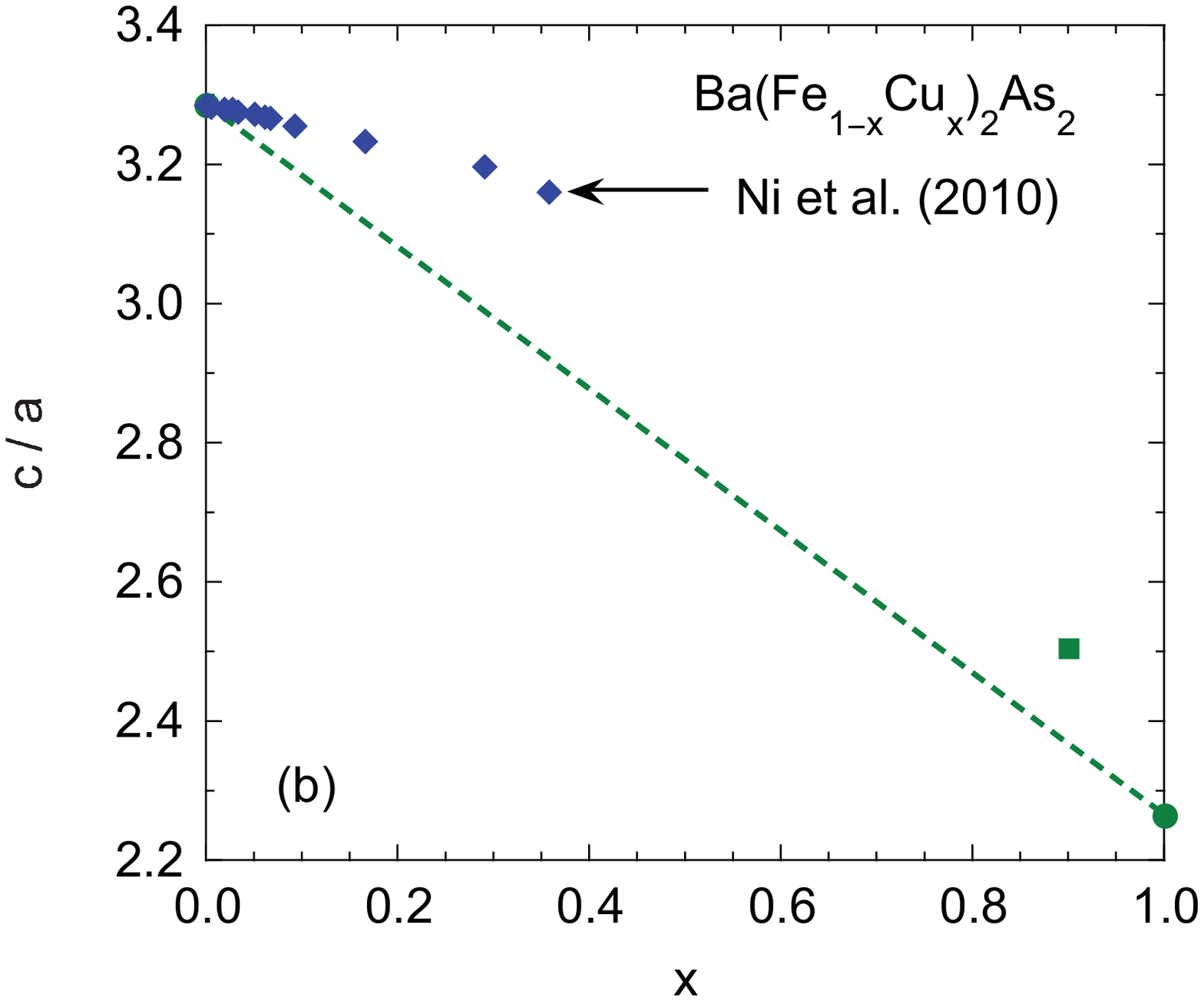}
\caption{(Color online) (a) Unit cell volume $V_{\rm cell}$ versus composition $x$ in Ba(Fe$_{1-x}$Cu$_x)_2$As$_2$ (upper sets of data) and for the two endpoints $x = 0$ and $x=1$ of Sr(Fe$_{1-x}$Cu$_x)_2$As$_2$.  The Ba(Fe$_{1-x}$Cu$_x)_2$As$_2$ data with blue diamond symbols are from Ni et al.\cite{Ni2010} and the green square at $x = 0.9$ is a preliminary result of the present authors.  The remaining data are from the literature.~\cite{Johnston2010} All compounds have the bct ${\rm ThCr_2Si_2}$-type structure.  The two dashed lines represent Vegard's law for the respective series of compounds.  (b) The $c/a$ ratio versus $x$ for the Ba(Fe$_{1-x}$Cu$_x)_2$As$_2$ system.  The data symbols have the same meaning as in (a).}
\label{Fig:Ba(Fe1-xCux)2As2_Vcell}
\end{figure}

Detailed structure refinements versus $x$ of Ba(Fe$_{1-x}$Cu$_x)_2$As$_2$ would shed light on the nature of the As--As bonding and the doping character of Cu, but such structure refinements have not been reported up to now.  However, Ni et al.\cite{Ni2010} reported the lattice parameters $a$ and $c$ and the unit cell volume $V_{\rm cell} = a^2c$ of Ba(Fe$_{1-x}$Cu$_x)_2$As$_2$ versus $x$ up to $x = 0.35$; their results for $V_{\rm cell}$ versus $x$ are shown in Fig.~\ref{Fig:Ba(Fe1-xCux)2As2_Vcell}(a).  Also included is our preliminary datum for a polycrystalline sample with $x = 0.9$ that was quenched from 900\,$^\circ$C, together with literature data for ${\rm BaFe_2As_2}$, ${\rm BaCu_2As_2}$, ${\rm SrFe_2As_2}$ and ${\rm SrCu_2As_2}$.\cite{Johnston2010}  These four compounds have the same tetragonal ${\rm ThCr_2Si_2}$-type structure.  Between the $x=0$ and $x=1$ end points of the Ba and Sr series, one sees from Fig.~\ref{Fig:Ba(Fe1-xCux)2As2_Vcell}(a) that the volume decreases in each series by about the same amount of $\approx 5$~\AA$^3$.  However, the data of Ni et al.\cite{Ni2010} for Ba(Fe$_{1-x}$Cu$_x)_2$As$_2$ show a qualitative deviation from Vegard's law, exhibiting a strong increase in $V_{\rm cell}$ with increasing $x$ instead of the expected decrease.  This anomalous behavior may indicate interesting changes in the As--As bonding, Cu oxidation state and doping behavior versus~$x$.  In view of Fig.~\ref{fig:fig_AT2X2_2}(b), we have plotted $c/a$ versus $x$ for the Ba(Fe$_{1-x}$Cu$_x)_2$As$_2$ system in Fig.~\ref{Fig:Ba(Fe1-xCux)2As2_Vcell}(b).  The trend of the data suggest that interlayer As--As bonds may be beginning to form by $x\sim0.35$. Detailed structural studies are clearly called for to clarify these characteristics versus Cu substitution concentration in FeAs-based ${\rm ThCr_2Si_2}$-type systems.

\section{\label{Conclusion} Summary and Conclusions}

We have presented experimental data on the crystallographic properties in Sec.~\ref{Crystallography} and physical properties in Secs.~\ref{SrCu2As2}--\ref{BaCu2Sb2} of single crystals of SrCu$_2$As$_2$, SrCu$_2$Sb$_2$ and BaCu$_2$Sb$_2$ and of aligned clusters of SrCu$_2$(As$_{0.84}$Sb$_{0.16}$)$_2$ crystals that were synthesized by the self-flux growth method.  Contrasting structures were observed for SrCu$_2$As$_2$ (ThCr$_2$Si$_2$-type), SrCu$_2$Sb$_2$ (CaBe$_2$Ge$_2$-type) and BaCu$_2$Sb$_2$ (a coherent intergrowth of ThCr$_2$Si$_2$-type and CaBe$_2$Ge$_2$-type unit cells, likely with a monoclinic distortion).  The $C_{\rm p}(T)$ and $\rho(T)$ measurements show metallic behaviors for all four compounds.  The $\chi(T)$ data exhibit nearly $T$-independent diamagnetic behaviors.  The $\chi$ of SrCu$_2$As$_2$ is found to be larger in the $ab$-plane than along the $c$-axis, whereas the $\chi$ values of SrCu$_2$Sb$_2$ and BaCu$_2$Sb$_2$ are larger along the $c$-axis. These differences in anisotropy appear to arise from the differences in the crystal structures between these compounds.  No evidence was observed from any of our measurements between 1.8 and~350~K for any phase transitions in any of the four compounds.

In Secs.~\ref{SrCu2As2}--\ref{BaCu2Sb2}, we estimated the different contributions to the intrinsic $\chi$ and analyzed the $C_{\rm p}(T)$ and $\rho(T)$ data within the frameworks of the Debye model of lattice heat capacity and the Bloch-Gr\"{u}neisen model of resistivity, respectively.  The $T$-dependence of $\rho$ is explained by electron-phonon scattering, as expected for an $sp$-band metal.  In particular, there is no clear evidence for a $T^2$ temperature dependence that would have indicated a significant contribution of $d$ orbitals to the density of states at the Fermi energy $E_{\rm F}$.  Furthermore, our data rule out the possibility that the Cu ions are present as Cu$^{+2}$ local magnetic moments with a $3d^9$ electronic configuration and spin $S = 1/2$.

Thus we conclude from the electronic and magnetic properties of these four compounds that they are all nonmagnetic $sp$-metals, as previously predicted by D.~J.~Singh from electronic structure calculations for ${\rm SrCu_2As_2}$ and ${\rm BaCu_2As_2}$ with the ${\rm ThCr_2Si_2}$-type structure.\cite{Singh}  These predictions are that the $3d$ bands of Cu are narrow and are centered about 3~eV below $E_{\rm F}$, so they are completely filled, and there is little admixture of the 3$d$ states into those at $E_{\rm F}$.  In chemical language this corresponds to a formal oxidation state of Cu$^{+1}$ with a nonmagnetic $3d^{10}$ electronic configuration.  As discussed in Sec.~\ref{Sec:PnicBonding}, for ${\rm SrCu_2As_2}$ the Cu$^{+1}$ oxidation state is consistent with the presence of the observed interlayer covalent As--As bonding, which together with the expected oxidation state Sr$^{+2}$ yields As$^{-2}\equiv[{\rm As-As}]^{-4}/2$.

On the other hand, for ${\rm SrCu_2Sb_2}$ with the ${\rm CaBe_2Ge_2}$-type structure, no Sb--Sb interlayer bonding is possible [see Fig.~\ref{fig:structure_fig}(b)] and the compound still shows the same physical properties as ${\rm SrCu_2As_2}$, again indicating a Cu$^{+1}$ formal oxidation state and $3d^{10}$ electronic configuration.  It appears that the concept of formal oxidation state loses its meaning for Sb in ${\rm SrCu_2Sb_2}$ because one cannot derive the Cu$^{+1}$ oxidation state from an assumed oxidation state Sb$^{-2}$ associated with the presence of Sb--Sb bonding.  In both ${\rm SrCu_2As_2}$ and ${\rm SrCu_2Sb_2}$, the nonmagnetic character of Cu and the $d^{10}$ electronic configuration evidently arise because of the stability of this electronic configuration.  In a band picture, the large Cu impurity potential causes the Cu $3d$ bands to be far below $E_{\rm F}$, inhibiting their participation in the electronic conduction.  The above dichotomy between the As and Sb compounds in fact suggests that the As--As covalent bonding observed in ${\rm SrCu_2As_2}$ is driven by the stability of the Cu$^{+1}$ oxidation state.  A band structure calculation for ${\rm SrCu_2Sb_2}$ and a comparison with that of ${\rm SrCu_2As_2}$, together with  measurements of the local electronic configurations of Sr, Cu and Sb in both ${\rm SrCu_2As_2}$ and ${\rm SrCu_2Sb_2}$, would help to clarify these issues.

As discussed in Sec.~\ref{Sec:CuDoping}, a topic that is currently being strongly debated is by how much, and even whether, transition metal substitutions on the Fe sites in $A{\rm Fe_2As_2}$ compounds change the conduction carrier concentration.  Early in the development of the iron arsenide high-$T_{\rm c}$ field, it appeared that substituting Co, Ni and Cu for Fe donates an additional one, two or three electrons to the conduction bands of the system, respectively,\cite{RevCanfield} but that expectation has been contested on several fronts.  Our results on ${\rm SrCu_2As_2}$ show that this compound is a collapsed tetragonal phase with Cu in the Cu$^{+1}$ oxidation state with a $d^{10}$ electronic configuration.  This electron configuration suggests that substituting Cu for Fe in (Ca,Sr,Ba)(Fe$_{1-x}$Cu$_x)_2$As$_2$ should result in hole-doping rather than the observed electron doping.  Thus the electronic character of the Cu dopant and the strength of the As--As interlayer bonding are both expected to drastically change between weakly Cu-substituted BaFe$_2$As$_2$ and pure BaCu$_2$As$_2$, perhaps via a first-order lattice instability such as a miscibility gap in the Ba(Fe$_{2-x}$Cu$_x)_2$As$_2$ system.  Detailed studies of the crystallography of (Ca,Sr,Ba)(Fe$_{1-x}$Cu$_x)_2$As$_2$ versus $x$ would help to clarify the changes in the As--As interlayer bonding and the nature of the carrier doping by Cu with increasing $x$.

\acknowledgments

We thank Paul Canfield, Rajendra Dhaka and David Singh for helpful comments.  This research was supported by the U.S. Department of Energy, Office of Basic Energy Sciences, Division of Materials Sciences and Engineering. ÊAmes Laboratory is operated for the U.S. Department of Energy by Iowa State University under Contract No.~DE-AC02-07CH11358.

\end{document}